\documentclass[11pt,a4paper]{article}                           

\usepackage{amsmath}
\usepackage{amssymb}
\usepackage[usenames]{color} 
\usepackage{multirow}
\usepackage{graphicx}
\usepackage{epstopdf}
\usepackage{array}
\usepackage{hhline}
\usepackage{cite}
\usepackage{microtype,colortbl}
\usepackage[bf]{caption}
\usepackage{color}
\usepackage[usenames,dvipsnames]{xcolor}
\usepackage{subcaption}
\usepackage{pstool}
\usepackage{bbold}

\usepackage{ifpdf}
\ifpdf
  \usepackage[pdftex,
    pdftitle={},
    pdfauthor={},
    pdfsubject={},
    bookmarksopen, bookmarksnumbered, bookmarksopenlevel=2]{hyperref}
\fi
\def\hybrid{\topmargin -20pt    \oddsidemargin 0pt
        \headheight 0pt \headsep 0pt
        \textwidth 6.25in       % A4 paper
        \textheight 9 in       % A4 paper
        \marginparwidth .875in
        \parskip 5pt plus 1pt 
          \jot = 1.5ex
   }
\hybrid
\numberwithin{equation}{section}
\numberwithin{table}{section}

\newcommand{\beq}{\begin{equation}}  \newcommand{\eeq}{\end{equation}}
\newcommand{\bal}{\begin{aligned}}   \newcommand{\eal}{\end{aligned}}
\newcommand{\bea}{\begin{eqnarray}}  \newcommand{\eea}{\end{eqnarray}}

\newcommand{\bmat}{\left(\begin{array}}
\newcommand{\emat}{\end{array}\right)}

\newcommand{\bbC}{\mathbb{C}}

\newcommand{\nn}{\nonumber}

\newcommand{\cO}{\mathcal{O}}

\newcommand{\cP}{\mathcal{P}}
\newcommand{\cC}{\mathcal{C}}

\newcommand{\cN}{\mathcal{N}}

\newcommand{\cH}{\mathcal{H}}

\newcommand{\cI}{\mathcal{I}}
\newcommand{\cJ}{\mathcal{J}}

\newcommand{\cM}{\mathcal M}
\newcommand{\cQ}{\mathcal Q}

\newcommand{\I}{\text{Im}}
\newcommand{\R}{\text{Re}}

\usepackage{cancel}

\newcommand{\be}{\begin{equation}}
\newcommand{\ee}{\end{equation}}

\newcommand{\half}{\frac{1}{2}}

\definecolor{Gray}{gray}{0.95}

\begin{document}

\baselineskip=14pt
\parskip 5pt plus 1pt

\vspace*{-1.5cm}
\begin{flushright}    % Publication numbers
  {\small MPP-2018-20
  
  }
\end{flushright}

\vspace{2cm}
\begin{center}        % Main title
  {\LARGE Infinite Distances in Field Space and Massless Towers of States}
\end{center}

\vspace{0.5cm}
\begin{center}        % Authors
{\large  Thomas W.~Grimm$^{1}$, Eran Palti$^2$, Irene Valenzuela$^{1}$}
\end{center}

\vspace{0.15cm}
\begin{center}        % Institutes
\emph{$^1$ Institute for Theoretical Physics \\
Utrecht University, Princetonplein 5, 3584 CE Utrecht, The Netherlands}\\[.3cm]
  \emph{$^2$Max-Planck-Institut f\"ur Physik (Werner-Heisenberg-Institut), \\
Fohringer Ring 6, 80805 Munchen, Germany}
             \\[0.15cm]
 
\end{center}

\vspace{2cm}

%%%%%%%%%%%%%%%%%%%%%%%%%%%%%%%%%%%%%%%%%%%%%%%
%%%%%%%%%%%%%%%%%%%%%%%%%%%%%%%%%%%%%%%%%%%%%%%
%%%%%%%%%%%%%%%%%%%%%%%%%%%%%%%%%%%%%%%%%%%%%%%
%%%%%%%%%%%%%%%%%%%%%%%%%%%%%%%%%%%%%%%%%%%%%%%
%%%%%%%%%%%%%%%%%%%%%%%%%%%%%%%%%%%%%%%%%%%%%%%
%%%%%%%%%%%%%%%%%%%%%%%%%%%%%%%%%%%%%%%%%%%%%%%
%%%%%%%%%%%%%%%%%%%%%%%%%%%%%%%%%%%%%%%%%%%%%%%
%%%%%%%%%%%%%%%%%%%%%%%%%%%%%%%%%%%%%%%%%%%%%%%

\begin{abstract}
\noindent
It has been conjectured that in theories consistent with quantum gravity infinite distances in field space coincide with an infinite tower of states becoming massless exponentially fast in the proper field distance. The complex-structure moduli space of Calabi-Yau manifolds is a good testing ground for this conjecture since it is known to encode quantum gravity physics. We study infinite distances in this setting and present new evidence for the above conjecture. Points in moduli space which are at infinite proper distance along any path are characterised by an infinite order monodromy matrix. We utilise the nilpotent orbit theorem to show that for a large class of such points the monodromy matrix generates an infinite orbit within the spectrum of BPS states. We identify an infinite tower of states with this orbit. Further, the theorem gives the local metric on the moduli space which can be used to show that the mass of the states decreases exponentially fast upon approaching the point. We also propose a reason for why infinite distances are related to infinite towers of states. Specifically, we present evidence that the infinite distance itself is an emergent quantum phenomenon induced by integrating out at one-loop the states that become massless. Concretely, we show that the behaviour of the field space metric upon approaching infinite distance can be recovered from integrating out the BPS states. Similarly, at infinite distance the gauge couplings of closed-string Abelian gauge symmetries vanish in a way which can be matched onto integrating out the infinite tower of charged BPS states. This presents evidence towards the idea that also the gauge theory weak-coupling limit can be thought of as emergent. 
\end{abstract}

\thispagestyle{empty}
\clearpage

\setcounter{page}{1}

%%%%%%%%%%%%%%%%%%%%%%%%%%%%%%%%%%%%%%%%%%%%%%%
%%%%%%%%%%%%%%%%%%%%%%%%%%%%%%%%%%%%%%%%%%%%%%%
%%%%%%%%%%%                 %%%%%%%%%%%%%%%%%%%
%%%%%%%%%%%  DOCUMENT BODY  %%%%%%%%%%%%%%%%%%%
%%%%%%%%%%%                 %%%%%%%%%%%%%%%%%%%
%%%%%%%%%%%%%%%%%%%%%%%%%%%%%%%%%%%%%%%%%%%%%%%
%%%%%%%%%%%%%%%%%%%%%%%%%%%%%%%%%%%%%%%%%%%%%%%
%%%%%%%%%%%%%%%%%%%%%%%%%%%%%%%%%%%%%%%%%%%%%%%

\newpage

\tableofcontents

\newpage

%%%%%%%%%%%%%%%%%%%%%%%%%%%%%%%%%%%%%%%%%%%%%%%
\section{Introduction}
\label{sec:intro}
%%%%%%%%%%%%%%%%%%%%%%%%%%%%%%%%%%%%%%%%%%%%%%%

Quantum field theory and gravity are notoriously difficult to combine at high energy scales close to the Planck mass $M_p$. However, at low energies, there might appear to be no consistency constraints limiting which effective quantum field theories can be coupled to gravity.\footnote{Mixed gauge-gravitational anomalies providing a notable exception.} This apparent freedom is deeply tied to the difficulty of obtaining universal predictions from string theory. In recent years there has been significant interest in proposals for such consistency constraints on effective field theories that can be coupled to quantum gravity. Quantum field theories which violate such constraints are termed to be in the Swampland \cite{Vafa:2005ui}. The most studied such proposed constraint is the Weak Gravity Conjecture \cite{ArkaniHamed:2006dz}. A different constraint, proposed in \cite{Ooguri:2006in}, is that in an effective quantum field theory that can can arise from string theory and therefore can be consistently coupled to quantum gravity, infinite distances in moduli space lead to an infinite tower of states becoming massless exponentially fast in the proper field distance. So if we consider two points in field space $P$ and $Q$, with a geodesic proper distance between them of $d\left(P,Q\right)$, then there should exist an infinite tower of states with characteristic mass scale $m$ such that 
\be
\label{SC}
\frac{m\left(P\right)}{m\left(Q\right)} \rightarrow e^{-\gamma d\left(P,Q\right)} \mathrm{\;as\;} d\left(P,Q\right) \rightarrow \infty \;.
\ee
Here $\gamma$ is some positive constant which depends on the choice of $P$ and $Q$ but which is not specified in generality. The conjecture (\ref{SC}) was referred to as the Swampland Conjecture in \cite{Klaewer:2016kiy}. Since there are more conjectures appearing recently to distinguish between the string landscape and the swampland, here we will rename this specific conjecture as the Swampland Distance Conjecture (SDC) to avoid confusion. This conjecture will form the focus of this paper.  One of the consequences of the conjecture is a limit on moduli space distances within any effective field theory which is consistent with string theory and has a finite cut-off. It is therefore of both formal and conceptual interest and of potential phenomenological importance in the context of large field inflation. 

The evidence for the conjecture is primarily based on case-by-case examples in string theory. There  is some evidence, which does not rely on string theory, relating the Swampland Distance Conjecture to the Weak Gravity Conjecture \cite{Palti:2017elp} and to black hole physics \cite{Klaewer:2016kiy}. It is also worth noting that the evidence appears to support a stronger statement, that the exponential behaviour of the mass of the states is reached at finite proper distance of order the Planck mass and that it holds for any scalar field not just moduli. This was denoted as the Refined Swampland (Distance) Conjecture in \cite{Klaewer:2016kiy}. In \cite{Palti:2015xra,Baume:2016psm} the behaviour of so-called closed-string monodromy axions in type IIA string theory was shown to be consistent with this stronger statement. Further evidence for the conjecture was found in \cite{Valenzuela:2016yny,Bielleman:2016olv,Blumenhagen:2017cxt} in the context of studying open-string monodromy axions, although the breakdown of the effective theory there manifests in a more subtle way. In \cite{Palti:2017elp} the Swampland Distance Conjecture was shown to hold for string moduli in the large volume or large complex-structure regime of Calabi-Yau compactifications for certain paths in moduli space defined by the variation of only a linear combination of the moduli. In \cite{Hebecker:2017lxm} further evidence was presented in the context of closed string axions belonging to the complex structure sector of certain Type IIB string theory flux compactifications. In \cite{Cicoli:2018tcq} a similar bound was found for the reduced K\"ahler moduli space obtained from Type IIB compactified on a certain type of Calabi-Yau threefolds. Further studies, over the full complex-structure moduli space of type IIB Calabi-Yau compactifications, will be reported in \cite{ralph}.

In this paper we will adopt a general approach to studying the Swampland Distance Conjecture where we do not rely on explicit example compactifications but rather on general properties of a large and rich class of moduli spaces in string theory. We will consider the complex-structure moduli spaces of Calabi-Yau (CY) manifolds in compactifications of type IIB string theory. These moduli spaces are excellent testing grounds for aspects of quantum gravity as they are known to encode highly non-trivial quantum gravity physics in their geometry. A CY complex-structure moduli space also has a very rich structure of loci that are at infinite distance. By this we mean points in the moduli space which are at infinite proper distance, as measured by the metric on the moduli space, along any path. There are also substantial mathematical tools for studying these moduli spaces which will allow us to show general results rather than a case-by-case analysis.  

Furthermore, most of the recent work has been focused on the parametric behaviour of the field metric, but very little is known about the nature of the tower of states becoming light. We will also focus on studying the properties of this tower of states, providing a candidate set of stable states which become massless at infinite distances. Specifically, we will consider the tower to be formed of BPS states, which in type IIB are D3-branes wrapping special Lagrangian three-cycles. Once this tower is identified, we will show that the exponential mass behaviour of the Swampland Distance Conjecture can be proven in generality due to a powerful mathematical theorem, termed the Nilpotent Orbit Theorem \cite{schmid}, which, among other things, gives a general expression for the asymptotic infinite distance form of the field space metric. Identifying the tower requires an understanding of the BPS state spectrum upon approaching infinite distance. The infinite distance point is singular and there is a monodromy upon circling it. It can be mathematically proven that this monodromy must be of infinite order. We will propose to identify an infinite tower of states by using the monodromy transformation acting on the states of the theory upon circling the infinite distance locus. By introducing significant further mathematical technology, particularly relating to Mixed Hodge Structures, we will be able to identify this tower quite precisely. 
Our analysis will be performed completely generally, for any CY moduli space and at any point in that moduli space. But we will restrict to one-parameter degeneration models, which means that the point of interest will belong to only one singular divisor, leaving more complicated configurations for future work. Because these are rich field spaces, possibly involving hundreds of coupled scalar fields, the analysis necessitates powerful mathematical machinery.  A significant part of the paper will therefore be dedicated to introducing these tools and how they can be used in this context. 

Our results can be summarised as follows. A locus of infinite distance in moduli space is labelled by an integer $d$ which can take the values $1$, $2$ or $3$. For $d=3$ loci we will identify quite precisely and generally a tower of BPS states which become massless exponentially fast in the proper distance. This is one of the central results of the paper. For $d< 3$ loci we will also propose candidates for the tower of BPS states. However, proving the existence of this tower can not be done with the same generality as for $d=3$ due to dependence on the global structure of the moduli space. Studying the generality of the results for such cases will require further work.

We will also provide evidence for a proposal for the underlying reason as to why the Swampland Distance Conjecture holds. We will show that integrating out the tower of BPS states induces a logarithmic distance divergence in the moduli space. Since it is well known that CY moduli spaces are quantum in nature, so that they already have integrated out the BPS states of wrapped branes, this divergence is naturally identified with the infinite distance in the moduli space. We therefore propose that this could be a general phenomenon, that infinite distances are quantum in nature and emerge from integrating out an infinite number of states.\footnote{This possibility was first mentioned in \cite{Ooguri:2006in}, and also a similar proposal was reached independently in \cite{HRR-toappear}.} Interestingly, the logarithmic divergence in the proper field distance requires that the number of stable BPS states grows as we approach the singularity, becoming infinite at infinite distance. By studying the distribution of walls of marginal stability for BPS states, we will show that the tower of states induced by the monodromy transformation exhibits precisely the right rate of increase in the stable states to match onto the integrating out requirements. It also implies that the cut-off due to quantum gravity physics decreases when we approach the singularity at a rate which coincides with the species bound relating to the tower of BPS states.

Our results also have natural interpretations relating to other general ideas about quantum gravity. We will show in generality that infinite distances are loci in field space where a global symmetry emerges. Since the effective theory entirely breaks down at the infinite distance singularities, the emergence of these global symmetries is blocked by string theory.  

Because the complex-structure moduli are in vector multiplets our results have a natural connection to the Weak Gravity Conjecture. We will show that at infinite distance the gauge couplings of the Abelian gauge fields in the vector multiplets vanish exponentially fast in the proper distance. This matches the proposal in \cite{Klaewer:2016kiy}. We will show that this behaviour can be recovered in detail in terms of integrating out a tower of charged BPS states. It therefore presents evidence that also the weak coupling limit is emergent in the same way as the infinite field distance. This emergence property will naturally tie into the Weak Gravity Conjecture. This matches general ideas proposed in \cite{Heidenreich:2017sim}.

The paper is organised as follows. In section \ref{sec:infinite_distance} we introduce the mathematical technology required to analyse infinite distance points in the moduli space of Calabi-Yau manifolds. We will introduce the Nilpotent Orbit Theorem of Schmid \cite{schmid} and how it can be used to characterise infinite distance loci and study them generally. In particular, we will introduce the relation between infinite distance loci and infinite order monodromy transformations about those loci. In section \ref{sec:BPS_states} we will introduce some relevant results about BPS states. We will discuss the relation between the monodromy transformation and the spectrum of BPS states. In particular, we will introduce the notion of an infinite monodromy orbit of massless BPS states, and show that this orbit forms a primary candidate for a subset of the spectrum of states which become massless on the monodromy locus. In section \ref{sec:orbits} we will introduce the technology of Mixed Hodge Structures and their utilisation in the $Sl_2$-orbit theorem of Schimd \cite{schmid}. This will then allow us to study when an infinite monodromy orbit through massless BPS states exists and to identify it quite precisely. We will present general results on this, and also study some particular examples. In sections \ref{sec:integrating_out} and \ref{sec:WGC} we will discuss some of the physics associated to our results. In particular, the relation between integrating out states, infinite distances, gauge couplings and global symmetries, as described above. We will also discuss the relation of the Swampland Distance Conjecture to the Weak Gravity Conjecture and the idea that they are both implied by the emergent nature of infinite field distance and weak gauge couplings. Finally, section \ref{sec:conclusions} contains our conclusions.

\section{Infinite distance divisors in Calabi-Yau moduli space} \label{sec:infinite_distance}

\begin{figure}[h!]
\begin{center} \includegraphics[width=6cm]{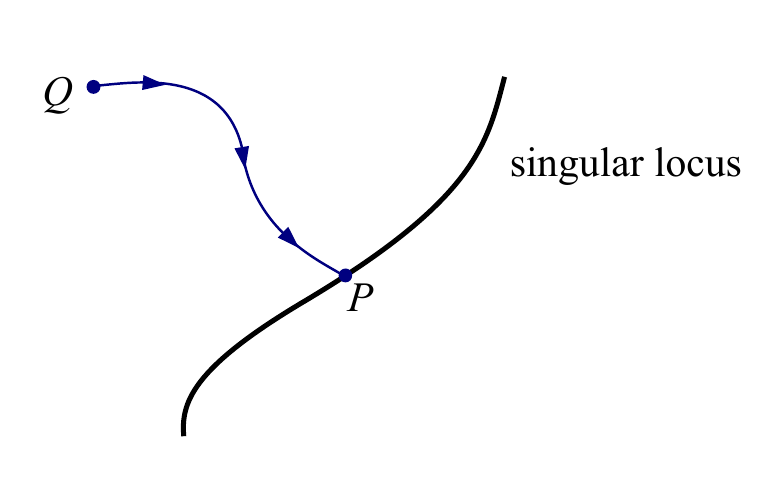} 
\vspace*{-.5cm}
\end{center}
\caption{Smooth path connecting a regular point $Q$ to a singular point $P$ which might be at infinite distance in moduli space.} \label{pathtodiv}
\end{figure}

In this section we introduce the mathematical concepts that allow us to study points in moduli space that are at infinite geodesic distance with respect to some specific metric $g$. 
%In this section we revisit the necessary conditions to have points in moduli space that are at infinite geodesic distance with respect to some specific metric $g$. 
We denote a point of infinite distance as one for which all paths $\gamma$ to such a point are infinitely long when measured with the metric $g$. 
Hence we want to make statements about the length of any smooth path $\gamma$ connecting $P,Q$
 given by
\beq \label{length_d}
    d_\gamma (P,Q) = \int_\gamma \sqrt{g_{IJ} \dot x^I  \dot  x^J } ds \ ,
\eeq
where $x^I(s)$ embeds the path and $\dot x^I = \partial x^I /\partial s$.
The key point will be to translate the information about being at infinite distance into a 
more algebraic statement. Firstly, we note that infinite geodesic distances can only occur when connecting 
a path to a singular point $P$ in moduli space as indicated in Figure \ref{pathtodiv}.  
 Secondly, we will see that such points are characterised by the 
existence of an infinite order monodromy matrix $T$ and  
by the action of the logarithm $N = \log (T)$ of the monodromy matrix  acting on the unique 
holomorphic three-form at this point. This will allow us to identify the universal asymptotic behaviour of the field metric $g$ when approaching such infinite distance points.

\subsection{Complex structure moduli space and monodromy}

To start with we recall some basic facts about the complex structure moduli 
space $\cM_{\rm cs}$ and introduce its natural metric, the Weil-Petersson metric $g_{\rm WP}$.
The complex structure moduli space for a Calabi-Yau manifold $Y_D$ of complex dimension $D$
is a $h^{D-1,1}(Y_D)$-dimensional K\"ahler manifold. Locally, it can be parametrised 
by coordinates $z^I$, $I=1,\ldots,h^{D-1,1} (Y_D)$, which are often called the 
complex structure deformation moduli. 
The metric on $\cM_{\rm cs}$ is determined by the holomorphic 
$(D,0)$-form $\Omega$. The metric $g_{\rm WP}$ 
is K\"ahler and locally obtained from the K\"ahler potential \cite{Andrianopoli:1996cm,Craps:1997gp}
\beq \label{Kpot_cs}
    K = - \log \Big[- i^{D} \int_{Y_D} \Omega \wedge \bar \Omega \Big]\ ,
\eeq
i.e.~one finds that $g_{\rm WP}$ has components $g_{I \bar J} = \partial_{z^I} \partial_{\bar z^J} K$.

The holomorphic $(D,0)$-form  $\Omega$ can be expanded into an appropriate real integral basis $\gamma_\cI$. 
It is a non-trivial task to identify such an `appropriate' integral basis $\gamma_\cI$. We refer to the literature discussing  
Calabi-Yau threefold and fourfolds for more details on its construction.
Furthermore, 
one can show that $\Omega$ depends holomorphically on the coordinates $z^I$. Hence, we write 
\beq   \label{Omega-exp}
    \Omega = \Pi^\cI(z) \, \gamma_\cI \equiv \mathbf{\Pi}^T \boldsymbol{\gamma} \ , \qquad \Pi^\cI = \int_{\Gamma_\cI} \Omega \ ,
\eeq
where $\gamma_\cI$ integrates to $\delta_\cI^\cJ$ over the cycle $\Gamma^\cJ$. The holomorphic functions 
$\Pi^I$ are called the \textit{periods} of $\Omega$. 
In order to rewrite intersection products it will be also convenient to introduce 
the intersection matrix $\eta$ with components
\beq \label{def-eta}
   \eta_{\cI\cJ} = \int_{Y_D} \gamma_\cI \wedge \gamma_\cJ \;,
\eeq
which is anti-symmetric for $D$ odd and symmetric for $D$ even. 
In Calabi-Yau threefolds, i.e.~$D=3$, the matrix $\eta_{\cI\cJ}$ is anti-symmetric and the basis $\gamma_\cI$ can be chosen to 
be symplectic. Hence, we can pick 
\beq \label{symplectic-basis}
   \gamma_{\cI} = (\alpha_L, \beta^K) \ ,\qquad  \int_{Y_3} \alpha_L \wedge \beta^K = \delta_L^K \ , \qquad \int_{Y_3} \alpha_L \wedge \alpha_K = \int_{Y_3} \beta^L \wedge \beta^K = 0 \ . 
\eeq
Let us stress that 
the coordinates $z^I$, periods $\Pi(z)^\cI$, and the basis $\gamma_I$ are adapted to the considered patch in 
$\cM_{\rm cs}$ and can very non-trivially change when moving to different regions in $\cM_{\rm cs}$.
With this definitions at hand we can write \eqref{Kpot_cs} as
\beq \label{Kpot_cs2}
  K = - \log \big[- i^{D}  \mathbf{\Pi}^T \eta  \mathbf{\bar \Pi} \big]\ . 
\eeq

It is crucial for our considerations that the complex structure moduli space $\cM_{\rm cs}$  is not generally 
smooth, but will admit special singular points. 
These can always be made to lie on divisors that intersect normally.\footnote{To be mathematically more precise, it was shown \cite{mumford}
that one can resolve the moduli space such that all special points are on divisors that intersect normally.} The periods $\mathbf{\Pi}$ are in fact multi-valued and experience 
monodromies along paths encircling such special divisors. 
To make this more precise, let us introduce local coordinates $z^I$, such that the considered singular divisor is given 
by $z^j = 0$ for some $j \in \{1,\ldots , h^{D,1}(Y_D)\} $. Note that we can consider several intersecting divisors. 
We encircle $z^j = 0$ by sending $z^j \rightarrow e^{2\pi i} z^j$. In general the 
periods will non-trivially transform with a matrix $T_i$ under this identification 
\beq \label{monodromy-trafo}
   \mathbf{\Pi}( ... ,e^{2\pi i} z^j,... ) = T_{j} \ \mathbf{\Pi}(...,z^j,...)\ .   
\eeq
Two facts about the 
$T_j$ will be important for us in the next sections \cite{landman,schmid}
\begin{eqnarray} 
 &\text{each\ } T_j \ \text{ is quasi-unipotent}: &\quad \exists \, m_j,n_j \in \mathbb{Z}: \quad (T^{m_j+1} - \text{Id} )^{n_j+1} = 0\ , \label{quasiuni} \\
& T_j \text{ locally arising at a point commute}: &\quad    [T_i , T_j] = 0 \ .   \label{commute}
\end{eqnarray}

Collecting all such $T_i$ throughout the moduli space $\cM_{\rm cs}$ one obtains a group $\Gamma$ known
as the monodromy group.\footnote{Strictly speaking the monodromy group can be defined by considering a certain representation 
$T_{I}^J$ of $\pi_1(\cM_{\rm cs}) $ acting on the period vectors.} In general, the elements of $\Gamma$ will not 
commute, and \eqref{commute} only holds for the elements $T_i$ at a point in a higher-dimensional moduli space. 
However, we will not need a detailed global understanding of the moduli space 
and therefore restrict most of our discussion to a local patch around such special points. 
It is important to 
note that the `infinite distance' in the metric $g_{\rm WP}$ will be picked up in such a local patch if the special point satisfied certain criteria. 
This will be discussed in the next subsections.

\subsection{The local K\"ahler potential and a necessary condition for infinite distance}

Our next goal is to find a local expression for the K\"ahler potential \eqref{Kpot_cs2} near the special points in 
moduli space with non-trivial monodromy matices $T_i$. 
In order to do that we again introduce an appropriate set $z^I$ of local coordinates. 
We first split them into two types  
\beq \label{local-coords}
   z^I = (z^j, \zeta^M)\ , 
\eeq
such that the special divisors are, as above, locally given by $z^j=0$.
The complex coordinates $\zeta^M$ will be included to keep the situation 
general. We will be interested in the point $P$ given by 
\beq \label{def-P}
   P :  \quad z^j = 0, \ \zeta^M = 0\ ,
\eeq
and expand the K\"ahler potential \eqref{Kpot_cs2} around this point. By definition 
this point lies on the special singular divisors $z^j= 0$. Here $\zeta^M$ can take any value, so we have chosen $\zeta^M = 0$ without loss of generality.

The coordinates $z^j$ are not yet convenient for our purposes. The reason 
for this is that part of the monodromy matrix $T_i$ introduced when circling 
$z^j \rightarrow e^{2\pi i} z^j$ as in \eqref{monodromy-trafo} will play no important role when 
evaluating the distance. To identify this part, we note that the 
property \eqref{quasiuni} implies that each $T_i$ can be decomposed as 
\beq
    T_i = T^{(s)}_i \cdot T^{(u)}_i\ , \label{Tdecom}
\eeq
with $T^{(s)}_i$ and $T^{(u)}_i$ having the following special properties. 
Each matrix $T^{(s)}_i$ is of finite order, i.e.~there exists an integer $m_i$ such that 
$(T^{(s)}_i)^{m_i-1}\neq \text{Id}$ and $(T^{(s)})^{m_i} = \text{Id}$. In contrast, for each matrix $T^{(u)}_{i}$ 
is unipotent, i.e.~$T^{(u)}_i$ is either the identity matrix or there exists an integer $n_i>0$ such that 
\beq \label{unipotent}
   (T^{(u)}_i-\text{Id})^{n_i}\neq 0\ , \qquad  (T^{(u)}_i-\text{Id})^{n_i+1}= 0\ .
\eeq
If $T^{(u)}_i$ is not the identity an unipotent matrix will be of infinite order, i.e.~there exists no $k_i$ such that 
$T_i^{k_i} = T_i$. This property will be of crucial importance below. 
The precise form of $T^{(s)}_i$ will not be relevant in the 
local expression of the K\"ahler potential. To avoid including factors of $m_i$ all over the 
place we therefore redefine
\beq
   z^j \ \rightarrow (z^j)^{m_j} \ ,
\eeq
but name, by an abuse of notation, the resulting coordinate $z^j$.\footnote{In mathematical terms this transformation 
corresponds to a base change.} In terms of this new coordinate, the monodromy matrix is only given by its infinite order part.

We are now ready to display the local form of the periods and the K\"ahler potential. To do that 
we first define 
\beq \label{def-Ni}
    N_i = \log T^{(u)}_i  = \sum_{k=1}^{\infty} (-1)^{k+1} \frac{1}{k} (T^{(u)}_i -\text{Id})^k= \frac{1}{m_i} \log T_i^{m_i}\ ,
\eeq 
which should be read as a matrix equation. One can easily check that the so-defined matrix $N_i$ is nilpotent with 
\beq 
    N_i^{n_i} \neq 0 \ , \qquad N_i^{n_i+1}=0\ , \label{nidef}
\eeq
where $n_i$ was already introduced in \eqref{unipotent}. Using these nilpotent matrices it was shown 
by Schmid \cite{schmid} that locally around the point $P$ given in \eqref{def-P} the periods take the form
\beq \label{nilpot_orbit}
\boxed{  \rule[-.7cm]{0cm}{1.4cm} \quad \mathbf{\Pi}(z,\zeta) =  \text{exp} \Big[ \sum_j \frac{1}{2\pi i} (\log z^j ) N_j  \Big] \mathbf{A}(z,\zeta)\ , \quad}
\eeq
with $\mathbf{A}$ being holomorphic at $P$. In other words, the non-trivial part of this statement is 
that a crucial part of the information about the singularity is in the matrices $N_j$. The vector $\mathbf{A}(z,\zeta)$
is regular at $P$ and admits an expansion
\beq
    \mathbf{A}(z,\zeta) = \mathbf{a}_0(\zeta) +  \mathbf{a}_{j} (\zeta)  z^j +  \mathbf{a}_{jk} (\zeta)  z^j z^k+  \mathbf{a}_{jkl} (\zeta)  z^j z^k z^l + \ldots 
\eeq
The name `nilpotent orbit' refers to the approximation
$\text{exp} \Big[ \sum_j \frac{1}{2\pi i} (\log z^j ) N_j  \Big] \mathbf{a}_0(\zeta)$ of the full period $\mathbf{\Pi}(z,\zeta)$.\footnote{Schmid gives an estimate how well this expression approximates the full period.}

To display the K\"ahler potential we note that the monodromies $T_i$ preserve $\eta$, i.e.~$T_i^T \eta T_i = \eta$, and therefore 
one has
\beq \label{property_Neta}
    N^T_i \eta = - \eta N_i\ . 
\eeq
 Using this fact and the expansion \eqref{nilpot_orbit} the K\"ahler potential \eqref{Kpot_cs2} takes the form 
\beq  \label{e-K1}
   e^{-K} =- i^{D}  \mathbf{A} ^T \eta \ \text{exp} \Big[- \sum_j \frac{1}{2\pi i} (\log |z^j|^2 ) N_j  \Big]   \mathbf{\bar A} \ . 
\eeq
To write this in a more practical form we define 
\beq \label{t-def}
   t^j = \frac{1}{2\pi i} \log z^j\ .
\eeq
Clearly, this redefinition implies that the point under consideration is now given by 
\beq
     P :  \quad t^j = i \infty, \ \zeta^M = 0\ .
\eeq
In these coordinates all $t$-dependence in $\mathbf{A}$ arises via exponentials $e^{2\pi i t}$. 
There are, however, polynomial terms in $t - \bar t$ arising in $e^{-K}$ due to the exponential
containing $N_i$. Inserting \eqref{t-def} in \eqref{e-K1} one finds 
\bea
    e^{-K} &=&- i^{D}  \mathbf{A}^T \eta \  \text{exp} \Big[  \sum_j (\bar t^j -  t^j ) N_j  \Big]  \mathbf{\bar A} \\
               & =& - i^{D}  \mathbf{a}_0^T \eta \  \text{exp} \Big[  \sum_j (\bar t^j -  t^j ) N_j  \Big]  \mathbf{\bar a}_0 + \cO(e^{2\pi i t})\ ,\\
               & \equiv& p(\text{Im}\, t^i , \zeta)  + \cO(e^{2\pi i t}) \label{Kpot3}
\eea
Due to the fact that the $N_j$ are nilpotent, $p$ is simply a polynomial in $ \text{Im} \, t^j $. 

This form of the K\"ahler potential allows us to make a couple of important observations. 
We note that near a singular point the 
K\"ahler potential $K$ depends on $\text{Im}\, t^j$ through a polynomial $p$ as well as exponentially
suppressed corrections $\cO(e^{-\text{Im}\, t^j})$. In contrast, the coordinates $\text{Re}\, t^i$ 
\textit{only} appear in the exponentially suppressed terms in \eqref{Kpot3}. This implies 
that if the degree of the polynomial $p$ is larger or equal to one in some variable $\text{Im}\, t^j$ then 
at large $\text{Im}\, t^j$ the K\"ahler potential enjoys an approximate shift symmetry 
\beq
    \text{Re}\, t^j \ \rightarrow \  \text{Re}\, t^j + c^j \ , \qquad c^j \in \mathbb{R}\ , 
\eeq
which is only broken by exponentially suppressed corrections to the discrete shifts with 
$c^j \in \mathbb{Z}$. In physical terms one thus would identify 
$ \text{Re}\, t^j $ as an axion. This agrees with the observation 
made in \cite{Garcia-Etxebarria:2014wla} that axion in complex structure moduli space arise at special points that have 
infinite order monodromy. Here we have shown this completely generally, by noting that $N_j$
only exists in these situations. Crucial is, however, that the polynomial has a non-vanishing 
degree.  To determine its 
degree we first introduce the integers $d_i$ such that 
\beq \label{def-d}
    N_j^{d_i}  \mathbf{a}_0 \neq 0 \ , \qquad  N_j^{d_i+1}  \mathbf{a}_0 = 0\ .
\eeq
One might be tempted to assert that the degree of $p$ in $ \text{Im} \, t^j $ is simply $d_j$, but this statement is far from trivial since  the degree could be lowered by a vanishing of the inner product involving $\mathbf{a}_0$ and $N_i^k \mathbf{a}_0$. 
%Unfortunately, it is not quite as simple as that, since the degree could be lowered by a vanishing of the inner product involving $\mathbf{a}_0$ and $N_i^k \mathbf{a}_0$. 
We will discuss in detail in subsection \ref{infinite_one} (and prove it in section \ref{mixed_hodge}), that at least for a one-dimensional degeneration the degree of the polynomial $p$ is indeed $d$.   

While it is hard to analyse infinite distance paths in the generality, it is possible to give a rather simple necessary 
criterion when the point $P$ is at infinite distance \cite{wang1}. Since the complex structure moduli space is 
K\"ahler, the length \eqref{length_d} of a path $\gamma$ is measured by the integral 
\beq \label{distance_complex}
   d_\gamma (P,Q) = \int_\gamma \sqrt{ 2 g_{I \bar J} \dot z^I  {\dot {\bar z}}^J} ds \ ,
\eeq
where $g_{I \bar J}$ is the K\"ahler metric derived from \eqref{Kpot3} evaluated along the path.
The necessary condition on $d_\gamma$ being infinite is 
\beq \label{necessary_c}
    \text{$P$ at infinite distance} \quad \Rightarrow\quad \exists N_i : \quad N_i  \mathbf{a}_0 \neq 0\ . 
\eeq
In other words, there has to be at least one monodromy $T_i$ of infinite order that allows us to define the $N_i$ satisfying \eqref{necessary_c}.
Furthermore this infinite order $T_i$ must act non-trivially on $\mathbf{a}_0$.
To show \eqref{necessary_c} it is easier to show the equivalent statement that if for all $N_i$ one has $N_i  \mathbf{a}_0 = 0$
then there is a path to the point which has a finite distance. So let us assume that this latter statement is true. Then the K\"ahler potential \eqref{Kpot3} has \textit{no} pure polynomial terms in $t^i$. Furthermore, one can use the tools presented in the next subsection to 
show that $p$ is constant in $t^i$ and does not vanish at $P$. Then one can focus on one specific path, namely the path 
$(t^i(s),\zeta^M(s)) = (i s,...,i s,0,...,0)$, and rather straightforwardly check that the integral \eqref{distance_complex} is finite 
when integrating from $s_0$ to $\infty$ \cite{wang2}. Since there is at least one path to the point which has finite distance, the point is said to be at finite distance in field space.

Let us end this section by pointing out that it is crucial to keep in mind that the arrow in \eqref{necessary_c}
points only in one direction. It was, however, conjectured in \cite{wang2} that the opposite direction is 
also true:
\beq \label{Conjecture}
   \text{Conjecture:}  \qquad \text{$P$ at infinite distance} \quad \Longleftrightarrow\quad \exists N_i : \quad N_i  \mathbf{a}_0 \neq 0\ . 
\eeq
The above sketch of a proof of \eqref{necessary_c} indicates why the other direction 
is much more involved. In the proof one has to construct one path which has finite length, which 
then implies that the point cannot be at infinite geodesic distance. Proving the opposite direction 
would require to study features of all possible paths, which is enormously complicated since 
many non-trivial cancellations can take place. 
In fact, this complication indicates a crucial problem that one is 
facing whenever one tries to analyse field space distances: there are many possible paths and it can 
be hard to disentangle features of a path from actual features of the metric. 

In the next subsection we will see that in the case in which the point $P$ lies on only a single 
special divisor, we can supply the relevant mathematics and study the paths in more detail. 
In fact, the conjecture \eqref{Conjecture} can then be proved and we will comment on the details 
of this proof. The assumption of $P$ being only on a single divisor reduces the problem to a 
complex one-dimensional problem in which it becomes tractable.

\subsection{Infinite distance paths in one-parameter degenerations} \label{infinite_one}

In the following we like to continue our study of infinite paths in the complex structure 
space $\cM_{\rm cs}$. In order to do this in detail we will impose restrictions on the 
points $P$ we consider. Our restriction will be that they only lie on a single 
special divisors. In this case we can study infinite distance paths in detail and indicate how the 
conjecture \eqref{Conjecture} is proved. 

In the considered special case, we can pick local complex coordinates \eqref{local-coords} 
that are now split as $z^I = (z, \zeta^M)$,
in which the special singular divisor is locally given by $z=0$. This local coordinates 
are also chosen, such that the point of interest is put at the origin $P: z= 0 ,\zeta^M = 0$.
Since we are dealing with a single divisor only, there is only a single monodromy matrix $T$ around 
$z=0$, and a single nilpotent matrix $N$ \eqref{def-Ni}. Recall that $N$ is given by 
\beq \label{N=logT}
   N = \log T^{(u)} = \frac{1}{m} \log T^m \ ,
\eeq
where $T^{(u)}$ is the unipotent part of $T$. As above we perform a coordinate transformation $z \rightarrow z^m$
to remove factors of $m$ and work only with $T^{(u)},N$. For completeness, let us also display the 
periods near $P$ again
\beq \label{simple_periods}
 \mathbf{\Pi}(z,\zeta) =  \text{exp} \Big[\frac{1}{2\pi i} (\log z) N  \Big] \mathbf{A}(z,\zeta)\ ,
\eeq
with
\beq \label{simple_periods_coeff}
     \mathbf{A}(z,\zeta) = \mathbf{a}_0(\zeta) +  \mathbf{a}_{1} (\zeta)  z +  \mathbf{a}_{2} (\zeta)  z^2+  \mathbf{a}_{3} (\zeta)  z^3 + \ldots \ ,
\eeq
where the $ \mathbf{a}_{j}$ are holomorphic functions of $\zeta^M$ near $P$. 
In the coordinate $t = \frac{1}{2\pi i} \log z$ this implies that the K\"ahler potential takes 
the form
\beq \label{eKexp_sing}
e^{-K}  = - i^{D}  \mathbf{a}_0^T \eta \  \text{exp} \Big[ - 2i\ \I \,t  \, N  \Big]  \mathbf{\bar a}_0 + \cO(e^{2\pi i t},\zeta)\,\equiv p(\text{Im}\, t, \zeta) + \cO(e^{2\pi i t},\zeta)\ .
\eeq
The first term on the right is a polynomial in $\text{Im}\, t$ and has again been denoted by $p(\text{Im}\, t, \zeta)$.

We are interested in the degree of the polynomial $p$. Therefore, as in \eqref{def-d}, we 
first define an integer $d$ as
\beq \label{def-d}
   N^d \mathbf{a}_0 \neq 0 \ , \qquad  N^{d+1} \mathbf{a}_0 = 0\ , 
\eeq
and note that since $N$ is nilpotent with $N^{n+1}= 0$ one has $d \leq n$. 
Recall from \eqref{necessary_c} that if $d=0$ one has that the point $P$ is at \textit{finite} 
distance. We are interested, however, in infinite distance points and therefore 
want to determine the degree of $p$ in $\I\, t$.
In order to do that, we expand the exponential in $p$ and write
\beq \label{expansion-p}
  p =  - i^D \sum_{j=0}^{d} \frac{1}{j!} (- 2i\ \I \,t)^{j}   \ S_{j} (\mathbf{a}_0, \mathbf{\bar a}_0)   \ ,
\eeq
where we defined the inner product 
\beq \label{def-Sj}
   S_{j} (\mathbf{a},\mathbf{b}) \equiv  \mathbf{a}^T \eta N^j  \mathbf{ b}\ .
\eeq
Note that we will sometimes denote $S_0\left(.,.\right)$ as $S\left(.,.\right)$.
In order to study the degree we need to determine the highest $j$ such that $S_j(\mathbf{a}_0,\mathbf{\bar a}_0)$ is non-vanishing. It will turn out that this highest $j$ is precisely $d$ defined in \eqref{def-d}. In fact, we will discuss in section \ref{mixed_hodge} how one proves 
\beq \label{Sd>0}
   i^{D-d} S_d(\mathbf{a}_0 ,  \mathbf{\bar a}_0) > 0\ . 
\eeq
It is then possible to  
prove \cite{wang1,lee}
\beq \label{One-modulus-relation}
 \boxed{  \rule[-.3cm]{0cm}{.8cm} \quad  \text{Theorem:}  \qquad  \text{$P$ at infinite distance} \quad \Longleftrightarrow\quad N  \mathbf{a}_0 \neq 0\ . \quad}
\eeq
In other words, there is some $d> 0$ for which \eqref{def-d} is satisfied.
Let us stress that this is a special case of the conjecture \eqref{Conjecture}. One can prove \eqref{One-modulus-relation}, 
since the one-modulus case with only $\I \, t$ appearing in \eqref{expansion-p} allows one to avoid issues related 
to path-dependence. Let us remark, though, that we are requiring the point to belong to a single singular divisor, but this divisor can be embedded in a higher dimensional moduli space parametrised by additional coordinates $\zeta^M$. Our results will, therefore, be also valid for Calabi-Yau manifolds of complex dimension $D$ with $h^{D,1}(Y_D)> 1$.

In order to determine the properties of $S_j(\cdot,\cdot)$ acting on $\mathbf{a}_0$, and subsequently proving \eqref{Sd>0}, \eqref{One-modulus-relation}, 
we need to dive further into mathematics and introduce so-called mixed Hodge structures. This will be done 
in subsections \ref{math-intro} and \ref{mixed_hodge}. But before that, we can already discuss the physical states becoming massless at the singular divisors and show the exponential mass behaviour of these states when approaching infinite distance points in moduli space, as stated by the Swampland Distance Conjecture. 
%The central ideas of this paper will therefore be presented in the next section, while the mathematical proofs of our statements are collected in section \ref{sec:orbits}.

%%%%%%%%%%%%%%%%%%%%%%%%%%%%%%%%%%%%%%%%%%%%%%%%%%%%%%%%
\section{Massless BPS States\label{sec:BPS_states}}
%%%%%%%%%%%%%%%%%%%%%%%%%%%%%%%%%%%%%%%%%%%%%%%%%%%%%%%%

Having established the formalism for the structure of the moduli space around points of infinite distance, in this section we consider the physical states near such points. The Swampland Distance Conjecture implies that we expect an infinite number of exponentially light states near such points of infinite distance. In general, it is not clear precisely which types of states should be becoming massless. However, for the specific setting of the complex-structure moduli space of Calabi-Yau threefolds there is a very natural class of states which are candidates. If we consider type IIB string theory then there are physical states corresponding to D3 branes wrapped on special Lagrangian three-cycles whose mass depends on the complex structure moduli. These are BPS states. In this paper we propose that an infinite number of such BPS states becomes exponentially light and eventually massless as we approach a locus at infinite distance. 

A mass scale is dimensionful and so we should define it relative to a reference mass. In specifying the mass of a state we will do so relative to the Planck mass $M_p$. Therefore, by states becoming massless we mean that the ratio of the physical mass of a state to the Planck mass goes to zero. In compactifications of type IIB string theory on Calabi-Yau three-folds there is a decoupling, due to ${\cal N}=2$ supersymmetry, between the complex-structure moduli space and the dilaton and K\"ahler moduli space. The latter two parameterise the string scale relative to the Planck scale. This means that our results will be decoupled from such scales, or in other words, rescaling the volume of the Calabi-Yau or the string coupling will not affect our results on the complex-structure moduli space. It would only modify the reference mass scale $M_p$. Therefore, we can trust our results even in exotic limits of complex-structure moduli space. Further, the mass of BPS states, which form the focus of our study, is given precisely by the central charge at all loops. This means that loop corrections to the mass only feed in through the corrections to the elements in the central charge. These properties will give us good control over the states. 

We will sometimes move between this type IIB setting and the mirror picture in type IIA. There, the relevant states are bound states of D0-D2-D4-D6 branes on even dimensional cycles whose mass depends on the K\"ahler moduli. More precisely, the branes are objects in the derived category of coherent sheaves. Note that one universal point of infinite distance in K\"ahler moduli space is the large volume limit. It may appear a little strange to propose that branes wrapping infinitely large cycles become massless in this limit. However, the way we define the mass as relative to the Planck mass implies that this is perfectly consistent and can be understood as the statement that the Planck mass diverges faster than the BPS mass of some states. The more exotic seeming behaviour is due to the fact that in type IIA the K\"ahler moduli control both the BPS mass of states and the ratio of the String scale to the Planck scale. So the decoupling we have in type IIB between the moduli space controlling BPS masses and the moduli space controlling the string scale is not present. This property means that one is able to probe quite exotic physics in type IIA string theory by using mirror symmetry with the relatively straightforward type IIB setting. Indeed, some of the BPS states which will become massless in the IIA setting will actually still be infinitely heavier than the string scale. 
%since  there is a sense in which they do. More precisely, we call a state massless if the ratio of its mass to the Planck mass goes to zero. Fixing our units such that the Planck mass is unity, this implies its mass vanishes. Then a simple physical way to understand why the wrapped branes become massless is that decompactifying, while keeping the string scale fixed, decouples gravity sending $M_p \rightarrow \infty$ which upon going to Planck units implies that the physical mass of the brane states vanishes. 
The fact that the states become massless in the sense of keeping $M_p$ finite matches the idea that the Swampland Distance Conjecture is gravitational in origin. Indeed, one infinite distance locus is the so-called geometric engineering limit where the massless spectrum reduces to that of a gauge theory with a finite number of states. However, this is only true in the sense that one simultaneously decouples gravity $M_p \rightarrow \infty$. 

The mass of particles corresponding to wrapped D3 branes is given by the volume of the special Lagrangian cycle that the D3 branes are wrapping. Because the setting has ${\cal N}=2$ supersymmetry these states are BPS which means that their mass is also given by the central charge. Either way, the mass formula is 
\be
M_{\bf q} = \left|Z_{\bf q}\right|  = e^{\frac{K}{2}}\left| S\left({\bf q},{\bf \Pi }\right) \right|\;. \label{BPSmass}
\ee
Here ${\bf q}$ is an integer vector specifying the charges of the particle under the $U(1)$ symmetries in the vector multiplets of the complex-structure moduli. In the geometric formulation in terms of special Lagrangian cycles it corresponds to the homology class of the special Lagrangian in the symplectic three-cycle basis. The mass of the particle is $M_{\bf q}$ and the central charge is $Z_{\bf q}$. The symplectic inner product $S\left({\bf q},{\bf \Pi }\right)$ is define in (\ref{def-Sj}) (as $S_0\left({\bf q},{\bf \Pi }\right)$). 

The mass formula (\ref{BPSmass}) gives us a powerful handle on the BPS states. However, it only tells us what the mass of a would-be BPS state of a given charge is. It does not tell us if such a state is actually present in the theory at a given point in complex-structure moduli space. Geometrically, since special Lagrangian cycles are not classified topologically, the presence of such a cycle in a given homology class depends on the value of the complex-structure moduli.\footnote{It also depends on the K\"ahler moduli. For example, special Lagrangian cycles can be identified explicitly as fixed loci of isometric anti-holomorphic involutions. The isometric condition depends on the K\"ahler moduli, though it was shown in \cite{Palti:2009bt} that there always exists a choice of K\"ahler moduli which renders any anti-holomorphic involution also isometric.} The dependence of the spectrum of BPS states on the complex-structure moduli is framed in the context of the stability of BPS states upon variations in complex-structure moduli. Over certain loci in moduli space, termed curves of marginal stability, some BPS states become unstable to decay to others. After crossing such a threshold line the would-be BPS state which decayed is no longer present in the theory. In other words, the state of that charge is no longer BPS but some unstable configuration of two other stable BPS states. Therefore, a given point in the charge lattice specified by ${\bf q}$ may or may not support an actual BPS state. We would like to identify an infinite number of BPS states which are becoming exponentially light upon approaching the point of infinite distance. The primary challenge in doing so is determining which charges support BPS states near the infinite distance point. Explicitly determining the special Lagrangian cycles geometrically is not possible with current technology. Instead, we will utilise the monodromies to gain some insights into the BPS spectrum. This will be a two-step process, with the first step discussed in section \ref{sec:bpsmond} and the second in section \ref{sec:BPS}.

\subsection{Monodromy orbits and massless states}
\label{sec:bpsmond}

In this section we consider properties of would-be BPS states near a monodromy locus. So we will work directly with the charges ${\bf q}$ without specifying if there is a BPS state of that charge in the theory. The point is to identify certain sets of charges by their properties near the monodromy locus. In section \ref{sec:BPS} we will then relate these sets of charges to the actual BPS spectrum.  

Consider the monodromy transformation (\ref{monodromy-trafo}). We will restrict to a one-parameter degeneration for now so that the transformation is
\be
{\bf \Pi} \left( z e^{2 \pi i}\right) = T  \;{\bf \Pi} \left( z \right) \;.
\ee
It is simple to check using  (\ref{Kpot_cs2}) that the moduli space metric is invariant under this transformation. It therefore appears like a redundancy in our description of the system which hints at an underlying gauge symmetry origin. Indeed, the monodromies are discrete remnants of higher dimensional continuous local symmetries. In type IIB they are embedded in higher dimensional diffeomorphisms. Since a gauge transformation can not change any physical properties of the theory neither should the monodromies. However, the BPS mass formula (\ref{BPSmass}) is not invariant under a monodromy transformation on the period vector and so the monodromy leads to a physical change in the mass of a state. 

The only way to make the physical change in the mass of states due to the monodromy action consistent with the idea that a monodromy action should not lead to a physical change in the theory is to propose that the monodromy action also rearranges the states in the theory. So while the mass of one state changes, there is another state which takes its place and so the full spectrum of states remains unchanged. To deduce the relation between these two states consider the  BPS mass formula (\ref{BPSmass}).  We see that its transformation allows one to associate a monodromy action on the charges 
\be
M_{\bf q} = e^{\frac{K}{2}}\left|S\left({\bf q},{\bf \Pi }\right) \right| \xrightarrow{T}  e^{\frac{K}{2}}\left|S\left({\bf q},T{\bf \Pi }\right) \right|  = e^{\frac{K}{2}}\left|S\left(T^{-1}{\bf q},{\bf \Pi }\right) \right|   \;. \label{bpsmonorb}
\ee
Therefore, if the theory contains a BPS state of charge ${\bf q}$, then after the monodromy there must be a BPS state of charge $T {\bf q}$ of mass equal to the original state. This defines an action of the monodromy on the charges. We can keep applying the monodromy transformation which generates a \textit{monodromy orbit} through the charge space. Specifically, for a monodromy matrix $T$, and a representative charge element ${\bf q}_s$ inside the orbit, the orbit is defined as
\be
{\cal O}_{T}[{\bf q}_s] \equiv \left\{ {\bf q} \in H^{3}(Y_3,\mathbb{Z}) : {\bf q} = T^m {\bf q}_s\; \mathrm{for\;some} \;m \in \mathbb{Z} \right\} \;. \label{mondordef}
\ee  
It will be later convenient for us to study the difference between a charge vector ${\bf q}_s$ and its image under $T^k$, we therefore introduce the notation 
\beq \label{def-deltaq}
   \delta_k \mathbf{q} =  T^k  \mathbf{q} -\mathbf{q} \ .
\eeq
Note that if we consider $T^{(u)}= e^N$ one finds  $\delta_k  \mathbf{q} = k N  \mathbf{q} + \cO(N^2)$.

An important point is that a given set ${\cal O}_{T}[{\bf q}_s]$ can have an infinite number of charge elements. We denote a monodromy orbit that is infinite as ${\cal O}^{\infty}_{T}[{\bf q}_s]$. A necessary condition for the existence of an infinite orbit ${\cal O}^{\infty}_{T}[{\bf q}_s]$ is that $T$ is of infinite order. This relates naturally to infinite distance since the result \eqref{One-modulus-relation} implies that at infinite distance points the monodromy matrix $T$ is of infinite order.\footnote{This can be inferred by noting that the existence of a non-zero nilpotent $N$ implies that $T$ contains a unipotent part that is not the identity matrix. Therefore, it can be split as in (\ref{Tdecom}), $T=T^{(u)}\cdot T^{(s)}$, with $T^{(u)}$ being unipotent. Furthermore, any non-trivial unipotent matrix is of infinite order. Note that for $T^{(u)} = e^N$ one finds that $T^{(u)}$ is unipotent if and only if $N$ is nilpotent. Further, we have introduced 
coordinates $z$ and $t$ after \eqref{N=logT} such that the corresponding monodromy is the unipotent $T^{(u)}$. } \footnote{After publishing this work we became aware that in \cite{Cecotti:2015wqa} some similar ideas were proposed in the context of supergravity theories with more than 8 supercharges.}
We can also state a sufficient condition for a monodromy orbit ${\cal O}_{T}[{{\bf q}_s}]$ to be infinite, by requiring\footnote{Noting that a unipotent matrix has all eigenvalues equal to $1$ and a nilpotent matrix all eigenvalues equal to $0$, this equation can equally be stated by demanding that $\mathbf{q}_s$ is not an eigenvector. }
\be
 T^{(u)} \mathbf{q}_s \neq \mathbf{q}_s  \quad \Leftrightarrow\quad  N {\bf q}_s \neq 0 \;. \label{transcond}
\ee
Therefore, a monodromy orbit at infinite distance ${\cal O}_{T}[{\bf q}_s ]$ is either infinite, if (\ref{transcond}) is satisfied, or is composed solely of ${\bf q}_s$ and its finitely many images under $T^{(s)}$.

Having defined the set of charges in a monodromy orbit, we now consider another set of charges which are those associated to a vanishing BPS mass on the monodromy locus. It is important to note that at this stage what we mean by this are charges ${\bf q}$ which lead to a vanishing central charge, and therefore the mass of a would-be BPS state, on the monodromy locus. We do not consider yet if there is a BPS state with that charge. Consider the BPS mass formula (\ref{BPSmass}) around the monodromy point. We can evaluate this by utilising the mathematical tools introduced in section \ref{sec:infinite_distance} and, in particular, in subsection \ref{infinite_one}. We restrict our considerations to points in moduli space that are characterised by a single monodromy matrix $T$ and corresponding $N$. As in subsection \ref{infinite_one} we choose coordinates $t = \frac{1}{2\pi i} \log z$ such that the infinite distance point is reached at $\I \,t \rightarrow \infty$. The moduli space can be multi-dimensional with additional directions parametrised by other coordinates $\zeta^M$. While 
expansion coefficients, such as the vectors $\mathbf{a}_0$ in \eqref{simple_periods_coeff}, in general can depend on the $\zeta^M$, we will suppress this dependence in the notation.  
We can use the nilpotent orbit theorem to expand the period vector as in (\ref{simple_periods}) with \eqref{simple_periods_coeff}, and the leading behaviour of the K\"ahler potential from \eqref{eKexp_sing}, (\ref{expansion-p}). Together, they imply the form for the central charge
\be
Z_{\bf q} = \frac{ \sum_{j=0}^{d} \frac{1}{j!} \, t^j\, S_j\left({\bf q},{\bf a}_0  \right)}{\left(i\sum_{j=0}^{d} \frac{1}{k!} (- 2i\ \I \,t)^{k}   \ S_{k} (\mathbf{a}_0, \mathbf{\bar a}_0)  \right)^{\frac{1}{2}}}  + {\cal O}\left(e^{2\pi i t} \right)\;. \label{bpsmassimt}
\ee
%where we have dropped terms exponentially suppressed terms containing $e^{2\pi i t}$.
This expression simplifies further if we focus only on the leading terms. Recall that $d$ is defined as the maximum integer for which $N^{d} {\bf a}_0 \neq 0$ and that \eqref{Sd>02} ensured that $i^{3-d}S_{d} (\mathbf{a}_0, \mathbf{\bar a}_0) > 0 $. Hence, in the denominator of 
\eqref{bpsmassimt} it is simply the term proportional to  $(\I\, t)^{d/2}$ that dominates. In the numerator we realise that
the holomorphic structure, and the fact that we restrict to singularity defined by a single coordinate $t$, allows us to 
drop the $\R\, t$-terms. This is because they are always dominated by the $\I\, t$-terms with the same coefficient. Therefore, we 
find that the leading terms in the central charge in the limit $\I\, t \rightarrow \infty$ are 
\beq
Z_{\bf q}^{\rm lead}=\frac{\sum_{j=0}^{d} \frac{1}{j!} \,  (i\, \I\, t)^j\, S_j\left({\bf q},{\bf a}_0  \right)}{\left(- \frac{2^d i^{3-d}}{d!} (\I \,t)^{d}   \ S_{d} (\mathbf{a}_0, \mathbf{\bar a}_0)  \right)^{\frac{1}{2}}}   + {\cal O}\left(e^{2\pi i t} \right)
\ , \label{Mlead}
\eeq
where we have indicated the appearance of the exponential terms relevant if $S_j\left({\bf q},{\bf a}_0  \right) =0$ for all $j$.
Finally, the leading behaviour of the mass is simply extracted through $M_{\bf q}^{\rm lead} = \left|Z_{\bf q}^{\rm lead}\right|$.

Using the expression for the mass of a state of charge ${\bf q}$ (\ref{Mlead}) we can determine the set of charges which lead to vanishing BPS mass on the monodromy locus $\I\,t \rightarrow \infty$. Let us denote this set of charges as $\cM$, i.e.~we define 
\beq
   \cM \equiv \left\{ \mathbf{q} \in H^3(Y_3,\mathbb{Z}) : \quad S_j\left({\bf q},{\bf a}_0  \right) = 0 \;,\; \mathrm{for\;all}\; j \geq \frac{d}{2} \right\} \;. \label{masslesscondI1} 
\eeq 
Note that this space forms a sublattice of the full space of charges. The central charge of these charges will either vanish polynomial or exponentially fast. We want to distinguish these two cases. 
The first set of charges which lead to massless states are elements of ${\cal M}_{\rm I}$ and are denoted as type I. We specify 
\be
  \cM_{\rm I} \equiv \left\{ \mathbf{q} \in \cM:\quad S_j\left({\bf q},{\bf a}_0  \right) \neq 0 \;\mathrm{for\;some}\; j < \frac{d}{2}\right\} . \label{masslesscondI2}
\ee
Note that this space is not properly a sublattice of $\cM$. %However, we can still think of it as a lattice with some extra condition. 
States of type I become massless at the monodromy locus due to the denominator in the first term of (\ref{Mlead}). Their mass decreases as a power law in $\I\,t$.
The space of charges with exponentially vanishing central charge will be denoted by $\cM_{\rm II}$. This space is clearly a
sublattice $\cM_{\rm II} \subset \cM$ and explicitly specified by 
\beq
  \cM_{\rm II} \equiv \left\{ \mathbf{q} \in \cM:\quad S_j\left({\bf q},{\bf a}_0  \right) = 0 \;,\; \mathrm{for\;all}\; j \geq 0 \; \right\} \ ,  \label{masslesscondII}
\eeq  
and calling these states to be of type II. 
For states of type II the first term in (\ref{Mlead}) vanishes identically and they therefore become massless exponentially fast\footnote{Do not confuse this exponential growth with the exponential mass behaviour stated by the Swampland Distance Conjecture. The latter is in terms of the proper field distance, while this one is in terms of the coordinate $\text{Im}\,t$. In fact, we will see in section \ref{sec:mass} that the states of type I are the ones exhibiting the mass behaviour expected by the Swampland Distance Conjecture.} in $\I\;t$. 
Clearly, with these definitions of the two subsets $\cM_{\rm I}$ and $\cM_{\rm II}$ one can decompose
\be
\cM = \cM_{\rm I} \oplus \cM_{\rm II} \; ,
\ee
with a summation performed with integer coefficients. 

In the next section we will consider BPS states and we will propose to use the monodromy transformations to identify a candidate set of stable BPS massless states at the singularity. Hence, we need to combine the concepts introduced in this section and identify monodromy orbits within the sets of massless charges $\cM$, $\cM_{\rm I}$, and $\cM_{\rm II}$. First note that 
\be
S_j\left(N{\bf q},{\bf a}_0 \right) = - S_{j+1}\left({\bf q},{\bf a}_0  \right) \; , \label{iterel}
\ee
as can be deduced from $S(N\, \cdot, \cdot ) + S(\cdot,N\, \cdot)=0$, already given in \eqref{property_Neta}. This implies that if ${\bf q}_s \in \cM$ then ${\cal O}_{T}[{\bf q}_s ]\subset \cM$. Further, if ${\bf q}_s \in \cM_{\rm I,II}$ then ${\cal O}_{T}[{\bf q}_s ] \subset \cM_{\rm I,II}$. In words, if a state is massless then all the states in its orbit are massless. If a state becomes massless exponentially fast in $\I\;t$, then all the states in its orbit become massless exponentially fast. If a state becomes massless as a power law, then all the states in its orbit become massless as a power law. 

Furthermore, the mass difference between two states $\bf{q}_s\in \cM$ and $T\bf{q}_s\in \cM$ is
\beq \label{DeltaZ}
\Delta Z= \frac{S_0(N\mathbf{q}_s,\mathbf{a}_0)}{\left(- \frac{2^d i^{3-d}}{d!} (\I \,t)^{d}   \ S_{d} (\mathbf{a}_0, \mathbf{\bar a}_0)  \right)^{\frac{1}{2}}}  + {\cal O}\left(e^{2\pi i t} \right)
\eeq
where $S_0(N\mathbf{q}_s,\mathbf{a}_0)=-S_1 (\mathbf{q}_s,\mathbf{a}_0)$ due to \eqref{iterel}. Therefore, states satisfying $S_1 (\mathbf{q}_s,\mathbf{a}_0)=0$ have a mass difference visible only at exponential order.

Our considerations in the next section suggest that it is natural 
to introduce an equivalence relation on the set of massless states $\cM$. 
Namely, we like to identify two charges ${\bf q}_1 \simeq {\bf q}_2$ if ${\bf q}_1-{\bf q}_2 \in \cM_{\rm II}$.
In mathematical terms this defines a quotient space 
\beq
\cM_Q = \cM / \cM_{\rm II}  \;,
\label{quotient}
\eeq
which is identical to using the equivalence relation on $\cM_I$. The elements of this 
quotient space are equivalence classes, which we denote by 
\beq
   [\mathbf{q}] = \{\mathbf{q}' \in \cM\,: \ \mathbf{q}' - \mathbf{q} \in \cM_{\rm II}  \}\ .
\eeq
We will propose that this quotient carries non-trivial information about the 
presence of stable BPS massless states at the singularity. We note that 
one can write $[T \mathbf{q}] = T [\mathbf{q}]$ and $[N \mathbf{q}] = N[\mathbf{q}]$, 
since $N,T$ map states of $\cM_{\rm II}$ into $\cM_{\rm II}$.

Combining the quotient construction with the 
construction of the monodromy orbit (\ref{mondordef}), we next introduce 
the  {\it  quotient monodromy orbit}.
We will denote this orbit by
\bea
{\cal Q}_{T}[{\bf q}_s] &\equiv& \left\{ [{\bf q}] \in \cM_Q :\ \  [{\bf q}] = T^m [{\bf q}_s]\; \mathrm{for\;some} \;m \in \mathbb{Z} \right\} \\
      &\equiv &\left\{ [{\bf q}] \in \cM_Q :\ \   {\bf q} \in \cO_T[{\bf q}_s]\right\} \;.  \nn
%\equiv \I \big( {\cal O}_{T,{\bf q}_s} \cap \cM  \ \rightarrow \  \cM_Q \big) = \{ [q] \in \cM_{Q}: \quad [q] = T^m [q_s]    \; ,
\label{Qorbit}
\eea
%where we consider the image of the map $\cM \rightarrow \cM_Q$.
This %involved looking 
definition simply means that the quotient monodromy orbit is defined by first restricting to massless charges in $\cM$ and then identifying two elements in the orbit if they differ by a type II charge.  Again, if this restriction has infinite elements then we denote it by ${\cal Q}_{T}^{\infty}[{\bf q}_s]$. It is important to note that even if the monodromy orbit $\cO_{T}[{\bf q}_s]$ has infinite elements the  quotient monodromy orbit need not.

We will then propose that if a given class $[{\bf q}] \in {\cal Q}_{T}[ {\bf q}_s ]$ actually contains a charge vector corresponding to a 
BPS state in the theory, then at the monodromy locus each class of the monodromy quotient 
orbit will contain a BPS state. Therefore, we need to identify an infinite massless quotient monodromy orbit ${\cal Q}_{T,{\bf q}_s}^\infty$ satisfying \eqref{transcond} at the infinite distance points. We will show in section \ref{sec:mass} that these states indeed exhibit an exponential mass behaviour in terms of the proper field distance, which further motivates their identification as the infinite tower of states of the Swampland Distance Conjecture. However, in order to identify this orbit, we need to introduce further mathematical tools involving the $SL_2$ orbit theorem and the so called mixed Hodge structures. This will be the topic of section \ref{sec:orbits}, although the results will already be outlined at the end of the next section. 
 
\subsection{The monodromy orbit and BPS states\label{sec:BPS}}

\begin{figure}[h!]
\begin{center} \includegraphics[width=\textwidth]{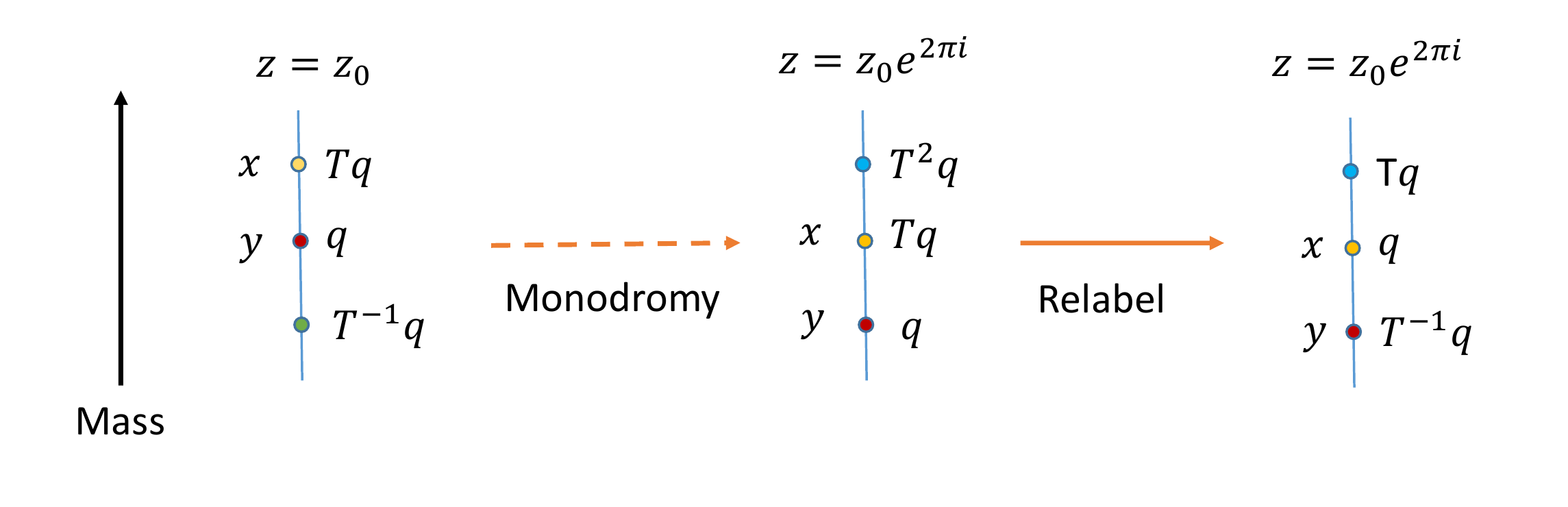} 
\vspace*{-.5cm}
\end{center}
\caption{Figure showing the states in the theory at $z=z_0$ and $z=z_0 e^{2 \pi i}$. The labels $x$ and $y$ track each physical state. The state $y$ of charge ${\bf q}$ is BPS at $z=z_0$ and the state $x$ of charge $T{\bf q}$ is BPS at $z=z_0 e^{2 \pi i}$. In the last step the charges of all the states are simultaneously relabelled so that the spectrum matches the one before the monodromy.} \label{fig:mon1}
\end{figure}

In the previous section we introduced the relevant structures in the charges, we now go on to discuss their relation to the BPS spectrum. In (\ref{bpsmonorb}) we introduced the natural action of the monodromy matrix $T$ on the charge vector ${\bf q}$. The underlying gauge symmetry nature of monodromy transformations implies that upon circling the monodromy locus the full spectrum of states should remain unchanged up to a possible global re-labelling of the charges. Let us be explicit about what this implies. We consider a monodromy about a locus $z=0$. Then the statement is that the theory should be the same at $z=z_0$ and at $z=z_0 e^{2 \pi i}$, where $z_0$ is an arbitrary reference value for $z$. Then say we have a BPS state of charge ${\bf q}$ at $z=z_0$. We should have a BPS state of the same mass at $z=z_0 e^{2 \pi i}$. Using (\ref{bpsmonorb}) this means that this BPS state should have charge $T {\bf q}$. So far we have only referred to a single state at each value of $z$. But there is another natural state in the spectrum at each point, specifically the state of charge ${\bf q}$ at $z=z_0 e^{2 \pi i}$ and the state of charge $T{\bf q}$ at $z=z_0$. All these states, and the action of the monodromy and the re-labelling are shown in figure \ref{fig:mon1}.

The data we have about the BPS spectrum at $z=z_0$, specifically that state ${\bf q}$ is BPS, allows us to deduce the existence of one BPS state at $z=z_0 e^{2 \pi i}$, of charge $T {\bf q}$, by utilising the gauge transformation nature of the monodromy. However, we can also explicitly track how the BPS state of charge ${\bf q}$ behaves as we send $z \rightarrow z e^{2 \pi i}$. The crucial question is whether this state remains BPS throughout the path in moduli space. If that is the case then we can deduce the existence of two BPS states at $z=z_0 e^{2 \pi i}$, of charges (before relabelling) ${\bf q}$ and $T{\bf q}$. Since the theory is invariant under the monodromy, this holds equally at $z=z_0$. Therefore, in such a situation we know that both ${\bf q}$ and $T{\bf q}$ are BPS states. If this also holds for the full infinite orbit generated by $T$, denoted ${\cal O}^{\infty}_{T}[{\bf q}]$, then we can deduce the existence of an infinite number of BPS states. This will be our strategy. 

To determine the fate of the BPS state over the path $z \rightarrow z e^{2 \pi i}$ we need to determine if it crosses a wall of marginal stability along this path. Let us review first some statements about such walls, see for example \cite{Denef:2000nb,Denef:2001xn,Douglas:2000qw,Douglas:2000gi,Aspinwall:2001zq,Aspinwall:2002nw,Jafferis:2008uf,Aspinwall:2009qy,Andriyash:2010yf}. Consider three BPS states of charges ${\bf q}_A$, ${\bf q}_B$ and ${\bf q}_C$. We have the relation between state $A$ and the anti-state $\bar{A}$ such that ${\bf q}_A=-{\bf q}_{\bar{A}}$. Now let the charges be related as 
\be
{\bf q}_C={\bf q}_B+{\bf q}_{\bar{A}} \;. \label{charrel}
\ee
Because the central charge is linear in the charges, the masses of BPS states of these charges are related through an inequality
\be
M_{{\bf q}_C} \leq M_{{\bf q}_B} + M_{{\bf q}_{\bar{A}}} \;.  \label{trianinequ}
\ee
Let us define the grade $\varphi$, introduced in \cite{Douglas:2000gi}, associated to a state labelled $A$ which is the phase of the central charge
\be
\varphi\left(A\right) = \frac{1}{\pi} \mathrm{Im} \log Z_{{\bf q}_A} \;. \label{defgrade}
\ee
Note that $\varphi\left(A\right) = \varphi\left(\bar{A}\right)+1$, and that we have the identification $\varphi\left(A\right) \sim \varphi\left(A\right)+2$. The inequality (\ref{trianinequ}) is saturated for $\varphi\left(B\right) - \varphi\left(A\right) = 1$. This is termed a curve of marginal stability. For $A$ and $B$ co-prime this is  a co-dimension one locus in moduli space, otherwise it is the full moduli space. The central point is that in crossing such curves the spectrum of BPS states can change by a BPS state decaying to two other BPS states. In this case the state $C$ can decay into $B$ and $\bar{A}$. Conversely, a BPS state remains stable over a continuous path in moduli space if the path does not intersect a curve of marginal stability. 

We can now apply this to the set of states near the monodromy locus by utilising the asymptotic form of the central charge (\ref{Mlead}). In particular, we are interested in the stability of BPS states within the monodromy orbit ${\cal O}_{T}[{\bf q}]$. More specifically, in the orbit associated to a charge which leads to a vanishing BPS mass on the monodromy locus, so that ${\cal O}_{T}[{\bf q}] \subset \cM$. First note that, as we approach the monodromy locus, charges in $\cM$ are much lighter than any charges not in $\cM$ and so can never decay to them. So given a BPS state in $\cM$, corresponding to state $C$ in the discussion above, over a monodromy path we need to consider if it crosses a curve of marginal stability with respect to decay to two other states in $\cM$, which are states $B$ and $\bar{A}$ above. At this point the splitting of the states into type I and type II in section \ref{sec:bpsmond} manifests strongly. This is because in (\ref{Mlead}) circling the singularity corresponds to $\R\;t \rightarrow \R\;t + 1$. Under this we see that the grade, as defined in (\ref{defgrade}), for states of type I remains invariant to leading order while for states of type II it undergoes a full transformation 
\be
\varphi_{\rm I} \rightarrow \varphi_{\rm I} + {\cal O}\left(\frac{1}{\I\; t}\right) \;,\;\; \varphi_{\rm II} \rightarrow \varphi_{\rm II} + 2 + {\cal O}\left(\frac{1}{\I\; t}\right) \;. \label{gradeIItrans}
\ee
We therefore observe that if states $B$ and $\bar{A}$ are of type I and type II respectively then the state $C$ must cross a curve of marginal stability upon circling the monodromy locus. This tells us that the curve of marginal stability for states that can be written as a sum of a type I and type II state intersects the monodromy locus. On the other hand, if $B$ and $\bar{A}$ are both of type I then the BPS state $C$ will not cross a curve of marginal stability along the monodromy path. 

Therefore, as we circle the monodromy locus, with a radius arbitrarily close to the monodromy locus, a type I BPS state will cross a curve of marginal stability to decay to a type I and a type II state,  but not to two type I or two type II states.\footnote{The latter can also be easily seen by noting that type I states are exponentially heavier than type II states, and that the possible decay can only be marginal.}\footnote{In the case of a decay to more than two states the condition for marginal stability becomes only stronger than the two-body decay, and so the conclusion that it is not possible to decay to two type I states still holds.} This gives us a physics interpretation of the quotient space $\cM_{Q}$ in (\ref{quotient}). Since in this space two type I states which differ by a type II state are identified, we have that each equivalence class of type I states in the quotient is stable against decay over the monodromy path. A decay will only move between different representatives in that class.

We can now return to our original motivation and determine the fate of the BPS state with charge ${\bf q}$ at $z=z_0$ over the monodromy path. We know that it will remain stable within its class in $\cM_Q$. Therefore, the only thing to check is that the state $T {\bf q}$ is not in the same equivalence class. But this is precisely the case for states corresponding to separate elements in the quotient monodromy orbit ${\cal Q}_{T}[{\bf q}]$ defined in (\ref{Qorbit}). So within this quotient monodromy orbit, we determine that inequivalent elements ${\bf q}$ and $T{\bf q}$ are both BPS states, sufficiently close to the monodromy locus, as long as one of them is. The same conclusion holds for the full orbit, including the case when it is infinite. Therefore, if there exists an infinite quotient monodromy orbit ${\cal Q}^{\infty}_{T}[{{\bf q}}]$, and if it contains at least one BPS state then, sufficiently close to the monodromy locus, we deduce the existence of an infinite number of BPS states spanning the full orbit. 

Note that the existence of an infinite quotient monodromy orbit ${\cal Q}_{T}[{\bf q}]$, and a single BPS state in that orbit, is a sufficient condition to identify the infinite tower of BPS states, but it is not a necessary one for the existence of such a tower. In particular, the quotient by all type II charges can be too strong since some of those charges may not contain BPS states. Even more generally, it is possible that there is an infinite tower of BPS states which become massless at infinite distance that is not related to any monodromy orbit.  

Note also that at finite distance away from the monodromy locus the infinite quotient monodromy orbit will not fully consist of BPS states. More explicitly, under $n$ monodromy transformations the grade (\ref{defgrade}) for type I states transforms as
\beq
\label{nfincha}
\varphi_{\rm I}\rightarrow \varphi_{\rm I}-\frac{n}{\pi\text{Im}\, t} + {\cal O}\left( \frac{1}{\left(\I\;t\right)^2}\right)\;.
\eeq
%Therefore, it remains approximately constant only for states with charge ${\bf q}\in {\cal O}_{T}[{\bf q}_s]$ satisfying ${\bf q}=T^n{\bf q}_s$ with $n \ll \text{Im}\, t$. States higher up in the tower are not stable, since their grade can substantially change and therefore cross a curve of marginal instability when circling the singularity. %This provides a cut-off scale for each state to be stable.  
Therefore, say we consider the initial BPS state as state ${\bf q}_C$ in (\ref{charrel}), and take ${\bf q}_B$ and ${\bf q}_{\bar A}$ as type I states. After $n$ monodromy transformations, at finite distance, the relative phase of ${\bf q}_B$ and ${\bf q}_{\bar A}$ will change as in (\ref{nfincha}). When $n \sim \I\;t$ this relative phase change is of order one and so ${\bf q}_C$ will cross a line of marginal stability. Hence, the number of states which are stable BPS particles, by the argument presented, behaves as $\I\;t$ and only diverges at the monodromy locus $\I\;t \rightarrow \infty$. In section \ref{sec:integrating_out} we will integrate out only the part of the tower which correspond to stable states to recover, this way, the logarithmic divergence in the field distance when approaching the singularity. Interestingly, this linear growth on the number of stable BPS states is the same that we will obtain from imposing the species bound, i.e, by considering only states which lie below the cut-off scale above which gravity becomes strongly coupled.

In summary, we deduce that if there exists an infinite quotient monodromy orbit ${\cal Q}^{\infty}_{T}[{{\bf q}}]$ and if it includes at least one BPS state, then sufficiently close to the monodromy locus the full infinite orbit is populated by BPS states. These are the candidate infinite number of BPS states which become massless on the monodromy locus. We will make some comments in section \ref{sec:orbits} regarding the existence of a single BPS state in the orbit. However, it is essentially an assumption of our construction. On the other hand, we will study in great detail in section \ref{sec:orbits} the first condition relating to the existence of an infinite orbit. Before that, we can already note a relation to infinite distances. If we consider a monodromy locus at finite distance, then $d=0$. Inspecting the BPS mass (\ref{Mlead}) we see that only type II states become massless on the monodromy locus. Therefore, there are no type I states $\cM_{\rm I} = \emptyset$. Hence, the quotient massless set ${\cM_Q}$ and the associated quotient monodromy orbit  ${\cal Q}^{\infty}_{T} [{{\bf q}}]$ must also be empty. So there does not exist an infinite massless quotient monodromy orbit at finite distance points in moduli space. 

To show the existence of an infinite quotient monodromy orbit at infinite distance will require introducing significant mathematical machinery. This is performed in section \ref{sec:orbits}. However, the results can be summarised concisely and so we do so here. We find that there exists an infinite massless quotient monodromy orbit ${\cal Q}_{T}^\infty[{\bf q}]$, where we specify precisely the possible charges ${\bf q}$, for the case of a monodromy with maximum nilpotency order $n=d=3$. This forms a central result of the paper and, when combined with the analysis of this section, forms strong evidence for the existence of an infinite number of massless BPS states on such loci. 

The quotient monodromy orbit is empty for $d<3$ and any $n$. More precisely, the orbit generated by the monodromy associated to the infinite distance locus is empty. This is a striking result which appears to go against the Swampland Distance Conjecture in some ways. However, it is important to state that the existence of an infinite quotient monodromy orbit is only a sufficient, but not necessary, condition for having an infinite number of massless BPS states. More generally, the quotient space $\cM_{Q}$ forms a good candidate for containing an infinite number of BPS states. We know from the monodromy action on the central charge that states in $\cM_{Q}$ have no walls of marginal stability around the monodromy locus. So they are promising candidates for BPS states.\footnote{In the case of $d=3$ we have that type I states with $S_1\left({\bf q},{\bf a}_0 \right) \neq 0$ are stable within a region of order $\I\;t$, while states with only $S_0\left({\bf q},{\bf a}_0 \right) \neq 0$ are stable within an exponentially large range. This suggests that, since the former states decay to the latter at polynomial distance from the monodromy locus, while the latter are stable at exponential distances, type I states with $S_1\left({\bf q},{\bf a}_0 \right) = 0$  are very good candidates for BPS states even without being part of a monodromy orbit.} The quotient space is also infinite for any infinite distance locus because $N^d{\bf a}_0$ is non-vanishing and is in $\cM$ but not in $\cM_{II}$.

In fact, we will show that there could still be an infinite set of charges in $\cM_Q$ induced by a monodromy transformation. However, this transformation is associated to a different locus which intersects the infinite distance locus. Due to the global nature in moduli space of this mechanism we are not able to show that this happens generally, but will show it for interesting examples, as well as discuss possible counter-examples, in section \ref{sec:intersection}.

\subsection{The exponential mass behaviour\label{sec:mass}}

The Swampland Distance Conjecture implies that the mass of the states should be exponentially decreasing in the proper distance upon approaching the infinite distance point. Note that this naturally assumes a geodesic approach towards the point. It is not practical to identify such geodesics in general Calabi-Yau moduli space, but the asymptotic approach to infinity is a one-parameter variation $\mathrm{Im\;}t \rightarrow \infty$ which makes the analysis feasible. Note that this assumes that we are approaching a generic point on an infinite distance locus. Special points, where multiple infinite distance divisors intersect, are more complicated to analyse because then two parameters are approaching infinity. While the results of section \ref{sec:bpsmond} about the massless states at infinity will hold generally, in this section we only consider generic points. With this assumption we proceed to show that the mass of the BPS states which become massless decreases exponentially in the proper distance upon approaching such points, in accordance with the conjecture.

As we revisited in section 2, the nilpotent orbit theorem implies that the K\"ahler potential takes the following simple form near a singular point in one-parameter models,
\beq
e^{-K}=p(\text{Im} \, t)+\mathcal{O}(e^{2\pi i t}) \;,
\eeq
where $p(\text{Im} \, t)$ is a polynomial of degree $d=\text{max}\{l|N^la_0\neq 0\}$. The Weil-Petersson metric is then given by
\beq
g_{t\bar t}=\partial_t\partial_{\bar t} K= \frac14 \frac{p'^2-p''}{p^2}+\mathcal{O}(e^{2\pi i t})=  \frac14\frac{d}{\text{Im} \, t^2}+\frac{\#}{\text{Im} \, t^3}+\dots+\mathcal{O}(e^{2\pi i t}) \;.
\eeq
While the subleading terms are sensitive to the particular structure of the moduli space near the singularity (encoded in the explicit form of  $p(\text{Im} \, t)$), the leading term is universal and only depends on the degree $d$. This universal term is quadratic in   $1/\text{Im} \, t$ implying that the proper field distance grows logarithmically
 \beq
 \label{onemodlog}
 d_\gamma(P,Q)=\int_Q^P\sqrt{g_{t\bar t}}\,|dt|\sim \frac{\sqrt{d}}{2}\, \text{log}(\text{Im} \, t)\,|_Q^P\rightarrow \infty
 \eeq
  for any smooth path $\gamma$ connecting $P,Q$  and diverges when approaching the singularity at $\text{Im} \, t\rightarrow \infty$. Hence, the singularity is at infinite distance if $d\neq 0$.  
 %Here the point $P$ is taken to be infinitely close to the singularity but not there yet, in order to see the logarithmic divergence arising. 
 This proves the Theorem \eqref{One-modulus-relation} for one-parameter models and the result is completely general for any one-parameter K\"ahler-Einsten manifold of any space-time dimensionality \cite{wang1}.

For a Calabi-Yau compactification preserving $\mathcal{N}=2$ supersymmetry, we have seen in \eqref{Mlead} that BPS states becoming massless at the singularity have a mass going as
\be
M_{\bf q} \simeq \frac{\sum_j \frac{1}{j!}(\text{Im} \, t) ^j S_j\left({\bf q},{\bf a}_0  \right)}{(2^d/d!)^{1/2}\,(\text{Im} \, t)^{\frac{d}{2}}}  \;,\quad \text{ with }
 0\leq j < \frac{d}{2} \;, \label{massexp}
\ee
for large $\text{Im} \, t$. In section \ref{sec:BPS_states} we motivated candidates for these states as those belonging to an infinite quotient monodromy orbit ${\cal Q}^\infty_{T,{\bf q}}$, which we argued to exist at infinite distance singularities. However, the behaviour of the mass (\ref{massexp}) is universal for any BPS states becoming massless, and the results of this section will therefore also be universal.

If we compare the effective theory at two different points $P,Q$ in the moduli space along a path $\gamma$ approaching the singularity, the mass of these BPS states decreases exponentially fast in terms of the proper field distance between the two points,
\be
\label{oneparaexpmas}
\frac{M_{\bf q}(P)}{M_{\bf q}(Q)} \simeq \frac{(\text{Im} \, t)^{s}|_Q}{(\text{Im} \, t) ^{s}|_P}\simeq e^{-\lambda \,d_\gamma(P,Q)} \;,%S_{j_{\text{max}}}\left({\bf q},{\bf a}_0  \right)
\ee
where $\lambda=s\sqrt{2/d}$ with $s=1,1/2$ for $d$ even or odd respectively. This is precisely the exponential mass behaviour predicted by the Swampland Distance Conjecture.

The above result can be generalised to higher-dimensional moduli spaces if the special point belongs only to one singular divisor, or in other words, it is a generic point of the singular locus.  %Suppose then that $P\in E_1$ and $P \not\in E_j$ where $E_1$ is an infinite divisor located at $\text{Im} \, t\rightarrow \infty$, meaning $N_1a_0\neq 0$. The rest of the coordinates of the higher dimensional moduli space will be denoted by $\zeta$.
%As explained in section 2, the periods take the form
%\beq
%\Pi(t,\zeta)=exp(tN_1)A(t,\zeta)
%\eeq
%where $A(t)=a_0(\zeta)+h(t,\zeta)$. Here $a_0(\zeta)$ is a polynomial function on the rest of the coordinates $\zeta$ while $h$ includes all exponentially suppressed terms.  This implies
As explained in section \ref{infinite_one}, the K\"ahler potential reads
\beq
e^{-K}=p(\text{Im} \, t,\zeta)+  \cO(e^{2\pi i t},\zeta) \;,
\eeq
where $\zeta$ denotes additional spectator coordinates. For convenience, we will denote the expansion of the polynomial as $p(\text{Im} \, t,\zeta)=\sum_{l=0}^d f_l(\zeta)(\text{Im} \, t)^l$. We now split the index range for the metric as $1$ denoting the $t$ coordinate, and $i,j$ denoting the other directions in field space. The different components of the metric read
\bea
g_{11} &=&\frac{d}{(\text{Im} \, t)^2}-\frac{2f_{d-1}}{f_d(\text{Im} \, t)^3}+\dots \;, \nn \\
g_{1j} &=& %\frac{\partial_j f_{d-1} f_d+\partial_j f_d f_{d-1}}{(f_d (\text{Im} \, t) +d_{d-1})^2}=
\frac{C_j}{(\text{Im} \, t)^2}-\frac{2f_{d-1}f_dC_j}{(\text{Im} \, t)^3}+\dots \;, \mathrm{\ where\ } C_j= \frac{\partial_j f_{d-1} f_d+\partial_j f_d f_{d-1}}{f_d^2} \;, \nn \\
g_{i\bar j}&=&\frac{-(\partial_j f_{d-1} + \partial_j f_{d} (\text{Im} \, t))^2 + (\partial^2_j f_{d-1} + \partial^2_j f_{d} (\text{Im} \, t)) (f_{d-1} + f_d (\text{Im} \, t))}{(f_{d-1} + f_d (\text{Im} \, t))^2} \;.
\eea
Using the above metric, it is possible to prove \cite{lee} that any real curve approaching a generic point of the divisor has infinite length,
\beq
\int_\gamma ds\geq \int \frac{\sqrt{d-\epsilon M}}{(\text{Im} \, t)}d(\text{Im} \, t)\pm \text{finite terms}\rightarrow \infty \;,
\eeq
as stated in the Theorem \eqref{One-modulus-relation}. Here $M=h^{D,1}(Y_D)-1$ and $\epsilon$ is picked small enough such that $d-\epsilon M>0$.
The BPS mass formula is slightly modified when including the dependence on the spectator moduli $\zeta$, but this can all be absorbed in the coefficients $\bf{a}(\zeta)$. 
%\be
%M_{\bf q} \simeq \frac{\sum_j \frac{1}{j!}(\text{Im} \, t) ^j S_j\left({\bf q},{\bf a}_0(\zeta)  \right)}{(2^d/d!)^{1/2}\,(\text{Im} \, t)^{\frac{d}{2}}}  \quad \text{ with }
% 0\leq j < \frac{d}{2} \;
%\ee
Therefore, it is again satisfied that the mass of these BPS states decreases  exponentially fast in the proper field distance when $\text{Im} \, t\rightarrow \infty$,
\be
\frac{M_{\bf q}(P)}{M_{\bf q}(Q)} \sim \,\frac{S_{j_{\text{max}}}\left({\bf q},{\bf a}_0(\zeta_P)  \right)}{S_{j_{\text{max}}}\left({\bf q},{\bf a}_0(\zeta_Q)  \right)} e^{-\lambda\,d_\gamma(P,Q)} \;,
\ee
as stated by the Swampland Distance Conjecture. Here $j_{\text{max}}=\{j\,|\,j<d/2\}$ and $\lambda\leq s\sqrt{2/(d-\epsilon M)}$ with $s=1,1/2$ for $d$ even or odd respectively. 

To summarise, in this section we have shown that BPS states becoming massless at an infinite distance locus have a mass which decays exponentially fast in the proper field distance when approaching the singularity. This mass behaviour is due to the universal behaviour of the field metric near infinite distance singularities. Our results therefore show that upon establishing candidate BPS states that become massless, the exponential behaviour of the Swampland Distance Conjecture will be present.

%%%%%%%%%%%%%%%%%%%%%%%%%%%%%%%%%%%%%%%%%%%%%%
\subsection{Microscopic physics for BPS stability} \label{sec:micBPS}
%%%%%%%%%%%%%%%%%%%%%%%%%%%%%%%%%%%%%%%%%%%%%%

The argument presented in the previous section for the stability of the BPS states over the monodromy orbit only utilised ${\cal N}=2$ supersymmetry. There is a deeper, more microscopic understanding of the relation between monodromy and BPS states which we discuss in this section. In particular, we can track more precisely what happens when a BPS state does encounter a curve of marginal stability over the monodromy path. The subtlety is that, while at finite distance monodromy loci the analysis appears to capture the correct physics, its application to infinite distance is less clear. However, we will present some evidence that at least the aspects of it most relevant to this paper may also hold at infinite distance. 

Consider the states $A$, $B$ and $C$ satisfying the charge relation (\ref{charrel}). In string theory these are D3 branes wrapping special Lagrangian three-cycles. The branes can form bound states. If we consider the charges of the states we see that $C$ is potentially a bound state of an anti-$A$ and $B$ while $B$ is potentially a bound state of $A$ and $C$. Then the statement of \cite{Douglas:2000gi} is that, in general, branes  $B$ and anti-$A$ form a bound state if
\be
\varphi\left(B\right)-\varphi\left(A\right) < 1 \;. \label{massstring}
\ee
We can utilise this to see how the bound state spectrum changes under monodromy. Let us consider a setup where brane $A$ is massless, or at least much lighter than branes $B$ and $C$. The prototypical example is being close to the conifold locus in moduli space where a brane becomes almost massless, and then considering the monodromy action on massive branes $B$ and $C$. Since brane $A$ is massless, we can parameterise how its central charge behaves as we circle the monodromy locus by an angle $\theta$ as $Z\left(A\right) \rightarrow  \left|Z\left(A\right) \right|e^{-i\pi\theta}$. Now, importantly, branes $B$ and $C$ are massive, and therefore their central charge angles will remain approximately constant upon circling the monodromy locus. This then implies that the bound states are stable for the values
\bea
&C& \mathrm{stable\;for\;} \theta >\left.\varphi\left(B\right)\right|_{\theta=0} - \left.\varphi\left(A\right)\right|_{\theta=0} - 1  \;, \\
&B& \mathrm{stable\;for\;} \theta <  \left.\varphi\left(B\right)\right|_{\theta=0} - \left.\varphi\left(A\right)\right|_{\theta=0}  \;.
\eea
Therefore, as we circle the monodromy locus $B$ becomes unstable and $C$ becomes stable. 

It is important to note that the fact that $M_{{\bf q}_C}\gg M_{{\bf q}_A}$ and $M_{{\bf q}_B} \gg M_{{\bf q}_A}$ was crucial in the above analysis. It implied that upon circling the monodromy locus one inevitably crosses a curve of marginal stability. We therefore find, for this case, the picture that states related by a monodromy transformation which differ by the charge of a massless state are mutually unstable, in the sense that only one of them can be stable at any point in moduli space. This picture has been generalised to having more than one but still a finite number of massless particles at the singularity in terms of multicentered BPS solutions \cite{Andriyash:2010yf}. In terms of the discussion of section \ref{sec:BPS}, we see that when we have a monodromy action mapping a type I charge to a charge which differs only by a type II charge from the initial one, then the initial BPS state is replaced by a BPS state which is a bound state of the initial state and the type II state. 	

The physics behind (\ref{massstring}) is that the mass of an open string stretching between anti-brane $\bar{A}$ and brane $B$ is given by $m^2 \sim \varphi\left(B\right)-\varphi\left(A\right) - 1$ \cite{Douglas:2000gi,Aspinwall:2002nw}. If the string is tachyonic then the branes form a bound state. Indeed, if the singularity inducing the monodromy is due to only the state $A$ becoming massless, then the monodromy action on the branes can be understood rather explicitly in terms of strings stretching between them. Consider the monodromy transformation on a massive BPS state of charge ${\bf q}_B$ such that it maps to ${\bf q}_C = T {\bf q}_B$. For such simple cases, we can relate the monodromy action to the charges
\be
\label{triangle}
{\bf q}_C = T {\bf q}_B = {\bf q}_B - S\left({\bf q}_A,{\bf q}_B \right) {\bf q}_A \;. 
\ee
The appearance of the inner product $S\left({\bf q}_A,{\bf q}_B \right)$ in (\ref{triangle}) can be understood in three ways. The first is in terms of mutual locality of the states $A$ and $B$. It amounts to the statement that a state should only undergo a monodromy transformation about a state which is not mutually local to it. The prototypical example being an electron-monopole pair. The second way to understand it is geometrically. Recall that the charge vectors are the homology classes of the three-cycles that the branes are wrapping, and $S\left({\bf q}_A,{\bf q}_B \right)$ is their intersection number. It therefore amounts to the statement that a cycle undergoes a monodromy only if it intersects the cycles which vanishes on the monodromy locus. The third way is by noting that these intersections between cycles are associated to the strings stretching between the branes wrapping them.  

So far this presents quite a coherent picture of the underlying microscopic physics. However, at infinite distance where an infinite number of states become massless, this can not be the full story. If we consider how charges transform under infinite distance monodromies $
T^k {\bf q} = {\bf q} + \delta_k {\bf q}$, as already defined in \eqref{def-deltaq},
then one can show that charges exists such that the monodromy action satisfies $S\left(\delta_k {\bf q}, {\bf q} \right)=0$. Therefore, for such charges, the monodromy action can never be written in the form (\ref{triangle}). This implies that the physics at infinite distances has some qualitative differences to physics at finite distances. However, in the analysis of section \ref{sec:BPS} we motivated the quotient monodromy orbit, where charges which differ by a type II charge are identified, by considering the curve of marginal stability for a type I state which can be written as a sum of a type I and type II states. In terms of the monodromy action (\ref{triangle}) this amounts to taking $B$ and $C$ as type I states and $A$ as type II. In section \ref{sec:orbits} we will show that in such a setup $S\left(\delta_k {\bf q}, {\bf q} \right) \neq 0$, while in appendix \ref{ap:qdq} we show that  $S\left(\delta_k {\bf q}, {\bf q} \right) = 0$ if $B$, $C$ and $A$ are type I states. Therefore, at least with respect to this type of interaction between BPS states and monodromy, the microscopic physics picture described in this section may indeed hold. If this is the case, then we can determine that the quotient by type II states is physically mapped to the statement that only one state in each equivalence class is a stable BPS state at any point in moduli space.  

It is informative to consider some examples of BPS spectra which support our proposals. Consider a simple model studied in \cite{Douglas:2000qw} of type IIA on a non-compact Calabi-Yau given by a bundle over $\mathbb{P}^2$. This is mirror to our type IIB setting, but the physics is the same. It is a one-parameter moduli space, parameterising the volume of the $\mathbb{P}^2$ and the value of the Neveu-Schwarz B-field, with a complex modulus $z$. The large volume limit is at $z=0$ with an associated infinite order monodromy. There is a $\mathbb{Z}_3$-orbifold limit at $z=\infty$ with an associated finite $\mathbb{Z}_3$ monodromy. The most directly relevant region is the large volume limit which is at infinite distance. The brane spectrum is given by coherent sheaves but for $\mathbb{P}^2$ these are always just bundles. So in the large volume regime we can consider the brane spectrum as corresponding to D4 branes wrapping the $\mathbb{P}^2$ and supporting stable bundles. The infinite monodromy action corresponds to integer shifts of the Neveu-Schwarz B-field which is equivalent to changing the bundle on the D4 by tensoring it with a line bundle. The bundle stability condition is unchanged by this action and therefore, given a stable bundle or brane state, the infinite monodromy generates an infinite orbit through BPS states in the theory in the large volume limit. 

We can also see where the quotient construction is important. At the conifold locus there is a massless state and a massive monodromy orbit where the difference between the charges in the orbit is given by the charge of the massless state. The states in this infinite orbit all decay to a product of states, including the massless state, at the conifold point.

The orbifold point is physically different to the other two because there are no massless states associated to the monodromy. It is consistent with the monodromy orbit corresponding to BPS states in the following sense. Consider the region in moduli space near the orbifold locus. Moving away from the orbifold point corresponds to a particular resolution of the orbifold singularity which breaks the $\mathbb{Z}_3$ symmetry. It was shown in \cite{Douglas:2000qw} that for each state which is present at the orbifold locus, there exists a path moving out from the orbifold point in moduli space on which it is stable. The angle of the path corresponds to the stable state on that path, and the $\mathbb{Z}_3$ monodromy action rotates between three angles and thereby the three different stable states. At the orbifold point all of the different states which were permuted become stable and the resulting theory contains a $\mathbb{Z}_3$ permuting different BPS states. It is interesting, however, as an illustration of why the quotient monodromy orbit can be a constraint which is too strong. The three different BPS states are type I states which differ by a type II charge. So the quotient monodromy orbit would identify them. However, there is no BPS state corresponding to the type II charge, it would be a massless state at the orbifold locus. Therefore, all the states in the full monodromy orbit are BPS.

\section{Infinite monodromy orbits and mixed Hodge structures\label{sec:orbits}}

In this section we introduce a mathematical machinery that appears to be 
tailor made to analyse the setting outlined in sections \ref{sec:infinite_distance} and \ref{sec:BPS_states}.
One of its most foundational results is the Nilpotent Orbit theorem, already introduced 
in section \ref{sec:infinite_distance}, but we will see that it goes far beyond that. 
The most important fact that we will use is that there is 
a natural `split' of the forms in the middle cohomology of any manifold $Y_D$ 
near the singularity that is finer than the normal $(p,q)$-decomposition and allows us 
to analyse the behaviour of the metric, central charge, and Hodge norm in detail. 
While we have a first glance at this structure in subsection~\ref{math-intro},
we will introduce the precise definition of the underlying limiting mixed Hodge structure in subsection~\ref{mixed_hodge}.
We will apply the results to Calabi-Yau threefolds in subsection \ref{sec:orbits_results}, thereby showing 
the statements about the quotient monodromy orbit of candidate BPS states summarised at the end of subsection \ref{sec:BPS}. 
It is important to stress that most of our discussion will consider one-modulus degenerations in 
moduli space. A small glance on what can happen in multi-moduli degenerations will 
be given in subsection \ref{sec:intersection}.

\subsection{A coarse introduction to the refined Hodge structure} \label{math-intro}

Before introducing the precise mathematical machinery to discuss the periods at singularities in moduli 
space, we use this introductory section to give a more intuitive overview of the appearing structures 
hopefully useful to physicists who worked on Calabi-Yau compactifications. For clarity we will restrict 
our attention to Calabi-Yau threefolds $Y_3$ and 
hence concentrate on the middle cohomology $H^{3}(Y_3,\mathbb{C})$.  
Given a fixed complex structure this middle cohomology splits by the Hodge-decomposition 
\beq \label{H3-decomp}
    H^{3}(Y_3,\mathbb{C}) = H^{3,0} \oplus H^{2,1} \oplus H^{1,2} \oplus H^{0,3}\ , 
\eeq
where the spaces $H^{p,q}$ are complex and spanned by $(p,q)$-forms that are closed but not exact. 
The dimensions $h^{p,q} = \text{dim}_{\mathbb{C}} H^{p,q}$ are the Hodge numbers. For a Calabi-Yau threefold 
one has $h^{3,0}=1$, while $h^{2,1}$ is not a priori fixed.
It turns out to be useful to define the spaces 
\begin{align} \label{pure-filtration}
    &F^3 = H^{3,0}\ , \qquad 
    &&F^{2} = H^{3,0} \oplus H^{2,1}\ ,&\\
    &F^{1}=H^{3,0} \oplus H^{2,1} \oplus H^{1,2}\ ,   &&F^0=H^{3,0} \oplus H^{2,1} \oplus H^{1,2} \oplus H^{0,3} \ .&\nn 
\end{align}
They form a filtration
$ F^3  \subset  F^2  \subset  F^1  \subset F^0 $.
On a smooth manifold the decomposition \eqref{H3-decomp} 
defines a so-called `pure polarized Hodge-structure' and \eqref{pure-filtration} a `pure Hodge filtration' as we discuss in more detail in subsection \ref{pureHodge}. One can show that the $F^i$ vary holomorphically in the complex structure deformations $z^I$. Furthermore, 
the derivatives of $F^3$ with respect to 
the fields $z^I$ yield an element of the lower $F^i$, since one shows that $\partial_{z^I} F^p \subset F^{p-1}$.
The vector spaces $F^p$ and their variation over the space $\cM_{\rm cs}$ give us a more
abstract way of thinking about the variations of $\Omega$ with respect to $z^I$.
This implies, 
in particular, that varying the complex structure  keeping $Y_3$ smooth, 
one can define the non-degenerate and positive-definite Weil-Petersson metric $g_{\rm WP}$, introduced after \eqref{Kpot_cs}, on the moduli space of complex structure deformations using $(2,1)$- and $(3,0)$-forms. 

On a singular space $Y_3$ this simple structure ceases to be sufficient to capture what happens with the metric. 
This is clear, for example, from the periods \eqref{simple_periods}, which diverge at the singular loci and hence force 
the $(3,0)$-form to develop singularities. At the singularity the information about the split \eqref{H3-decomp}
seems lost. However, we have seen in section~\ref{sec:infinite_distance} that the crucial elements in the behaviour of the metric at the singular loci are the monodromy matrix $T$, or rather the nilpotent matrix $N$ defined in \eqref{N=logT}, and the leading coefficient $\mathbf{a}_0$ in the expansion of the periods \eqref{simple_periods}. 
The underlying mathematical structure is captured 
by a so-called `limiting mixed Hodge-structure' first introduced by Schmid in \cite{schmid}, building on Deligne's work \cite{Deligne1974}. 
Focusing as in subsection \ref{infinite_one} on one-parameter degenerations $z \rightarrow 0$, the important objects are the spaces 
\beq
    F^p_\infty =\lim_{z\rightarrow 0}\ \text{exp} \Big[-\frac{1}{2\pi i} (\log z) N  \Big] F^p\ . \label{Finfdef}
\eeq
Despite the fact that this removes the overall divergent factor, 
it turns out that the vector spaces $F^p_\infty$ and the corresponding $H^{p,q}_\infty$ 
are no longer a Hodge filtration and Hodge structure for the full space $H^3(Y_3 ,\bbC)$.

The basic idea of the mixed Hodge structure is to add some finer structure capturing the 
influence of the matrix $N$. More precisely, one further splits up the $H^{p,q}$ in \eqref{H3-decomp} near the 
singularity, after removing the singular terms, to define new $I^{r,s}$ with a broader allowed index structure.
This splitting is called Deligne splitting \cite{Deligne1974} and will be discussed in more detail below. 
While the $H^{p,q}$ have $p+q=3$, the $I^{r,s}$ have $r+s \in \{0,...,6\}$, but still span 
$H^{3}(Y_3 , \mathbb{C}) = \bigoplus_{p,q} I^{p,q}$.
For example, the $H^{3,0}$  `splits' at the singularity to have contributions in potentially the following spaces 
\beq
    H^{3,0} \rightarrow   \quad \{ I^{3,3}\ ,\ I^{3,2} \ ,\ I^{3,1}\ ,\ I^{3,0}  \}\ .
\eeq
To determine where the original form $\Omega \in H^{3,0}$ actually resides in the limit depends on the type of 
singularity. We will discuss this in more detail in subsection \ref{mixed_hodge}. 
In fact, introducing the dimensions $i^{p,q} = \text{dim}_{\mathbb{C}} (I^{p,q})$ one finds 
\beq
    \sum_{q} i^{p,q}  = h^{p,3-p}  \ . 
\eeq
The spaces $I^{p,q}$ capture the non-trivial information about the nilpotent matrix $N$. In particular, 
they are constructed such that 
\beq \label{NI=I}
      N I^{p,q} \subset I^{p-1,q-1}\ . 
\eeq

One can work with forms in $I^{p,q}$ to some extend analogously to the standard $(p,q)$-forms. In particular, 
we will see in more detail below that 
\beq \label{simple-orth}
    S(I^{p,q},I^{r,s}) = 0\ , \quad \text{unless} \quad p+r = 3\ \text{and} \ q+s = 3\ . 
\eeq 
This condition corresponds to the statement that one only can integrate a top form of weight $(3,3)$
to a non-vanishing number. The conditions \eqref{simple-orth} are the implied orthogonality relations. 
More non-trivial is are the statements of when the inner product of two $(p,q)$-elements \textit{does not vanish}.
In order to give such a criterion one needs to identify a subset of so-called primitive forms 
$P^{p,q}  \subset I^{p,q}$ for $p+q\geq3$,
by demanding that all elements in this space satisfy $N^{p+q-2} P^{p,q} = 0$. One 
can then show that each $I^{p,q}$ admits a decomposition into a direct sum of the spaces $N^j P^{p-j,q-j}$, where we 
point out that this respects \eqref{NI=I}. For these primitive forms one then finds the positivity condition 
\beq \label{positive-v}
   v \in P^{p,q}\ , \ v \neq 0 \quad \Rightarrow\quad  i^{p-q} S_{p+q-3} (v,\bar v) > 0\ ,
\eeq
with $S_j(\cdot ,\cdot ) \equiv S(\cdot , N^{j} \cdot)$ as introduced in \eqref{def-Sj}.
While these properties are all similar to standard $(p,q)$-forms, there is a crucial difference 
between the two notions. Namely, in general one finds $\overline{I^{p,q}}\neq I^{q,p}$, but 
rather that $\overline{I^{p,q}}$ yields in addition to $I^{q,p}$ also elements in the lower $I^{r,s}$ with $r<q$ and $s<p$.  
Hence, the $I^{p,q}$, defined at the singular locus, are \textit{not} a standard Hodge decomposition.

Consider now, for example, an element $\mathbf{v} \in I^{3,d}$. By construction one can use the identity  
\eqref{NI=I} to conclude $N^{d+1} \mathbf{v}=0$. 
Furthermore, it is immediate that $\mathbf{v}$ is in $I^{3,d}_{\rm prim}$, following simply from the definition.
Hence, we can apply \eqref{positive-v} to conclude that its inner product $i^{3-d} S_{d}(\mathbf{v}, \mathbf{\bar v})$
is non-vanishing and positive.  This will be precisely what we need in order to address the properties of the 
$\mathbf{a}_0$-coefficient appearing in \eqref{simple_periods}. 
The Calabi-Yau condition on $Y_3$ restricts the possible splits into a mixed Hodge structure significantly.
Since $h^{3,0}=1$ there are only $4$ cases to consider 
\beq \label{a0-choices}
   \mathbf{a}_0 \in I^{3,d}=P^{3,d}\ , \quad d=0,1,2,3 \ .
\eeq
For each of these cases one can study how $h^{2,1}$ can split into $i^{2,q}$, which leads to a classification of possibly 
allowed Hodge diamonds $i^{p,q}$  \cite{Kerr2017}.

\subsection{Mathematical machinery of mixed Hodge structures} \label{mixed_hodge}

In this subsection we introduce in more detail the mathematical machinery to define and study
the mixed Hodge structure $\cH^{p,q}$, and associated Deligne splitting $I^{p,q}$, on the middle cohomology $H^{D}(Y_D,\mathbb{C})$
of a Calabi-Yau $D$-fold.  The reader feeling sufficiently informed by subsection \ref{math-intro} or 
already familiar with these mathematical structures can safely skip to subsection \ref{sec:orbits_results}.

\subsubsection{Polarized pure Hodge structures} \label{pureHodge}

To start with a more familiar concept let us first recall some facts about 
a \textit{pure} Hodge structure and Hodge filtration. 
A pure Hodge structure of weight $w$ is defined on a vector 
space $V_\bbC$, if it admits a Hodge
decomposition 
\beq \label{Hodge-decomp}
   V_\bbC = \cH^{w,0}  \oplus  \cH^{w-1,1}  \oplus \ldots  \oplus  \cH^{1,w-1}  \oplus \cH^{0,w}\ ,
\eeq
with the subspaces satisfying $\cH^{p,q} = \overline{\cH^{q,p}}$. The weight $w$ is the sum of 
the $p,q$ for the summands in \eqref{Hodge-decomp}. Using the $\cH^{p,q}$ one can also  
define a Hodge filtration as $F^p = \oplus_{i\geq p} \cH^{i,w-i}$. It is called a filtration 
since 
\beq \label{F-filtration}
   V_\bbC = F^0 \ \supset \ F^1 \ \supset\  \ldots\ \supset\   F^{w-1}\ \supset\   F^w = \cH^{w,0}\ , 
\eeq
and is required to satisfy $\cH^{p,q} = F^p \cap \bar F^q$. Clearly, the existence of such $F^p$ is 
equivalent to the existence of a pure Hodge structure \eqref{Hodge-decomp}. 
A prominent examples of a pure Hodge structure and Hodge filtration arises on the
middle cohomology $ V_\bbC \equiv H^D(Y_D ,\bbC)$ of a smooth manifold $Y_D$.
The weight of this pure Hodge structure is then $w=D$ and the spaces $\cH^{p,q} = H^{p,q}$ are the cohomology 
groups of $(p,q)$-forms spanning $H^{p,q}$. 

In a next step we introduce the notion of a \textit{polarized} pure Hodge structure. This concept essentially 
states that there is an appropriate bilinear form $S(.,.)$ on $V_\bbC$. More precisely, one 
demands that there exists an $S$ such that: 
\begin{eqnarray} \label{Hodge-Riemann}
 &(1)&  S(\cH^{p,q}, \cH^{r,s}) = 0\ , \quad \text{for}\quad p \neq s,\ q \neq r;  \\
 &(2)&  v \in \cH^{p,q}\quad v \neq 0, \quad i^{p-q} S(v,\bar v) > 0\ .  \nn
\end{eqnarray}
Let us note that 
this implies that it makes sense to introduce the Hodge norm 
\beq \label{def-Hodge-norm}
   || v ||^2 =  S(C v ,\bar v)\ , \qquad v \in V_{\bbC}\ ,
\eeq
where $C$ is a linear operator acting as $i^{p-q}$ on elements of $\cH^{p,q}$.
The familiar example for a polarized pure Hodge structure is again the middle cohomology $H^D(Y_D,\bbC)$ 
for which the bilinear form is given by 
\beq \label{non-degenerateform}
   S(\alpha, \beta) = \int_{Y_D} \alpha \wedge \beta\ .
\eeq 
The operator $C$ is nothing but the familiar Hodge-star in this case. 

On a smooth Calabi-Yau manifold we can identify $\cH^{p,q} = H^{p,q}$ 
use the $(D,0)$-form $\Omega$ as a representative of $F^D$. Its derivatives with respect to 
the fields $z^I$ yield an element of the lower $F^i$, since one shows that $\partial_{z^I} F^p \subset F^{p-1}$.
The vector spaces $F^p$ and their variation over the space $\cM_{\rm cs}$ give us a more
abstract way of thinking about the variations of $\Omega$ with respect to $z^I$.
The problem is to follow the Hodge structure $H^{p,q}$ filtration $F^p$ to the singular divisor $z=0$. Clearly, as we have seen from the example of 
$\Omega$ with periods \eqref{simple_periods} the periods generally diverge in the limit $z\rightarrow 0$. Nevertheless, one 
can define an appropriate limiting value of the $F^p$, denoted by $F^p_\infty$ as in \eqref{Finfdef}.
These limiting values still give a filtration 
\beq
 H^D(Y_D ,\bbC)  = F^0_\infty \ \supset \ F^1_\infty \ \supset\  \ldots\ \supset\   F^{D-1}_\infty\ \supset\   F^D_\infty  \ .
\eeq
However, it turns out that the vector spaces $F^p_\infty$ and the corresponding $H^{p,q}_\infty$ 
are no longer a Hodge filtration and Hodge structure for the full space $H^D(Y_D ,\bbC)$.
In particular, $S(\cdot,\cdot)$ has not the above non-degeneracy on $H^{p,q}_\infty$. 

\subsubsection{Monodromy weight filtrations and mixed Hodge structures\label{sec:filt}}

The properties of $S_{j}(\cdot,\cdot)= S(\cdot,N^j \cdot)$ are important when making contact 
with subsection \ref{infinite_one} and section \ref{sec:BPS_states}.  To address the problem of degeneracy and orthogonality we thus 
want to mix this information with the structure that $N$ induces on 
the space $V_\mathbb{C} \equiv H^D(Y_D ,\mathbb{C})$.
The main fact that we will exploit is that the nilpotent operator $N$ acting on a vector 
space $V_{\mathbb{C}}$ induces a unique monodromy weight filtration $W_{j} (N)$, 
which consists of complex vector subspaces of $V_{\mathbb{C}}$. 
These form a filtration 
\beq 
W_{-1}\equiv 0\ \subset\  W_0\ \subset\ W_1\ \subset\ ...\  \subset\ W_{2D-1}\  \subset \ W_{2D} = V_{\bbC}\ .
\label{filtration}
\eeq
This filtration becomes unique if one imposes that the following defining properties
\begin{align}
 &  1.) \quad N W_i \subset W_{i-2} &\\
   & 2.) \quad  N^j : Gr_{D+j} \rightarrow Gr_{D-j}\ \ \text{is an isomorphism,}  \hspace*{6cm}& \label{iso-prop}
\end{align}
where we have defined the graded spaces 
\beq \label{def-Gri}
  Gr_{j} \equiv W_{j}/W_{j-1}\ .
\eeq
Note that the uniqueness of this filtration can be inferred from analysing the Jordan form of $N$. 
The quotient in \eqref{def-Gri} indicates that in order to construct $Gr_i$ one considers 
elements of $W_i$ and takes them to be in the same equivalence class if they only differ by an element of $W_{i-1}$.
We have used a similar quotient construction in \eqref{quotient} for $\cM_Q = \cM/\cM_{\rm II}$. 

Let us discuss some of the properties of the $W_i$. Firstly, we note that there is a simple representation 
of the $W_i$ in terms of the kernels $\text{ker}\, N^j$ and images $\text{im}\,N^j$ as 
\bea
   W_0 &=& \text{im}\,N^D\ ,\quad W_{1} = \text{im}\, N^{D-1} \cap \text{ker}\,N\ ,\nn \\
    W_{2} &=& \text{im}\, N^{D-2} \cap \text{ker}\,N \oplus \text{im}\, N^{D-1} \cap \text{ker}\,N^2\ ,\ \ldots \ ,\quad
    W_{2D-1} = \text{ker} N^D\ .
\eea
This implies immediately that if the unipotency index is smaller than the complex dimension of the manifold, $n<D$, some of the previous subspaces will be empty. In particular,  for all $j>n$  we have $W_{D+j}=W_{D+n}$ and $W_{D-j}=0$. Such that the filtration looks like
\beq \label{restricted_filtration}
   0=W_0=W_1 =\ldots = W_{D-n-1}\ \subset\ ...\  \subset\ W_{D+n}=  W_{D+n+1}  =\ldots =  W_{2D} = V_{\bbC}\ ,
\eeq
and the non-trivial information about the filtration is in the vector spaces $W_{D-n},\ldots,W_{D+n}$.
Using the uniqueness of the filtration it is also not difficult to study orthogonality relations among the $W_i$
as 
\beq \label{orthog}
     S(W_i , W_{2D - i - j}) =0\ ,\qquad j>0 \  ,
\eeq
with the bilinear form $S(.,.)$ introduced in \eqref{def-Sj} having components $\eta$. 

We now have the required background to state a main result of Schmid \cite{schmid} for the one-modulus case.  
Namely, Schmid proved the Sl$_2$-orbit theorem, which has as one of its consequences that the induced Hodge filtration $F^p_\infty$ defined in \eqref{Finfdef} 
and the monodromy weight filtration $W_p$ defined after \eqref{filtration} form 
a \textit{mixed Hodge structure $(W,F_\infty,N)$} on the vector space $H^D(Y_D,\mathbb{C})$.\footnote{It is important to 
stress that one wants to restrict considerations to forms that are 
primitive with respect to the K\"ahler form $J$ on the Calabi-Yau $D$-fold. 
For the middle cohomology $H^D(Y_D,\mathbb{C})$ these are the forms that are trivial upon wedging with $J$.
For Calabi-Yau threefolds this condition is trivial and one finds the whole space $H^3(Y_3,\bbC)$, while 
for fourfolds this gives a non-trivial restriction to a subspace $H^4_{\rm prim}(Y_4,\bbC)$. While we work 
with these K\"ahler primitive forms in the following, we will abuse notation and drop the subscript `$\rm prim$'.} 
This structure is well-defined 
for $Y_D$ being singular. Such mixed Hodge structures have numerous applications in mathematics \cite{peters2008mixed}. 

The crucial feature of this data is that each $Gr_{j}$ defined in \eqref{def-Gri} 
admits an induced Hodge filtration 
\beq \label{Hodge-filtration}
    F^p Gr_j \equiv ( F^p_\infty \cap W_j )/ ( F^p_\infty \cap W_{j-1})\ .
\eeq
This implies 
that on a singular space $Y_D$ 
we can deal with a \textit{pure} Hodge structure of weight $j$ when restricting to the 
spaces $Gr_j$.\footnote{Note that the filtration $\left\{F^p_\infty\right\}$ in (\ref{Finfdef}) is not invariant under rescalings of the parameter $z$ and so there is no canonical choice for it. In fact this rescaling freedom can be used to set some components of elements in one of the $\left\{F^p_\infty\right\}$ to zero. However, the restriction of the filtration to the $Gr_j$ is invariant under such a rescaling. It is this restriction which has a good geometric meaning.} In other words, in the notation of \eqref{Hodge-decomp} we have to 
set
\beq \label{Grj-split}
  V^j_{\bbC} = Gr_j  = \bigoplus_{p+q=j} \cH^{p,q} \ ,\qquad   \cH^{p,q} =  F^p Gr_j \cap \overline{F^q Gr_j}\ ,
\eeq
where we recall that $w=p+q$ is the weight of the corresponding Hodge structure. 
The operator $N$ is a morphism among these pure Hodge structures. Since $N(F^{j}_\infty) \subset F^{j-1}_\infty$ and $N(W_j) \subset W_{j-2}$
one finds 
\beq \label{N-morphism}
   N Gr_j \subset Gr_{j-2}\ , \qquad N \cH^{p, q} \subset \cH^{p-1,q-1}\ .
\eeq
Note that this induces a jump in the weight of the pure Hodge structure by $-2$.
However, the mixed Hodge structure is preserved by $N$.  

Finally, it will be important for us to use the fact that $(W,F_\infty,N)$ is actually a \textit{polarized mixed Hodge structure} \cite{schmid}. 
While giving the relevant definitions here briefly, we will introduce the for us relevant form of this fact in a slightly different reincarnation 
and in more detail in subsection
\ref{D-splitting}.  
To identify a polarized mixed Hodge structure, one first introduces the primitive subspaces $\cP_i \subset Gr_i$, by setting 
$\mathcal{P}_{D+j} \equiv \text{ker}\{ N^{j+1} : Gr_{D+j} \rightarrow Gr_{D-j-2} \} \ , j \geq 0$ and 
$\mathcal{P}_{D+j} \equiv 0$, for $j < 0$. In fact, one shows that each space $Gr_j$ decomposes 
as $Gr_j = \bigoplus_{i \geq \text{max}(D-j,0)} \, N^i \cP_{j + 2 i}$. Importantly, the $\cP_j$ can be shown 
to carry a pure Hodge structure of weight $j$, \textit{polarized} with respect to the bilinear forms $S_{j-D}(.,.)$ 
introduced in \eqref{def-Sj}.

%The important property of the mixed Hodge structure for us is that the $F^p_\infty$ filtration induces a \textit{polarized} Hodge structure of 
%weight $j$ on the primitive part $\mathcal{P}_{j} \subset Gr_j$. Denoting the $(p,q)$-spaces by $\cH^{p,q}_{\rm prim}$ we split 
%\beq
%    \cP_j = \bigoplus_{p+q=j} \cH^{p,q}_{\rm prim}\ ,
%\eeq
%where we recall from \eqref{def-cP} that all $\cH^{p,q}_{\rm prim}$ with $j = p+q < D$ are trivial. 
%It is polarized with respect to the inner product $S_{j-D}(\cdot,\cdot)$. 
%More precisely, this means that the conditions \eqref{Hodge-Riemann} are satisfied, i.e.~one has
%\bea
%&(1)&  S_{j-D} (\cH^{p,q}_{\rm prim}, \cH^{r,s}_{\rm prim}) = 0 \qquad \text{for}\quad p \neq s,\ q \neq r,\ p+q=r+s=j;  \label{Sj-orthogonality1} \\ 
%  \label{polarization}
% &(2)&  v \in \cP_j\ , \quad v \neq 0\quad \Rightarrow \quad  S_{j-D} (C v,\bar v) > 0 \ ,
%\eea
%where $C$ is the Weil operator acting as $i^{p-q}$ on elements of $\cH^{p,q}_{\rm prim}$. 
%We note that this is particularly useful since due to \eqref{decgtop} one has
%\beq \label{decomp-cH}
%   \cH^{p,q}= \bigoplus_{j\geq 0} N^j \cH^{p+j,q+j}_{\rm prim}\ . 
%\eeq

\subsubsection{Deligne splitting} \label{D-splitting}

Having defined a mixed Hodge structure $(W,F,N)$,  
we can now introduce the finer split of the complex vector space 
$V_{\bbC} = H^{D}(Y_D,\mathbb{C})$. Deligne defined in \cite{Deligne1974} a splitting 
\beq \label{V-decomp}
   V_\bbC = \bigoplus_{p,q} I^{p,q} \ ,  
\eeq
where 
\beq \label{def-Ipq}
   I^{p,q} = F^p \cap W_{p+q} \cap \big( \bar F^{q} \cap W_{p+q} + \sum_{j \geq 1} \bar F^{q-j} \cap W_{p+q-j-1}  \big)\ .
\eeq
While complicated looking at first, it turns out that it is the unique splitting \cite{CKS} with the following  properties
\beq  \label{FpWl-split}
  F^{p}  = \bigoplus_s \bigoplus_{r\geq p} I^{r,s} \ ,\qquad W_{l} = \bigoplus_{p+q \leq l} I^{p,q} \ ,
\eeq
and
\beq \label{not-split}
     \overline{I^{p,q}} = I^{q,p} \ \text{mod}\ \bigoplus_{r < q,s<p} I^{r,s}\ .
\eeq
One can also easily infer in analogy to \eqref{N-morphism} that 
\beq \label{NI=I}
   N I^{p, q} \subset I^{p-1,q-1}\ .
\eeq
Note that the $I^{p,q}$ might be viewed as the analogs to the $\cH^{p,q}$. In fact, one can show 
that there exists an isomorphism identifying these spaces. The $I^{p,q}$ have the advantage 
that they yield a straightforward decomposition \eqref{V-decomp} of $V_\bbC = H^{D}(Y_D,\bbC)$, 
but have the disadvantage (in contrast to the $\cH^{p,q}$) that they only satisfy \eqref{not-split}.
In other words, for the $I^{p,q}$-decomposition the usual rules for complex conjugation of 
$(p,q)$-forms are not satisfies. This complicates the identification of real elements. 
If a splitting satisfies $ \overline{I^{p,q}} = I^{q,p}$ it 
is called \textit{split over $\mathbb{R}$}. Remarkably, as was shown in \cite{CKS}, the 
is always a unique map of $\delta$, with properties described in \cite{CKS}, such 
that $(W,e^{i\delta} F,N)$ is admitting a Deligne splitting that is split over $\mathbb{R}$.  
In the following we will not work with the $\mathbb{R}$-split case, but it can be useful to 
keep in mind that such a transformation always exists. 

To study the positivity properties of elements in $I^{p,q}$, we next introduce the primitive subspaces ${P}^{p,q} \subset I^{p,q}$ by defining 
\beq \label{def-Ppq}
    {P}^{p,q} =  I^{p,q} \cap \text{ker} N^{p+q-D+1}\ .
\eeq 
One can now check that the $I^{p,q}$ can be decomposed in terms of the $P^{p,q}$ as
\be
 I^{p,q} = \bigoplus_{i \geq 0} \, N^i (P^{p+i,q+i})   \;. \label{P-decomp}
\ee
The $P^{p,q}$ inherit a polarization relation if the underlying mixed Hodge structure is polarized. 
Concretely, one has (see e.g.~\cite{CattaniKaplan})
\bea
  S_l(P^{p,q}, P^{r,s}) &=& 0  \qquad \text{for}\quad r+s=D+l =p+q;\ (p,q) \neq (s,r)\ , \label{Sj-orthogonality1-I}\\
  i^{p-q} S_l (v , \bar v) &>& 0 \qquad \text{for} \quad v \in P^{p,q}\ , \ v\neq 0\ .  \label{polarizingSj}
\eea
These conditions use the forms $S_{l}(\cdot ,\cdot) = S(\cdot,N^l \cdot)$  introduced in \eqref{def-Sj}. 
While \eqref{Sj-orthogonality1-I} describes the orthogonality relations among $(p,q)$-forms and $(r,s)$ with $p+q=r+s$ 
one can also study the orthogonality if this condition is violated. Using the definition \eqref{def-Ipq}, the property \eqref{NI=I}, and 
the orthogonality \eqref{orthog} one finds 
\beq \label{Sj-orthogonality2-I}
    S_{j} (I^{p,q}, I^{r,s}) = 0 \quad \text{unless} \quad p+r-j= D\ \text{and} \ q+s-j = D\ ,
\eeq
which essentially states that integrals like \eqref{non-degenerateform} defining $S(\cdot,\cdot)$ can only be performed 
over top-forms. 

Let us stress that the Deligne splitting is a finer split of the usual Hodge structure 
on $H^{D}(Y_D,\bbC)$. In fact, one finds that the Hodge numbers $h^{p,D-p} = \text{dim}\, H^{p,D-p}$ are related 
to the dimensions  of  $I^{p,q}$ by 
\beq \label{HD-numbers}
    \sum_{q=0}^D i^{p,q} = h^{p,D-p}\ , \qquad i^{p,q} = \text{dim}\, I^{p,q}\ . 
\eeq
The numbers $i^{p,q}$ are sometimes referred to as Hodge-Deligne numbers and form 
a Hodge diamond as familiar from the $h^{p,q}$. The described construction implies that they satisfy the conditions 
\beq \label{i-relations}
    i^{p,q}  =  i^{q,p} =  i^{D-p,D-q}\ .  
\eeq
Clearly, for a Calabi-Yau manifold one has $h^{D,0}=1$, such 
that \eqref{HD-numbers} are further constraint for these geometries. 
A detailed account of these facts can be found in \cite{Kerr2017}, where also Calabi-Yau threefolds
are discussed in much detail. We will only summarise some relevant facts about $a_0$ 
introduced in \eqref{simple_periods_coeff}.

\subsubsection{Properties of ${\bf a}_0$}

Having introduced the mathematical machinery of mixed Hodge structures and the associated Deligne splitting, 
we are now in the position to apply them to the coefficients in the nilpotent orbit \eqref{FpWl-split}. 
Note that the mixed Hodge structure under consideration is $(W,F_\infty,N)$, i.e.~the limiting mixed 
Hodge structure at the singular locus. Since $a_0 \in F^3_\infty$ we can use \eqref{FpWl-split}
to infer
\beq
   a_0 \in   I^{D,0} \oplus I^{D,1} \oplus ... \oplus I^{D,D}\ .
\eeq
The Calabi-Yau condition $h^{D,0}=1$ together with \eqref{HD-numbers} implies that only one of these spaces can be non-trivial.
In fact, it follows from \eqref{def-d} that 
\beq \label{a0location}
   a_0 \in I^{D,d} = P^{D,d}\ . 
\eeq 
To see this, we note that $N^{d} {\bf a}_0 \neq 0$ implies that 
$\mathbf{a}_0$ has non-trivial parts in $I^{D,d} \oplus ... \oplus I^{D,D}$,
since otherwise all components of $\mathbf{a}_0 $ would be shifted to zero by $N^d$ due to the property \eqref{NI=I}.
Furthermore, the condition $N^{d+1} {\bf a}_0 = 0$ implies that $\mathbf{a}_0$ is trivial in $I^{D,d+1} \oplus ... \oplus I^{D,D}$.
To see this, note that by the definition \eqref{def-P} and \eqref{NI=I} one has $I^{D,d+i} = P^{D,d + i}$ for $i\geq 0$.
However, the polarization condition \eqref{polarizingSj} implies that $a_0$ has to be trivial in $P^{D,d+1} \oplus ... \oplus P^{D,D}$, since 
otherwise one contradicts $N^{d+1} {\bf a}_0 = 0$.

Having identified the location of ${\bf a}_0 \in I^{D,d}=P^{D,d}$, we can evaluate its properties when inserted into
 $S_{l}$. Firstly, 
note that the polarization condition \eqref{polarizingSj} directly implies 
\beq \label{Sd>02}
     i^{D-d}S_{d}( \mathbf{a}_0 , \mathbf{\bar a}_0 ) > 0 \ . 
\eeq
This result can be used in \eqref{expansion-p} to conclude that the degree of the polynomial $p$ 
is actually exactly $d$ and that the coefficient of the leading monomial is positive. 
Using these facts it is not hard to show \eqref{One-modulus-relation} and \eqref{necessary_c}, 
i.e.~one can derive the metric and check that an infinite distance point implies $d>0$. 
These results can be readily shown to hold for any $D$ and $n$. 
One might also wonder about the inner product $S( \mathbf{a}_0 , \mathbf{\bar a}_0 ) $.
Naively applying the intuition for $(p,q)$-forms suggest that it should vanish. However, this 
is not the case in general (unless the $I^{p,q}$ are split over $\mathbb{R}$ as discussed in subsection \ref{D-splitting}),
since by \eqref{not-split} one has 
\beq
    {\bf \bar a}_0 \in I^{d,D}  \bigoplus_{r < d,s<D} I^{r,s} \ . 
\eeq
Hence, one has to evaluate $S( \mathbf{a}_0 , \mathbf{\bar a}_0 )$ using all  lower $I^{r,s}$, which implies 
that the vanishing conditions \eqref{Sj-orthogonality1-I} and \eqref{Sj-orthogonality2-I} are in 
general violated and the inner product can be non-vanishing. 

Let us stress again that the mathematical machinery introduced in this 
section is, on the one hand, crucial to show \eqref{One-modulus-relation}, and, on the other hand, has to be 
employed to determining 
the monodromy orbits in subsections \ref{sec:bpsmond} and \ref{sec:BPS}. We will focus on the latter in the next subsection. 
It might appear, however, rather involved when approached in these abstract terms. Therefore, we supplement 
a detailed appendix \ref{app:explicit_examples} in which we discuss simple examples.

\subsubsection{Growth of the Hodge norm\label{sec:grownorm}}

To close this mathematical section we will state yet another result that will tie in nicely with the discussion of the 
gauge coupling function later in section \ref{sec:WGC}. More precisely, we will discuss the growth of the 
Hodge norm 
\beq
\label{hodgenorm}
    || v||^2 = \int_{Y_D} v \wedge *  \bar v \ ,
\eeq 
for a complex $D$-form $v$, when moving along a path in moduli space. This norm has already been introduced in \eqref{def-Hodge-norm}, $||v||^2 = S(C v , \bar v)$. In the following we discuss its behaviour in the local geometry when approaching 
the singular locus. These results are non-trivial and follow from the SL$_2$ orbit theorem \cite{schmid}.

To begin with we recall that we can consider a variation of Hodge structures, i.e.~how the standard $H^{p,q}$ change 
when varying the complex structure moduli. Packaged into the $F^p$ as given before \eqref{F-filtration} (see also \eqref{pure-filtration}), 
one thus defines the a bundle, with fibers varying holomorphically over $\cM_{\rm cs}$. 
As in the previous discussion we will consider a small variation in the 
local coordinates $t,\zeta^M$ used in subsection \ref{infinite_one}. Now one can pick any $D$-form $v(t,\zeta)$, which comprises 
a flat section of the above bundle.  
The growth of $v(t,\zeta)$ is in direct correspondence with the property of $v$ having support in certain spaces $W_j$. Namely 
one has $v(t) \in W_j$ if and only if the forms behaves near the singularity as 
\beq \label{vt-growth}
     || v(t) ||^2 = c_{j} (\I\, t)^{j-D} + c_{j-1} (\I\, t)^{j-D-1} + \ldots+c_0 (\I\, t)^{-D} + \cO(e^{- \I\, t}) \ ,
\eeq
for $\I\, t \rightarrow \infty$, $\R\, t$ fixed. Note that the coefficients $c_j$ can be zero in this expression, which 
is in accordance with $W_{j-1} \subset W_{j}$. The form \eqref{vt-growth} can be readily used 
to determine the growth of an element in $I^{p,q}$ with $p+q=j$, since the definition \eqref{def-Ipq} of these spaces contains an 
intersection with $W_{p+q}$ or a lower $W_{i},\, i<p+q-1$. 
Formulated in terms of the $Gr_j$ defined in \eqref{def-Gri}, one has
\beq \label{growth-theorem}
   \boxed{  \rule[-.3cm]{0cm}{.8cm} \quad  
 || v(t) ||^2 = \tilde c_{j} (\I\, t)^{j-D} + \ldots +\tilde c_0 (\I\, t)^{-D} + \cO(e^{- \I\, t}) \quad \Longleftrightarrow  \quad v(t) \in Gr_j \ ,\quad }
\eeq
which now gives precisely the leading term of the growth near the singularity, i.e.~$\tilde c_j > 0$, while 
the dots indicate all sub-leading contributions. It is important to stress, that the $t$-dependence not only arises 
from $v(t)$, but also from the norm itself. 
To check that \eqref{growth-theorem} is 
compatible with what we said before, consider $v(t) = \Omega$. We have argued in analysing the polynomial $p$ in
\eqref{expansion-p} that the leading term in $i^D \int_{Y_D} \Omega \wedge \bar \Omega$ is $(\I\, t)^d$. This is precisely what follows 
from \eqref{growth-theorem} if $\Omega(t) \in Gr_{D+d}$ compatible with our identification 
of the location of $\mathbf{a}_0$.   

The growth theorem immediately implies that there are three cases for the growth of forms to consider
\begin{align}
  &(1) \quad v \in Gr_{D+i}   \ , i>0\quad &&\text{norm goes to infinity as}\quad \I\,t\rightarrow \infty \ , &\\
 \label{GrD} &(2) \quad v \in Gr_{D}  \ , \qquad &&\text{sub-leading terms are relevant}\ , & \\
  &(3) \quad v \in Gr_{D-i}\ , i>0 \quad  &&\text{norm goes to zero as}\quad \I\,t\rightarrow \infty \ , &
\end{align}
We will see in subsection \ref{sec:orbits_results} and section \ref{sec:WGC} that the split into the different $Gr_j$ or $I^{p,q}$ with $p+q = j$
can be used to define a natural split into electric and magnetic states.

\subsection{Infinite monodromy orbits at singular loci\label{sec:orbits_results}}

In the section \ref{sec:bpsmond} we discussed special sets in charge space defined by monodromy orbits ${\cal O}_{T}[{{\bf q}_s}]$ 
and by the condition for a vanishing BPS mass on the monodromy locus $\cM_{\rm I}$ and $\cM_{\rm II}$. 
These objects admit a finer structure induced by the mixed Hodge structure, 
introduced in subsection \ref{math-intro} and \ref{mixed_hodge}, on the monodromy locus. 
In this section we analyse the features of such monodromy orbits for the case of Calabi-Yau threefolds.
We discuss the four possible cases $d=0,1,2,3$ and confirm the statements made at the end of 
subsection \ref{sec:BPS}.

\begin{table}[h!]
\begin{center}
\begin{tabular}{|c|c c c c c c c|} 
\hline
  & \multicolumn{7}{c|}{\rule[-.4cm]{0cm}{1.0cm}$\bigoplus_{p+q=j} I^{p,q}$ }\\
\hline \hline
 \rule[-.3cm]{0cm}{.8cm} $ j = 6$& & & & $\color{blue} P^{3,3}$ && & \\
\hline
 \rule[-.3cm]{0cm}{.8cm} $j=5$  & & & $\color{blue} P^{3,2}$ \hspace{-1cm}& &  \hspace{-1cm}$P^{2,3}$ \hspace{-.5cm} & & \\
\hline
\rule[-.3cm]{0cm}{.8cm} $j=4$ & & $\color{blue} P^{3,1}$ & & $P^{2,2} \oplus N P^{3,3}$  & 
   &  $P^{1,3}$  & \\
\hline

 \rule[-.3cm]{0cm}{.8cm} $j=3$ & $\color{blue} P^{3,0}$ \hspace{0cm} & &\hspace{-.5cm} $P^{2,1} \oplus N P^{3,2}$ \hspace{-1cm} & &
 \hspace{-1cm} $P^{1,2} \oplus N P^{2,3}$ \hspace{0cm} & & $P^{0,3}$  \\
 \hline
 \rule[-.3cm]{0cm}{.8cm} $j=2$ & &  $ N P^{3,1}$ \hspace{-.5cm}& &\hspace{-.5cm} $N P^{2,2} \oplus N^2 P^{3,3}$ \hspace{-.5cm} & 
   & $N P^{1,3}$  & \\
   \hline
 \rule[-.3cm]{0cm}{.8cm} $j=1$  & & &\hspace{-.5cm} $ N^2 P^{3,2} $ \hspace{-1cm}& &  \hspace{-1cm}$N^2 P^{2,3}$ \hspace{-.5cm} & & \\
\hline
  \rule[-.3cm]{0cm}{.8cm}$j=0$ & & & & $ N^3 P^{3,3}$ && & \\
\hline
\end{tabular}
\end{center}
\caption{The table shows the general from of a Deligne splitting of the third cohomology $H^{3}(Y_3,\bbC)$,
induced by mixed Hodge structure, at the singular locus. The rows correspond to the decomposition of 
$(p,q)$-forms with $p+q=j$ into the primitive spaces $P^{p,q}$.
Note that the associated Hodge diamond with $i^{p,q} = \text{dim}_{\mathbb{C}}I^{p,q}$ 
decomposition is symmetric about middle row and the diagonal due to \eqref{i-relations}. 
We indicated in blue the possible locations for $\mathbf{a}_0$, i.e.~the limiting value of  $\Omega \in H^{3,0}$.
} 
\label{tab:hodgedec}
\end{table}

It is worth more explicitly evaluating the mixed Hodge structure, or rather the associated Deligne splitting $I^{p,q}$, in this case. This is shown in table~\ref{tab:hodgedec}. In relating to Hodge structures, both ${\bf q}$ and ${\bf a}_0$ represent three-forms
%, namely ${\bf q}^T \gamma$ and ${\bf a}^T_0 \gamma$, 
when using a real integral basis of \eqref{Omega-exp}, \eqref{symplectic-basis}. However, while ${\bf a}_0$ is generally complex, the charge vector ${\bf q}$ is quantised and real. We therefore embed it into the complex-vector space (\ref{V-decomp}) by taking the real part. This implies that in order to exploit the orthogonality relations \eqref{Sj-orthogonality1-I}, \eqref{Sj-orthogonality2-I} we have to decompose each ${\bf q}$ into a elements of $I^{p,q}$ plus its complex conjugate. It is crucial to realise that one thus has to account for the property \eqref{not-split}, i.e.~that complex conjugation of $I^{p,q}$ does not only lead to elements in $I^{q,p}$ but also forms with lower $(r,s)$-weight. 

It will also turn out to be convenient for us to introduce an adapted real symplectic basis $(\alpha_K,\beta^K)$, with properties as 
stated in \eqref{symplectic-basis}, adapted to the Deligne splitting. In the absence of a singularity the splitting only 
reduces to the middle row in table~\ref{tab:hodgedec}. At the singularity  
we have the refined split of table~\ref{tab:hodgedec}. Essentially we want to use some of the $\alpha_K$ to span the spaces
in the upper rows $I^{p,q}$, $p+q>3$, and some of the $\beta^K$ to span the spaces in the lower rows $I^{p,q}$, $p+q<3$, and the remaining 
$(\alpha_K,\beta^K)$ to span the middle row. Unfortunately, again the introduction of the $(\alpha_K,\beta^K)$ basis is complicated 
by the fact that it is real and one generally has \eqref{not-split}. We will introduce the appropriate bases for the following cases in turn. 

\subsubsection{The case ${\bf a}_0 \in I^{3,3}$}

Let us first consider the case $d=3$, as already noted in \eqref{a0-choices} this implies that ${\bf a}_0 \in P^{3,3}$. 
Since $P^{3,3}$ thus has to have complex dimension $1$, we find that  
that all other $P^{3,i}$ with $i\neq 3$ are empty and table \ref{tab:hodgedec} simplifies further.   
We would now like to construct the sets of massless charges $\cM$, $\cM_{\rm I}$ and $\cM_{\rm II}$. To do this we will assume that ${\bf a}_0$ is a generic element in $P^{3,3}$. This will lead to subsets of the full possible  $\cM$, $\cM_{\rm I}$ and $\cM_{\rm II}$, but it will suffice for our purposes. 
The orthogonality relations that identify massless charges are in (\ref{masslesscondI1}), (\ref{masslesscondI2}) and (\ref{masslesscondII}).
To determine the space $\cM^{d=3}_{\rm I} $ we thus impose $S_{3}({\bf q} ,{\bf a}_0 )=S_2({\bf q} ,{\bf a}_0)=0$ 
and use the orthogonality relations \eqref{Sj-orthogonality1-I},  \eqref{Sj-orthogonality2-I} of the Deligne splitting. 
Since $N^3 \mathbf{a}_0$ and $N^2 \mathbf{a}_0$ are of type $(0,0)$ and $(1,1)$, respectively, we have to exclude charges 
of type $(3,3)$ and $(2,2)$. Using table \ref{tab:hodgedec} the remaining choices are
\beq  \label{cM3}
   \cM^{d=3} = \text{Re} \big( P^{2,1} \oplus N P^{2,2}  \oplus N^2 P^{3,3}  \oplus N^3 P^{3,3} \big)\ ,
\eeq
where we indicated that the charges are real numbers and hence one has to consider elements in the 
space plus its complex conjugate. Note that one has to use \eqref{not-split} to 
evaluate the complex conjugate. This yields
\bea
  && \overline{I^{2,1}}=\overline{P^{2,1}} = P^{1,2}\ ,\\
  && \overline{I^{1,1}} \subset I^{1,1} \oplus I^{0,0} = N P^{2,2}  \oplus N^2 P^{3,3}  \oplus N^3 P^{3,3}\;,
\eea 
 where one uses $i^{3,2}=i^{0,1} = 0$.
The condition \eqref{masslesscondI2} defining the space $\cM_{\rm I}^{d=3}$ implies that the charge has support in at least one of the last three subspaces in \eqref{cM3}. Clearly, the condition \eqref{masslesscondII}, i.e.~$S_j({\bf q} ,{\bf a}_0 )=0$ for $j=0,1,2,3$, is more restrictive 
and we find 
\beq \label{MII3}
\cM^{d=3}_{\rm II} = \text{Re} \big( P^{2,1} \big) \;. 
\eeq
In section \ref{sec:BPS_states} we showed that a charge ${\bf q}_s$ will induce an infinite monodromy orbit through massless charges ${\cal O}^{\infty}_{T}[{\bf q}_s] $ if it is not annihilated by $N$. Utilising \eqref{NI=I} we see that a sufficiently generic element in the massless charges in (\ref{cM3}) will generate such an infinite orbit. Specifically, the component $N^2 P^{3,3}$ is not annihilated by $N$. This follows because we know that $N^3 {\bf a}_0 \neq 0$ and so $N^3 P^{3,3}$ is not empty. 

Having established the existence of an infinite massless monodromy orbit ${\cal O}^{\infty}_{T}[{\bf q}_s] $ we next need to study the quotient monodromy orbit ${\cal Q}^{\infty}_{T}[{\bf q}_s] $. Recall that the quotient construction ensures that the elements in the orbit are BPS states. Evaluating the difference between charges in the orbit $\delta_k {\bf q}_s$ as defined in \eqref{def-deltaq}, we have that 
\be
\delta_k {\bf q}_s \in  \text{Re} \big( N^3 P^{3,3} \big)\;.
\ee
The quotient of this by $\cM^{d=3}_{\rm II}$, as in (\ref{MII3}), is not empty. Indeed, the fact that the vector space is a direct sum decomposition in the $P^{p,q}$ implies that the quotient has infinite elements. The quotient monodromy orbit is therefore also infinite  ${\cal Q}^{\infty}_{T}[{{\bf q}_s} ]$, as claimed at the end of subsection \ref{sec:bpsmond}.\footnote{Note that we can also explicitly see that $S\left({\bf q},\delta {\bf q}\right)=0$ which, following the discussion in section \ref{sec:micBPS}, provides some further evidence for the stability of the orbit elements.} For concreteness: 
\beq \label{final-d=3}
 \boxed{  \rule[-.3cm]{0cm}{.8cm} \quad  
{\bf q}_s\in  \text{Re} \big( P^{2,1}\oplus N P^{2,2}  \oplus \underline{N^2 P^{3,3}}  \oplus N^3 P^{3,3} \big)\ \Rightarrow \ {\cal Q}^{\infty}_{T,{\bf q}_s}\neq 0\quad }
\eeq
where we have underlined $N^2 P^{3,3}$ to indicate that the charge ${\bf q}_s$ must have support in this subspace in order to generate an infinite quotient monodromy orbit.
This presents one of the primary results of the paper. We explicitly identified an infinite tower of BPS states which become massless exponentially fast in the proper distance upon approaching any infinite distance $d=3$ locus on Calabi-Yau threefolds. 

We can also present this result in an adapted real symplectic basis $(\alpha_K, \beta^K)$. Let us define this basis by 
first introducing 
\begin{align} \label{alpha-beta-d=3}
%   &\text{span}\{ \alpha_0 \} =  \R\, P^{3,3}\ ,\qquad &&\text{span}\{ \alpha_1 \} =\R\, N P^{3,3}\ , & \\
   &\text{span}\{ \beta^1 \} =  \R  \big( N^2 P^{3,3} \big)\ ,\qquad &&\text{span}\{ \beta^0 \equiv N\beta^1 \} =\R\big( N^3 P^{3,3}\big) \ ,&\\
   &\text{span}\{ \beta^\alpha \} =  \R  \big( N P^{2,2} \big) \ ,\qquad &&\text{span}\{ \beta^a \} =\R\big( P^{2,1} \big)\ ,& \nn
\end{align}
where $\alpha=1,\ldots, i^{2,2}-1$ and $a = 1,\ldots , i^{2,1}$ with $i^{p,q}=\text{dim}_\mathbb{C} I^{p,q}$. It is not hard to check that indeed $S(\beta^K,\beta^L) = 0$, due 
to the orthogonality relations \eqref{Sj-orthogonality1-I}, \eqref{Sj-orthogonality2-I} . 
 The basis elements $\alpha_K$ are then defined via 
the symplectic pairing $S(\cdot,\cdot)$, i.e.~they are those real elements that exactly obey \eqref{symplectic-basis}.
Note, however, that it is not easy, in general, to state the span of the $\alpha_K$, since $\R( I^{p,q})$, $p+q>0$ can contain the 
lower $I^{r,s}$.
In this adapted basis the charge vector \eqref{final-d=3} reads
\beq
\label{qs-d=3}
   \mathbf{q}_s = q_0 \beta^0 + q_1 \beta^1 + q_\alpha \beta^\alpha + q_a \beta^a\ , \qquad q_1 \neq 0 \ .  
\eeq
The orbit is then generated by the action of $T$, the differences \eqref{def-deltaq} are readily 
evaluated to be $\delta_k \mathbf{q}_s = k q_1\beta^0$, with $k \in \mathbb{Z}$.

\subsubsection{The case ${\bf a}_0 \in I^{3,2}$}

Let us next consider the case $d=2$. In this case we have ${\bf a}_0 \in P^{3,2}$ and the $P^{3,i}$ with $i\neq 2$ are empty. We find for the massless spaces
\bea  \label{cM2}
   \cM^{d=2} &=& \text{Re} \big( P^{2,2}  \oplus N P^{2,2}  \oplus N^2 P^{3,2} \big)\ , \\
   \label{cMII2}
   \cM_{\rm II}^{d=2} &=& \text{Re} \big( P^{2,2}  \oplus N P^{2,2} \big)\ .
\eea
Note that for $d<3$ we require that $S_j({\bf q} ,{\bf a}_0 )=0$ for $j=1,2,3$. Also note that a charge having support in $ \text{Re} \big(  N^2 P^{3,2} \big)$ has  $S({\bf q} ,{\bf a}_0 )\neq 0$ and so is of type I. There is an infinite monodromy orbit ${\cal O}^{\infty}_{T}[{{\bf q}_s}] $, however its elements differ by
\be
\delta_k {\bf q}_s \in  \text{Re} \big( N P^{2,2} \big) \subseteq \cM_{\rm II}^{d=2}\;.
\ee
Therefore, the quotient monodromy orbit contains only a single element. Therefore, for the case $d=2$, we find that there does not exist an infinite quotient monodromy orbit. Note that we have in this case $S(\delta_k {\bf q}_s ,{\bf q}_s )\neq 0$. This lends further evidence from the microscopic perspective, as discussed in section \ref{sec:micBPS}, for considering the quotient monodromy orbit. 

Again we can introduce an adopted basis $(\alpha_K,\beta^L)$ to preset the above 
result.  We first define
\begin{align} \label{alpha-beta-d=2}
   &\text{span}\{ \beta^1 \} =  \R  \big( N P^{3,2} \big)\ ,\qquad &&\text{span}\{ \beta^0 \equiv N \beta^1\} =\R\big( N^2 P^{3,2}\big) \ ,&\\
   &\text{span}\{ \beta^\alpha \} =  \R  \big( N P^{2,2} \big) \ ,\qquad &&\text{span}\{ \beta^a \} =\R\big( P^{2,1} \big)\ ,& \nn
\end{align}
where $\alpha=1,\ldots, i^{2,2}$ and $a = 1,\ldots , i^{2,1}-1$, and introduce basis elements $\alpha_K$ to obey \eqref{symplectic-basis}.
In this adapted basis a charge vector generating an infinite orbit takes the form
\beq \label{qs-d=2} 
\mathbf{q}_s = q_0 \beta^0 + q_\alpha \beta^\alpha + \tilde q^\alpha \alpha_\alpha\ ,   \qquad  \tilde q^\alpha \neq 0\ , 
\eeq
where we indicated that there is a component along $P^{2,2}$ parametrised by $\tilde q^\alpha$.\footnote{Note that 
$\R\big(P^{2,2}\big)$ also contains the lower $I^{r,s}$ with $r<2$, $s<2$. While $\alpha_\alpha$ generally has support in the 
$\R \big( P^{2,2}\big)$ it might require to include terms involving the $\beta^K$ to actually span $\R \big( P^{2,2}\big)$.}
However, using \eqref{def-deltaq} one has $\delta_k \mathbf{q}_s = k  \tilde q^\alpha N \alpha_\alpha$, which is trivial in the quotient $\cM/\cM^{d=2}_{\rm II}$.

\subsubsection{The case ${\bf a}_0 \in I^{3,1}$}

Let us also comment on the case $d=1$ which is the remaining case for which one finds infinite distance paths according to \eqref{One-modulus-relation}. In this case we have  ${\bf a}_0 \in P^{3,1}$ and the $P^{3,i}$ with $i\neq 1$ are empty. We find for the  spaces of candidate massless states
\bea  \label{cM1}
   \cM^{d=1} &=& \text{Re} \big( P^{2,2}  \oplus N P^{2,2}  \oplus N P^{3,1} \oplus P^{2,1} \big)\ , \\
   \label{cMII2}
   \cM_{\rm II}^{d=1} &=& \text{Re} \big( P^{2,2}  \oplus N P^{2,2}  \oplus P^{2,1} \big)\ .
\eea
There is an infinite monodromy orbit ${\cal O}^{\infty}_{T}[{\bf q}_s] $, however its elements differ by
\be
\delta_k {\bf q}_s \in  \text{Re} \big( N P^{2,2} \big) \subseteq \cM_{\rm II}^{d=1}\;.
\ee
Therefore, again, the quotient monodromy orbit contains only a single element. Also note that again $S(\delta_k {\bf q}_s,{\bf q}_s )\neq 0$.  

The adopted basis $(\alpha_K,\beta^L)$ is now defined by choosing 
\begin{align} \label{alpha-beta-d=1}
   &\text{span}\{ \beta^0 \} =  \R  \big( N P^{3,1} \big)\ ,\quad \text{span}\{ \beta^\alpha \} =  \R  \big( N P^{2,2} \big)  \ ,\quad
   \text{span}\{ \beta^a \} =\R\big( P^{2,1} \big)\ ,& 
\end{align}
where $\alpha=1,\ldots, i^{2,2}$ and $a = 1,\ldots , i^{2,1}$, and introduce basis elements $\alpha_K$ to obey \eqref{symplectic-basis}.
In this adapted basis a charge vector generating an infinite orbit takes the form
\beq
   \mathbf{q}_s = q_0 \beta^0 + q_\alpha \beta^\alpha + q_a \beta^a+\tilde q^\alpha \alpha_\alpha\ ,  \qquad \tilde q^\alpha \neq 0 \ .  
\eeq
where the same cautionary remark as in \eqref{qs-d=2} concerning the $\alpha_\alpha$ applies. 
However, one has $\delta_k \mathbf{q}_s =k \tilde q^\alpha N \alpha_\alpha$, which is trivial in the quotient $\cM/\cM^{d=1}_{\rm II}$.

\subsubsection{The case ${\bf a}_0 \in I^{3,0}$}

Finally, we include a brief discussion of the case $d=0$, in which the points on the singular locus are 
not at infinite distance. Clearly, we have  ${\bf a}_0 \in P^{3,0}$ with all other $P^{3,i}$ empty. In this 
case there is only one set of massless states
\beq  \label{cM0}
   \cM^{d=0} = \cM_{\rm II}^{d=0} =  \text{Re} \big( P^{2,2}  \oplus N P^{2,2} \oplus P^{2,1} \big)\ .
\eeq
The set $\cM_{\rm I}^{d=0} $ is empty, since all states in $\cM^{d=0}$ have exponentially vanishing central charge.
Note that this result trivialises further if $N=0$, i.e.~$n=0$, since then the mixed Hodge structure reduces to a pure 
Hodge structure and $P^{2,2}$ does not exist. 
The adopted basis $(\alpha_K,\beta^L)$ is  defined as 
\begin{align} \label{alpha-beta-d=1}
   &\text{span}\{ \beta^0 \} =  \R  \big(  P^{3,0} \big)\ ,\quad \text{span}\{ \beta^\alpha \} =  \R  \big( N P^{2,2} \big)  \ ,\quad
   \text{span}\{ \beta^a \} =\R\big( P^{2,1} \big)\ ,& 
\end{align}
where $\alpha=1,\ldots, i^{2,2}$ and $a = 1,\ldots , i^{2,1}$. The basis elements $\alpha_K$ are defined to obey \eqref{symplectic-basis}.
In this adapted basis a charge vector generating an infinite orbit takes the form
\beq
   \mathbf{q}_s =  q_\alpha \beta^\alpha + q_a \beta^a+\tilde q^\alpha \alpha_\alpha\ ,  \qquad \tilde q^\alpha \neq 0 \ .  
\eeq
It is obvious that there is no quotient monodromy orbit in this case.

This completes the analysis of the the quotient monodromy orbits for the different possible (generic points on) infinite distance loci. We find that only $d=3$ loci support such an infinite orbit. This result is not ideal because it is more difficult to identify an infinite number of massless BPS states near loci with $d < 3$. However, in the next section we show that one can still utilise other monodromies in the complex-structure moduli space to identify the BPS states.

For later use, we also provide here the growth of the charge vectors spanning the different spaces $\text{Re}(P^{p,q})$. The real symplectic basis is given in \eqref{alpha-beta-d=3},\eqref{alpha-beta-d=2} and \eqref{alpha-beta-d=1} for the different cases $d=3,2,1$ and the leading behaviour of the Hodge norm is given in the growth theorem in \eqref{growth-theorem}. We note that it is crucial for us to determine the 
highest $I^{p,q}$, $j=p+q$ in which the elements in the basis $(\alpha_K,\beta^L)$ have non-trivial support. This 
allows us to identify them as representatives of $Gr_j$ and then to apply \eqref{growth-theorem}. 
For example, for $d=3$ we have $\text{span}\{ \beta^0 \} =\R\big( N^3 P^{3,3}\big) $ so $j=p+q=0$ and $\beta^0$ has support in $Gr_0$, which implies $||\beta^0||^2 \sim (\text{Im} t)^{j-3} \sim (\text{Im} t)^{-3}$. The rest of the cases are given in Table \ref{tab:growth}. Notice that we do not include the growth for $(\alpha_a,\beta^a)$ since they belong to $Gr_3$ (case (2) in \eqref{GrD}) and therefore the growth is unknown since it is completely determined by the sub-leading terms. The results for the growth of these charge vectors will be used in section \ref{sec:WGC} when computing the leading order behaviour of the gauge kinetic function.
\begin{table}[h!]
\begin{center}
\begin{tabular}{| c | c | c | c | c | c | c | } 
\hline
 \rule[-.3cm]{0cm}{0.8cm} $d$ & $||\beta^0||^2$ & $||\beta^1||^2$ & $||\beta^\alpha||^2$ & $||\alpha_0||^2$ & $||\alpha_1||^2$ & $||\alpha_\alpha||^2$  \\
\hline
 \rule[-.3cm]{0cm}{0.8cm} $0$ & \text{unknown}   & \slash & $(\text{Im} t)^{-1}$& \text{unknown}   & \slash & $(\text{Im} t)^{1}$ \\
\hline
 \rule[-.3cm]{0cm}{0.8cm} $1$ & $(\text{Im} t)^{-1}$  & \slash & $(\text{Im} t)^{-1}$& $(\text{Im} t)^{1}$  & \slash & $(\text{Im} t)^{1}$ \\
\hline
 \rule[-.3cm]{0cm}{0.8cm} $2$ &  $(\text{Im} t)^{-2}$  & \text{unknown}  & $(\text{Im} t)^{-1}$ &  $(\text{Im} t)^{2}$  & \text{unknown}  & $(\text{Im} t)^{1}$  \\
\hline
 \rule[-.3cm]{0cm}{0.8cm} $3$ & $(\text{Im} t)^{-3}$  & $(\text{Im} t)^{-1}$ & $(\text{Im} t)^{-1}$ & $(\text{Im} t)^{3}$  & $(\text{Im} t)^{1}$ & $(\text{Im} t)^{1}$   \\
\hline
\end{tabular}
\end{center}
\caption{Table showing the leading growth behaviour of the charge symplectic basis $(\alpha_K,\beta^K)$.} 
\label{tab:growth}
\end{table}

The charge symplectic basis $(\alpha_K,\beta^K)$ clearly has a natural interpretation in terms of electric and magnetic states. We denote the states associated to $\beta$ charges as electric, while the states associated to $\alpha$ charges as magnetic. It is interesting to note that for $d < 3$ type II states are such that both electric and magnetic states become massless on the monodromy locus. They therefore lead to Argyres-Douglas type theories, though we expect that the theories in the infinite distance limit are even more exotic. 

Let us finally remark about the additional condition required beyond the existence of an infinite quotient monodromy orbit which is that there should be at least one BPS state in the orbit.\footnote{Note that we require this BPS state for a given value of the $\zeta^M$ as in (\ref{local-coords}).} It is clear that there is at least one BPS state becoming massless since the monodromy locus corresponds to a singular point of the moduli space. But we do not know how to prove that such a state resides in the monodromy orbit. However, we can motivate it in terms of the Completeness Hypothesis \cite{Polchinski:2003bq} and the Weak Gravity Conjecture \cite{ArkaniHamed:2006dz}. For the case $d=3$ above we see that this amount to requiring a BPS state with a charge that has a non-vanishing component in $ \text{Re} \big( N^2 P^{3,3} \big)$. Since the spaces $P^{p,q}$ form a direct product decomposition of the charge space, having a BPS state with this non-trivial restriction is implied by a requirement that the BPS states in $\cM$ should form charge vectors that are a non-degenerate basis on $\cM$. The condition of having a state (not necessarily BPS) for a charge vector of each space $P^{p,q}$ reminds to the Completeness Conjecture. The additional requirement that the state in  $ \text{Re} \big( N^2 P^{3,3} \big)$ is indeed BPS can be guaranteed if the state satisfies the Weak Gravity bound, which for supersymmetric theories corresponds to the BPS condition $M=|Z|$. Since the presence of this single state implies the presence of all states in its monodromy orbit, the WGC is satisfied for a whole tower of particles. This is similar in spirit that the Lattice Weak Gravity Conjecture \cite{Heidenreich:2015nta,Montero:2016tif,Heidenreich:2016aqi}, but our tower of states satisfying the WGC does not form a lattice. Similar ideas about a Tower WGC have recently appeared in \cite{Andriolo:2018lvp}. Further motivation for the existence of the infinite quotient monodromy orbit of BPS states will be given in sections \ref{sec:integrating_out} and \ref{sec:WGC} from integrating out these states and recovering the behaviour of the proper field distance and the gauge kinetic function.
% analysed in the rest of the paper.

\subsection{Monodromy intersection loci\label{sec:intersection}}

In the previous sections we saw that loci with $d < 3$, but where $d \neq 0$, are at infinite distance but do not have an associated quotient infinite monodromy orbit through massless BPS states. This is not a contradiction with the proposal that such loci support an infinite number of massless BPS states, just that it is not possible to identify these through the monodromy around the infinite distance locus. In this section we will argue that even in such cases it may be possible to identify an infinite monodromy orbit through BPS states. 

The idea is to establish an infinite monodromy orbit through massless states by using a different monodromy to the one around the infinite distance locus. So, for example, one considers the intersection locus of an $n=d=1$ locus, which we label as $\cC_2$, with a different monodromy locus with $n=d=3$ which we label $\cC_1$. Let us denote the sets of charges which lead to a vanishing BPS mass on $\cC_i$ as $\cM^i$, $\cM_{\rm I}^i$ and $\cM_{\rm II}^i$ with $i=1,2$. At this intersection point there are two monodromies acting $T_1$ and $T_2$, with the indices labelling their respective loci. We have shown that $T_2$ does not generate an infinite quotient monodromy orbit through massless BPS states on $\cC_2$.\footnote{Because $n=1$ it actually does not generate an infinite monodromy orbit at all inside $\cM^2$.} However, $T_1$ can generate such an orbit, which we denote $\cQ^{\infty}_{T_1,{\bf q}} \subset \frac{\cM^2}{\cM^2_{II}}$, in the patch around the intersection $\cC_1 \cap \cC_2$. If that is the case, then locally near the intersection point we have determined an infinite number of massless BPS states. We also expect that these states remain BPS as we move away from the intersection point along the locus $\cC_2$. The reason is that, by definition, the BPS mass of these states stays vanishing anywhere along this locus and so the states should not decay. However, there could be some subtleties if boson-fermion pairs of BPS states could be lifted (see e.g. \cite{Andriyash:2010yf}), so we cannot be completely sure that there will be an infinite orbit of BPS states far away from the intersection point.

%To show this more precisely we need to generalise the previous one-monodromy construction. We consider two monodromy loci $\cC_1$ and $\cC_2$ which intersect. Let us denote the sets of charges which lead to a vanishing BPS mass on $\cC_i$ as $\cM^i$, $\cM_{\rm I}^i$ and $\cM_{\rm II}^i$ with $i=1,2$. We have the two associated monodromy matrices $T_1$ and $T_2$. We take $\cC_1$ to be a curve with $n=d=3$ and $\cC_2$ to have $n=d=1$. Then the results of section \ref{sec:orbits_results} show that $T_2$ does not generate an infinite quotient monodromy orbit.\footnote{Because $n=1$ it actually does not generate an infinite monodromy orbit at all inside $\cM^2$.} However, $T_1$ may generate such an orbit, we denote $\cQ^{\infty}_{T_1,{\bf q}} \subset \frac{\cM^2}{\cM^2_{II}}$, in the patch around the intersection $\cC_1 \cap \cC_2$.  then if we move away from this locus along the curve $\cC_2$ these massless BPS states will remain in the spectrum. This is because the curves of marginal stability for such states do not intersect $\cC_2$, as argued in section \ref{}. 

There are two problems with arguing for infinite massless states using this method. The first is that it relies on the intersection structure of infinite distance loci which is global data of the moduli space. This means that we will not be able to show any results in generality. Instead, we can only give examples to motivate such a possibility. The second problem is that there are known isolated $n=d=1$ loci. Since these do not intersect any other monodromy locus such a construction can not be carried straightforwardly to them. We will discuss these examples cases and show that they do share some interesting similarities with the cases where the $n=d=1$ is not isolated, which leaves a possibility that they could be eventually understood in a similar way.

Let us first give an example of such a construction. We consider the manifold $\mathbb{P}^{\left(1,1,2,2,2\right)}$ studied in detail in \cite{Candelas:1993dm}. The moduli space is two (complex) dimensional and contains a curve with $n=d=1$ which we denote as $\cC_2$, and is denoted $C_{\infty}$ and $D_{(1,0)}$ in \cite{Candelas:1993dm}. There is another curve which is maximally unipotent with $n=d=3$ which we denote $\cC_1$ and is denoted $D_{(0,-1)}$ in \cite{Candelas:1993dm}. The two curves intersect at a point in the moduli space and this is the single Large Complex-Structure point. The associated monodromies are $T_1$ and $T_2$ with logarithms $N_1$ and $N_2$. In order to establish an infinite quotient monodromy orbit we need to show that such an orbit is generated by $T_1$ near the intersection point. The monodromy matrices are
\be
N_1 = 
\left (
\begin{array}{cccccc}
0 & 0 & 0 & 0 & 0 & 0 \\
1 & 0 & 0 & 0 & 0 & 0 \\
0 & 0 & 0 & 0 & 0 & 0 \\
-2 & -4 & 0 & 0 & 0 & 0 \\
0 & -8 & -4 & 0 & 0 & 0 \\
-\frac{22}{3} & 0 & -2 & 0 & -1 & 0 
\end{array}
\right) \;,\;\; 
N_2 = 
\left (
\begin{array}{cccccc}
0 & 0 & 0 & 0 & 0 & 0 \\
0 & 0 & 0 & 0 & 0 & 0 \\
1 & 0 & 0 & 0 & 0 & 0 \\
0 & 0 & 0 & 0 & 0 & 0 \\
-2 & -4 & 0 & 0 & 0 & 0 \\
-2 & -2 & 0 & -1 & 0 & 0 
\end{array}
\right) \;.
\ee
We denote ${\bf a}_0$ as the appropriate one for the one-parameter nilpotent orbit associated to $N_2$. So such that the general formulae for the one-parameter case, such as (\ref{bpsmassimt}), hold in this case. It takes the form
\be
{\bf a}_0 = \left (
\begin{array}{c}
1 \\ t_1 \\ 0 \\ -1 -2 t_1 -2 t_1^2 \\ -\frac{11}{3} - 4t_1^2 \\ \frac{1}{3}\left( -11 t_1 + 4 t_1^3 + 6 \xi \right) \\
\end{array}
\right) \;.
\ee
Here $t_1$ and $\xi$ are complex parameters which have specific geometric meanings in \cite{Candelas:1993dm}, but which are not important for our discussion. The chosen basis is such that $\eta$ takes the six-dimensional form of (\ref{etacan}). Then charges ${\bf q}$ which are in $\cM^2_{\rm I}$ have to satisfy $S_1\left({\bf q},{\bf a}_0 \right)=0$ and $S_0\left({\bf q},{\bf a}_0 \right)\neq 0$, while the charges in $\cM^2_{\rm II}$ have $S_1\left({\bf q},{\bf a}_0 \right)=S\left({\bf q},{\bf a}_0 \right)=0$ . Explicitly we see that such massless charges take the form
\be
\label{orbit_intersection}
\cM^2_{\rm I} \simeq 
 \left (
\begin{array}{c}
0 \\ 0 \\ q_2 \\ 0 \\ q_4 \\ q_5 \\
\end{array}
\right)  \;,\;\;  
\cM^2_{\rm II} \simeq \emptyset\;.
\ee
It is now manifest that $N_1$ acts non-trivially on states in $\cM^2_{\rm I}$ and that $S_1\left(N_1  {\bf q},{\bf a}_0\right)=0$. Therefore, the maximally unipotent monodromy $T_1$ generates an infinite orbit $\cO^{\infty}_{T_1,{\bf q}} $. Since $\cM^2_{\rm II}$ is empty this maps directly to an infinite quotient monodromy orbit $\cQ^{\infty}_{T_1,{\bf q}} $. 

As we mentioned there are examples where the locus with $n=d=1$ is isolated. In particular the one-modulus cases given in (\ref{TDcases}) all have $n=d=1$ loci which must be isolated since they are points. However, there is a sense in which they are quite similar to the $n=d=1$ locus in the two-parameter $\mathbb{P}^{\left(1,1,2,2,2\right)}$ model. This is most directly seen by considering the mirror manifolds. In the mirror type IIA setting the appropriate branes are given by coherent sheaves in the derived category. The large complex structure point where the two curves $\cC_1$ and $\cC_2$ intersect is mirror to the the large volume point. The geometry of the mirror is a $K3$ fibration over a $\mathbb{P}^1$ base (see for example \cite{Aspinwall:2002nw}). The generic point on the $n=d=1$ locus $\cC_2$ is mirror to the limit where the volume of the $\mathbb{P}^1$ goes to infinity while the volume of the $K3$ stays finite. In this limit any D2 brane wrapping a holomorphic curve in the $K3$ becomes physically massless, which directly identifies an infinite number of massless states. In terms of our monodromy orbits, the large complex-structure point  where $\cC_2$ and $\cC_1$ intersect is mirror to the large volume limit where both the $\mathbb{P}^1$ and $K3$ develop infinite volumes. The $T_1$ monodromy which generates the infinite orbit near the large complex-structure point is therefore naturally associated to this $K3$ fibration structure. 

Let us return to the one-parameter models with the isolated $n=d=1$ point. We consider explicitly the $\mathbb{P}^5\left[3,3\right]$ case following the analysis in \cite{DHT}. This corresponds to the complete intersection Calabi-Yau (see, for example, \cite{Walcher:2009uj} for a discussion)
\bea
& &\frac{x_1^3}{3} + \frac{x_2^3}{3} + \frac{x_3^3}{3} - \frac{1}{z^{\frac{1}{6}}} x_4 x_5 x_6 = 0 \;, \nn \\
& &\frac{x_4^3}{3} + \frac{x_5^3}{3} + \frac{x_6^3}{3} - \frac{1}{z^{\frac{1}{6}}} x_1 x_2 x_3 = 0 \;,
\eea
where the $x_i$ are coordinates on $\mathbb{P}^5$ and $z$ is the complex-structure modulus so that the monodromy point with $n=d=1$ is at $z=\infty$. This type of degeneration is called a Tyurin degeneration \cite{TYU}. At this point we see that the fibre splits into a union of two Fano three-folds and these actually intersect over a $K3$. We therefore see a $K3$ emerge, however, the mirror manifold cannot have a $K3$ fibration. It was nonetheless shown in \cite{DHT} that if we replace the mirror Calabi-Yau by its bounded derived category of coherent sheaves, which is the relevant object for the D-brane states, then one recovers what is called a non-commutative $K3$ surface in \cite{CT}. The similarity of the $K3$ structures between the one-parameter and two-parameter examples hints that perhaps even the isolated $n=d=1$ loci may have some similar structure to that found at the intersection locus between the $\cC_1$ and $\cC_2$ curves where the $T_1$ monodromy played a role. However, we leave a more detailed investigation along this direction for future work.

%%%%%%%%%%%%%%%%%%%%%%%%%%%%%%%%%%%%%%%%%%%%%
\section{Infinite distances from integrating out states\label{sec:integrating_out}} 
%%%%%%%%%%%%%%%%%%%%%%%%%%%%%%%%%%%%%%%%%%%%%

The work so far has focused on evidence for a relation between infinite distances in field space and towers of states which become exponentially light. However, an underlying microscopic fundamental physics explanation for this correlation is so far missing. In this section we propose such an explanation. We propose that the correlation exists because infinite distances in field space are a consequence of the infinite tower of states. Specifically, integrating out the infinite tower of states induces an infinite distance in the low-energy effective field theory.\footnote{In \cite{HRR-toappear} a similar proposal was reached independently.} We will present highly non-trivial evidence for this proposal in the context of the Calabi-Yau compactifications studied in this work by matching the results of integrating out the BPS states with the behaviour of the moduli space. 

A well-known and fascinating phenomenon is the ability of string theory to automatically include quantum effects in the low energy description of certain string compactifications. Therefore, moduli spaces in string theory are quantum in nature. In particular, singularities in the moduli space of string vacua can be explained by the existence of physical states which become massless at the singularity. The breakdown of the low energy effective theory arises from integrating out `wrongly' these states, and the divergence of some physical quantities near the singularity can be re-derived by computing the effect of the one-loop quantum corrections in a Wilsonian effective field theory approach.

The typical example is the conifold singularity of the moduli space of Calabi-Yau compactifications of Type II string theories \cite{Strominger:1995cz,Vafa:1995ta}. The logarithmic divergence of the metric at the conifold singularity can be obtained at one loop by integrating out a single charged hypermultiplet corresponding to a BPS state which becomes massless at the conifold point. 
Other examples are singularities in $N=2$ Yang-Mills theory which are resolved by the inclusion of massless BPS magnetic monopoles \cite{Seiberg:1994rs}, or orbifold singularities in K3 compactifications of Type II theory, where the massless states correspond to RR solitons of spin one \cite{Harvey:1995rn}. 

This means that the infinite tower of BPS states that we have explicitly identified at infinite distances in moduli space is already integrated out into the structure of the moduli space. Our proposition is therefore to identify the divergence in the field distance with the effect on the moduli space of this integrating out. As we will explain later, the one-loop contribution to the field distance of integrating out a single state is always finite, so a divergence can only appear if the number of states becoming massless at the singularity is indeed infinite \cite{Ooguri:2006in}. This explains why the conifold point, with only one state becoming massless, is still at finite distance in the moduli space. In this section, we show how the quantum corrections to the field metric coming from integrating out the infinite monodromy orbit of massless BPS states at one-loop yield indeed a logarithmic divergence in the field distance of a trajectory approaching the singularity. This supports the identification of the massless monodromy orbit of infinite order found in the previous section as the origin of infinite distance points in the moduli space of $N=2$ Calabi-Yau compactifications. 

The integrating out procedure can only be performed within the realm of an effective quantum field theory. We therefore can only integrate out the BPS states starting from some UV scale. We show that the natural UV scale is the so-called {\it species scale} (see for example \cite{ArkaniHamed:2005yv,Distler:2005hi,Dimopoulos:2005ac,Dvali:2007wp,Dvali:2007hz})
\be
\label{spesca}
\Lambda_{\mathrm{Species}} = \frac{M_p}{\sqrt{S}} \;,
\ee
where $S$ is the number of particles below the species scale. Actually, we will also match this scale onto the stability of the tower of BPS states. Integrating out the states from this scale precisely reproduces the logarithmic behaviour in moduli space. This nicely matches onto the ideas of emergence from that scale for the Weak Gravity Conjecture, as will be studied in section \ref{sec:WGC}, and as also proposed in \cite{Harlow:2015lma,Heidenreich:2017sim}.

\subsection{Field space corrections from integrating out states}
\label{sec:intgen}

Let us consider a four-dimensional effective theory with two scalar particles $h$ and $\phi$. We take $\phi$ to be massless and $h$ to have a mass $m$ which depends on $\phi$. The Lagrangian is 
\be
\mathcal{L} =  \half \left(\partial h \right)^2 + \frac12 \left(\partial \phi \right)^2+ \frac12 m\left(\phi\right)^2 h^2 \;.
\ee
We would like to work with an effective field theory with a cut-off below $m$ where the heavy scalar $h$ has been integrated out. In this effective theory the scalar propagator for $\phi$ will receive a one-loop correction from integrating out $h$ due to the cubic interaction originating from the mass term. Specifically, if we write $\phi = \left< \phi \right> + \delta \phi$ then the interaction term is 
\be
m(\phi)^2h^2 \supset 2\left[m \left(\partial_\phi m\right)\right]_{\left< \phi \right>} \delta \phi \;h^2 \;.
\ee
By computing this one-loop diagram, we obtain that the field space metric at the low energy effective theory involving only $\phi$ is given by
\beq
\label{metric_scalar}
 g_{\phi\phi}=\frac12 + \frac{(\partial_\phi m)^2}{8\pi^2}\left(\frac{2\pi}{3\sqrt{3}}-1\right) \;.
\eeq
Here, field space metric is such that the low-energy effective theory is
\be
\label{leee}
\mathcal{L} =  g_{\phi\phi} \left(\partial \phi \right)^2 \;.
\ee
The second term in the field space metric (\ref{metric_scalar}) is a one-loop quantum correction. In general, the metric will also receive higher order corrections. Further, even at 1-loop there are other corrections but they appear at higher powers of the coupling $\partial_\phi m$ so they are subleading and we will not consider them for simplicity here. The proper field distance between two points as measured by the quantum corrected field space metric is given by 
\beq
d(\phi_1,\phi_2) \simeq C \int_{\phi_1}^{\phi_2} (\partial_\phi m) \ d\phi= C(m(\phi_2)-m(\phi_1)) \;,
\label{distance1}
\eeq
where $C$ is the constant factor in \eqref{metric_scalar}. 
If we approach a point at which $m(\phi_1)=0$, the low energy effective theory involving only $\phi$ breaks down, but the proper field distance to this point is always finite due to the finiteness of $m(\phi_2)$. Note that the goal here is to show that the quantum correction of integrating out a single scalar particle can never generate a divergence on the field distance, so we have omitted the classical contribution for simplicity, although this one is not necessarily subleading. 

Let us repeat the procedure this time with a heavy Fermion $\psi$. We consider the theory 
\be
\label{uvpsith}
\mathcal{L} =  \bar\psi\partial_\mu\gamma^\mu \psi + \frac12 \left(\partial \phi \right)^2-  m\left(\phi\right) \bar\psi\psi \;.
\ee
We can then integrate out $\psi$ to obtain again the low energy effective theory only in terms of $\phi$. The field space metric at low energies is given by
\beq
\label{metrscapsi}
g_{\phi\phi}=g^{\mathrm{UV}}_{\phi\phi} + \frac{(\partial_\phi m)^2}{8\pi^2}\left(\log\frac{\Lambda_{UV}^2}{m^2}\right) \;.
\eeq
Here we have matched at the scale $m$ the correction to the scalar propagator in the full theory with the field space metric in the low energy effective field theory. The logarithmic term comes from the fact that the one-loop fermionic contribution, unlike the scalar one\footnote{Integrating out a scalar field can also give rise to a logarithmic running of the field space metric but only at order $(\partial_\phi m)^4$ and higher. }, yields a logarithmic running of the field space metric. The scale $\Lambda_{\rm UV}$ is the cut-off scale of the original theory with $\psi$ as in (\ref{uvpsith}). The value of the field space metric at the UV scale is $g^{\mathrm{UV}}_{\phi\phi}$. For studying the case of a single particle being integrated out, we can take this to be of order one.

The regime of interest for us is one where the quantum part of the field space metric dominates over the classical value. Keeping within a perturbative regime we require $\left( \partial_{\phi} m \right) \ll 1$ and so the scalar correction (\ref{metric_scalar}) does not naturally allow for such a setting. However, in the fermionic case (\ref{metrscapsi}) we see that as $\frac{\Lambda_{\rm UV}}{m} \rightarrow \infty$ we recover such a limit. In fact, this divergence is a singular locus in $\phi$ moduli space at $m\left( \phi \right) =0$, which in string theory is precisely the conifold locus. But it is important to remark that, even if the field metric diverges, such a divergence still leads to a finite proper distance up to the singular point. This can be computed analogously to \eqref{distance1}, obtaining a finite result for the proper field distance. Therefore, to find a quantum effect leading to infinite proper distance we need to consider a different regime where the quantum part dominates which is when there are many particles being integrated out.

Suppose then the four-dimensional low energy effective theory of a single scalar field $\phi$ arising from integrating out, not only one, but S heavy scalar fields $h_i$ whose masses are parametrised by $\phi$, so 
\be
\label{effthemulh}
\mathcal{L}= g^{\mathrm{UV}}_{\phi\phi} \left(\partial \phi \right)^2 + \sum_{i=0}^S \left[  \half \left(\partial h_i \right)^2 + \half m_i\left(\phi\right)^2h_i^2 \right]\;.
\ee 
Here we have specified the theory at a UV scale where the tower of heavy scalars is in the effective theory. As discussed below, we expect that this scale will involve quite exotic physics and so we keep the kinetic term for $\phi$ as unspecified at this point. We now integrate out at one loop the heavy scalars. If $\sum_{i=0}^S (\partial_\phi m_i)^2 \gg g^{\mathrm{UV}}_{\phi\phi}$ then the quantum part dominates the effective field space metric. 
In the following we will operate under the assumption that indeed $g^{\mathrm{UV}}_{\phi\phi}$ is always sub-dominant to the quantum corrections. We will return to this point in the next section. In this regime the proper distance between two points $\phi_1$ and $\phi_2$ is given by
\beq
d(\phi_1,\phi_2) \simeq C \int_{\phi_1}^{\phi_2} \sqrt{\sum_{i=0}^S (\partial_\phi m_i)^2} \ d\phi \;.
\label{distance}
\eeq
Here $C$ is some constant pre-factor which will not play an important role in our discussion. 

\begin{figure}[t]
\begin{center}
\includegraphics[width=7cm]{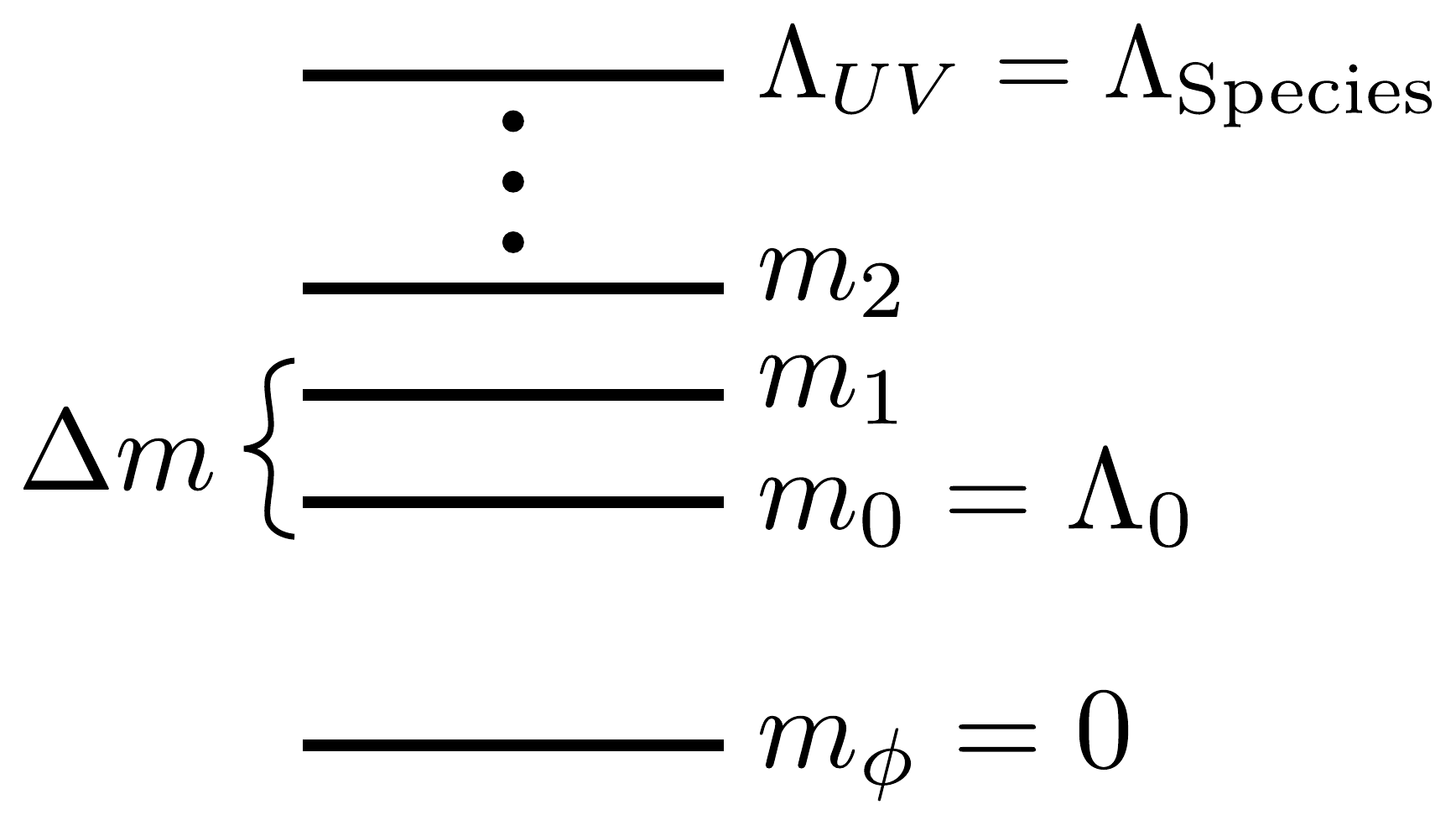}
\end{center}
\caption{Effective theory of one light scalar field $\phi$ and S heavy ones $h_i$ up to $\Lambda_{UV}$.}
\label{masses}
\end{figure}

This setup has two effective theory cut-off scales. The first, $\Lambda_{\rm UV}$ is the cut-off scale of the theory (\ref{effthemulh}). It determines how many states can be present in the theory and so fixes $S$. In particular there is an upper bound due to gravity for this cut-off which is the species scale (\ref{spesca}), above which gravity becomes strongly coupled and the effective theory entirely breaks down. We will take indeed this as the cut-off scale for (\ref{effthemulh}) so $\Lambda_{\rm UV} \sim \Lambda_{\rm Species}$. 
The second cut-off scale is for the low-energy effective theory where the massive states are integrated out (\ref{leee}). We denote this  $\Lambda_{0}$. It is set by the mass of the lightest massive state $\Lambda_{0} \sim m_0$. The scales are shown in figure \ref{masses}. 

Next we would like to rewrite $\Lambda_{UV}$ for the case where the massive states are such that they form a tower of states with increasing mass. So we consider the case where
\be
\label{masskto}
m_{k}\left(\phi\right) = m_{0}\left(\phi\right) + k \Delta m\left(\phi\right) \;.
\ee
Here $m_0$ is the mass of the lightest state and $\Delta m$ is the separation scale in the tower between states, which can depend on $\phi$. It is also shown in figure \ref{masses}. Assuming that $S \Delta m \gtrsim m_0$, which will hold in our cases of interest, we can relate the UV cut-off scale and the number of states to the mass separation in the tower as
\beq
\Lambda_{UV} \simeq \left(M_p^2 \Delta m\right)^{\frac13} \;, \;\; S \simeq \left(\frac{M_p}{\Delta m} \right)^{\frac23}\;.
\label{species}
\eeq 
With this result we see that the full structure of the integrating out procedure is determined by the properties of the spectrum of massive states. This is precisely what we have determined in this paper for a subset of the spectrum of BPS states. We therefore can calculate the effect of this spectrum, and this is done in the next section.

Before moving to this let us remark about the similar analysis for the fermionic case. In the case of $N$ fermions we have that the relevant expression is
\beq
\label{metrscapsiN}
g_{\phi\phi}=g^{\mathrm{UV}}_{\phi\phi} + \sum_i^S \frac{\left( \partial_\phi m_i\right)^2}{8\pi^2}\left(\log\frac{\Lambda_{UV}^2}{m_i^2}\right) \;.
\eeq
There are then two potentially large contributions, one from the number of particles $S$ and one from the logarithmic part. In the conifold like case where, $S=1$ and $\frac{\Lambda_{UV}}{m} \rightarrow \infty$ the logarithm was the important piece. However, now we will focus on the contribution from $S$. In fact we will see that as we approach infinite distance $S$ will diverge while the logarithm will flow to a constant. 

\subsection{Application to the monodromy loci}

In the previous section we studied in a toy model how the properties of a tower of massive states affects the proper field distance in an effective theory where the states are integrated out. In this section we would like to apply these ideas to the actual string theory setting of type IIB on Calabi-Yau threefolds with the D3-brane BPS states playing the role of the tower of states to be integrated out. There are a number of differences between this full model and the toy model of the previous section. A very useful difference is that due to ${\cal N}=2$ supersymmetry the correction from integrating out the states is one-loop exact in perturbation theory. This is crucial in order to trust a regime where the one-loop correction may dominate the classical one. The limit approaching a conifold singularity is an example of such a setting where the metric is dominated by the one-loop term. By contrast, a difference which makes the analysis much more involved is that we do not really have a UV theory to start from. In particular, we only have information on the massive BPS states from the effective IR theory. 

In this work we only perform an initial analysis of the integrating out procedure and so will adopt a simplified approach. We will not attempt to construct the UV theory and integrate out the states scale-by-scale. Instead we will apply the analysis of the toy model of section \ref{sec:intgen} directly to the masses of the BPS states as given by the IR theory. We believe that this captures the key physics in the following sense. The string theory setting implies that in the IR theory all the UV physics has been already integrated out. This manifests in the field space metric but also in the IR expression for the mass of the BPS states.\footnote{Unlike the case of the conifold, where the IR expression for the mass of the state only receives a sub-leading correction from integrating out the BPS state itself, approaching infinite distance the IR expression can be dominated by the corrections from integrating out the states.} Therefore, in utilising the IR expressions for the masses of the massive BPS states we are already accounting for the corrections to the masses from integrating out the BPS states. We therefore expect that applying the toy model analysis of section \ref{sec:intgen}, utilising these corrected IR mass expressions, captures the corrections from integrating out the states to the field space metric. This will be our working assumption, though we are aware that a full analysis of the integrating out procedure would start from a UV theory and integrate out the tower of states accounting for the corrections to the field space metric and the mass of the states on the same footing at each scale. We note, however, that should the UV theory be such that the one-loop corrections are not dominant over the classical expression then the analysis performed in this section becomes much more explicit. Since in such a case the IR expressions for the mass of the BPS states can be matched onto the expressions for their mass in the UV theory. 

Finally, a more trivial difference between the toy model and ${\cal N}=2$ analysis is that the BPS states are hypermultiplets which means they contribute as four real scalars and two fermions in the loop diagram. This essentially modifies the constant $C$ in the final result (\ref{onemodlog}), and so we will not account for this difference in detail.  

In the field space we have a one-parameter approach to infinite distance loci determined as $\I\;t \rightarrow \infty$. To match the previous discussion let us relabel $\phi = \I\;t$. Recall that any infinite distance locus can be classified by an integer $d$, defined in \eqref{def-d}, which ranges between one and three. The tower of states we will consider are the BPS states identified in the infinite quotient monodromy orbit in section \ref{sec:orbits}.\footnote{For $d<3$ we will consider massless BPS states which are not generated by the local monodromy, since as shown in section \ref{sec:orbits}, the local monodromy does not generate such a tower. The possible relation to a monodromy action would be through global aspects as discussed in section \ref{sec:intersection}.} Each state is specified by its charge ${\bf q}$. They have a mass given to leading order by \eqref{Mlead}
%Here we apply the general results of the previous sub-se\ction to the specific case of BPS states around monodromy loci. 
\be
M_{\bf q} \simeq \frac{\sum_j \frac{1}{j!}\phi ^j S_j\left({\bf q},{\bf a}_0  \right)}{(2^d/d!)^{1/2}\,\phi^{\frac{d}{2}}}  \quad \text{ with }
S_j\left({\bf q},{\bf a}_0  \right) = 0 \;,\; \mathrm{for\;all}\; j \geq \frac{d}{2} \;.
\ee
Denoting the $\phi$-independent coefficients as $c_j\equiv \sqrt{\frac{d!}{2^d}} \frac{1}{j!}S_j\left({\bf q},{\bf a}_0  \right)$, the mass formula reads
\be
M_{\bf q} \simeq  \frac{1}{\phi^{d/2}}\left( c_0+c_1\phi+\dots c_r \phi^r\right) \quad \text{with } r< d/2 \;.
\label{M5}
\ee
The quotient monodromy orbit consists of states with charges ${\bf q}$ which transforms under the monodromy transformation $T$ as $M_{T\bf q}\simeq M_{\bf q} +c/\phi^{d/2}$ in accordance with \eqref{DeltaZ}, where $c$ is a constant which depends on the specific properties of each example.\footnote{For instance, $c=c_1=\sqrt{3/4}\, S_1\left({\bf q},{\bf a}_0  \right)$ for the $n=d=3$ case. For $d < 3$ we argued that in terms of a monodromy transformation we must utilise a different monodromy to the one about the infinite distance locus, say $T_2$. In that case we have $c=\sqrt{d!/2^d}(S_0\left(T_2 {\bf q},{\bf a}_0 (\zeta) \right)-S_0\left({\bf q},{\bf a}_0 (\zeta) \right))$.} Connecting with the notation of subsection \ref{sec:intgen} we thus identify 
$m_i = M_{T^i \bf q_{s}}$, for some charge $\bf q_{s}$ seeding the orbit. 
Hence, we conclude that for the considered tower of BPS states one has 
\beq \label{m0Deltam}
  \Lambda_0 = m_0 \simeq \left. \begin{array}{c}  \phi^{-\frac12} \\ \phi^{-\frac{d}{2}} \end{array} \right\} \begin{array}{l}  d=3 \\ d<3 \end{array} \;,  \qquad \Delta m \simeq \phi^{- d/2}\ .
\eeq
Here we set $M_p=1$ and suppressed constant coefficients, which will not play a role in our discussion, for simplicity. Furthermore, let us 
stress that in order to obtain \eqref{m0Deltam} we use that we quotiented by type II states, 
since otherwise the density of states can be exponentially high $\Delta m \sim e^{-\phi}$.

The result \eqref{m0Deltam}  can now be used in the general analysis of subsection \ref{sec:intgen}. We obtain from (\ref{species}) that
\be
%\Lambda_0 = m_0 \simeq \left. \begin{array}{c}  \phi^{-\frac12} \\ \phi^{-\frac{d}{2}} \end{array} \right\} \begin{array}{l}  d=3 \\ d<3 \end{array} \;, \;\; \Delta m \simeq \phi^{-\frac{d}{2}} \;,\;\; 
S \simeq \phi^{\frac{d}{3}} \;,\qquad \Lambda_{UV} \simeq \phi^{-\frac{d}{6}} \;.
\label{expformodp}
\ee
This assumes that the tower of states consists of stable BPS states up to the $S^{\rm th}$ element. 
In section \ref{sec:BPS} we determined that the BPS states in the tower go up to element of order $\phi$. We see that this beautifully matches the maximum growth of $S$ in (\ref{expformodp}). 
Finally, we can now insert the results \eqref{m0Deltam} and (\ref{expformodp}) into the general formula for the proper field distance (\ref{distance}), using (\ref{masskto}), obtaining
\beq
\label{distinout}
d\left(\phi_1,\phi_2\right) \simeq  C \int_{\phi_1}^{\phi_2} \frac{d}{\sqrt{12}c}\frac{1}{\phi}d\phi= C\frac{d}{\sqrt{12}c} \log\left(\frac{\phi_2}{\phi_1}\right) \;.
\eeq
This precisely reproduces the logarithmic behaviour seen in the proper distance in field space when approaching infinite distance (\ref{onemodlog}). Equivalently, it implies the exponential behaviour of the mass of the states in the proper distance. The result therefore forms non-trivial evidence for our proposition for the tower of states as the origin of infinite distance. 

In the case when the tower of states which we integrate out are fermions we need to consider the expression (\ref{metrscapsiN}). The analysis proceeds in the same way with the only difference being the additional logarithmic factor. While for the finite distance case the logarithmic factor was divergent, in the infinite distance case it behaves as a constant in the following sense. The primary contributions to the quantum corrected metric in the expression for the proper distance (\ref{distinout}) come from heavy modes in the tower. Their mass behaves as $S \Delta m \sim \phi^{-\frac{d}{6}} \sim \Lambda_{UV}$. Therefore, we see that the logarithm behaves as a constant to leading order in the expression for the metric and we recover the same parametric behaviour as in \eqref{distinout}.

%There are some comments to make regarding the difference between the full ${\cal N}=2$ Calabi-Yau setting and the general toy model in section \ref{sec:intgen}. In reaching the result (\ref{distinout}) we have combined these two rather imprecisely. The BPS states are hypermultiplets which means they contribute as four real scalars and two fermions in the loop diagram. This essentially modifies the constant $C$ in the final result (\ref{onemodlog}). 
%
%A more fundamental difference is that due to ${\cal N}=2$ supersymmetry the correction from integrating out the states is one-loop exact in perturbation theory. This is helpful in order to trust a regime where the one-loop correction is dominant. Another difference is that in the analysis of section \ref{sec:intgen} we took $g_{\phi\phi}$ and $m\left(\phi\right)$ as independent quantities. However, maintaining supersymmetry implies that they are related through the Kahler potential. We can repeat a similar analysis but now taking the Kahler potential as an arbitrary function of $\phi$. So we have
%\be
%\Delta m \sim e^{\frac{K}{2}} \;.
%\ee
%Then (\ref{metrscapsiN}) leads to
%\be
%g_{\phi\phi} \simeq g^{\mathrm{UV}}_{\phi\phi} + \frac{1}{8\pi^2} \frac14 \left(\partial_{\phi} K \right)^2 \;.
%\ee
%This should then be compared to the ${\cal N}=2$ formula
%\be
%g_{\phi\phi} = \frac14 \partial^2_{\phi} K \;.
%\ee
%Taking the one-loop term to be dominant implies
%\be
%\partial^2_{\phi} K  \sim \left(\partial_{\phi} K \right)^2 \implies K \sim \log \phi \;.
%\ee
%This agrees with the asymptotic geometric result. 

The Calabi-Yau setting also has the subtlety that there may be other BPS states becoming massless at infinite distance than those identified through the monodromy orbit. In particular, for $d=3$ there may be an infinite tower of states with $\Delta m \sim \phi^{-\frac12}$. These would lead to sub-leading corrections upon integrating out. Let us comment that taking $\Lambda_{UV}$ as the species scale was a natural choice, but the logarithmic behaviour can be recovered from integrating out states from other (field dependent) cut-offs  \footnote{Indeed, for $d=3$ we find that taking the UV scale as $\Lambda_0$ also reproduces the result. Specifically, we consider $\Lambda_0$ at a certain point in field space, then move a little in field space and some states become light, we integrate them out and repeat again. This iterative process reproduces the logarithmic behaviour, but only for $d=3$. }, as long as they feature the same field dependence as in \eqref{expformodp}. This highlights the fact  that, even if an infinite tower of states is necessary to generate an infinite proper field distance, not every infinite tower will yield such a result as the structure of the tower matters in a crucial way. In particular, it is essential to have an increasing number of states (or equivalently a decreasing $\Lambda_{UV}$) as we approach the singular point. This reminds to the mathematical results in section \ref{sec:infinite_distance} for which a monodromy matrix of infinite order is a necessary but not sufficient condition to get infinite distance.

At this point we must return to the assumption of this analysis that the quantum corrections dominate over the 'classical' value of the field space metric at $\Lambda_{UV}$. This was the assumption utilised in order to work with the expression (\ref{distance}). First we note that the classical behaviour may also have an underlying logarithmic divergence in the distance. Our calculation is not precise enough to match the coefficient in front of the logarithm between the geometry and the integrating out. This in an important caveat to stress with regards to interpreting the logarithmic behaviour as due to the tower of states. Assuming that the geometry result is coming purely from the quantum part amounts to the statement that $g_{\phi\phi}^{UV}$ falls off faster than $\phi^{-2}$ as $\phi \rightarrow \infty$. In particular, this would mean that the proper distance as measured with such a classical metric would be finite and so the field space is compact at $\Lambda_{UV}$. While this possibility would amount to a truly emergent infinite distance, we should keep in mind that a less spectacular, but in some sense more robust, observation is that we could deduce that $g_{\phi\phi}^{UV}$  should behave like $\phi^{-2}$ if it should match the one-loop correction. Such a match between the UV and one-loop parts would match the ideas of strong coupling unification of \cite{Heidenreich:2017sim}. 

%Finally, let us note an important point. We obtained the logarithmic behaviour for the proper distance induced by integrating out states. However, there can always be an underlying classical behaviour which is also logarithmic. Our analysis is not precise enough to differentiate between a quantum element and a classical one. This therefore forms a caveat to interpreting the logarithmic behaviour as due to the tower of states.

\subsection{Relation to the Scalar WGC and to global symmetries}

The Swampland Distance Conjecture studied in this paper has an interesting relation to a Scalar Weak Gravity Conjecture formulated and explored in \cite{Palti:2017elp}. The conjecture states that for each scalar field $\phi^i$ there is a state with mass $m$ satisfying\footnote{It is most sharply stated for massless scalar fields. The generalisation to massive scalars is less clear and will receive corrections. See \cite{Lust:2017wrl} for some work on estimating such corrections.} 
\beq
\label{SWGC}
g^{ij}\left(\partial_i m\right)\left( \partial_j m \right)M_p^2 > m^2 \;,
\eeq 
where $g_{ij}$ is the metric on the scalar field space. In fact,  in ${\cal N}=2$ it was proven that this is true for all but one combination of charged BPS states and follows from the positivity of the scalar fields kinetic terms.\footnote{The combination of charged states which violate it is associated to the graviphoton which has no scalar partner \cite{Palti:2017elp}.} Using our results, it can be checked that (\ref{SWGC}) is always satisfied for any charges in the monodromy orbit of states becoming massless at the monodromy locus. The relation to the Swampland Distance Conjecture is that if we consider a canonically normalised field $g^{ij}=\delta^{ij}$ then, assuming the same state satisfies (\ref{SWGC}) as $\phi$ varies over arbitrarily long distance, its mass must be exponential in $\phi$.

The formulation (\ref{SWGC}) is a direct analogue to the electric WGC statement $g^2 q^2 M_p^2 \geq m^2$, where $g$ is the gauge coupling of the gauge field and $q$ is the charge of the particle. Similarly, it is a statement about the mass of a single particle. The magnetic WGC makes a statement about the cut-off of the effective field theory $\Lambda \sim g M_p$, where $g$ is the gauge coupling. The Swampland Distance Conjecture is also a statement about the cut-off of the effective theory which is at the mass scale of the tower of states. In this sense, it is most naturally interpreted as a Magnetic Scalar WGC.

The magnetic WGC is related to the appearance of a global symmetry \cite{ArkaniHamed:2006dz}. This is because as $g \rightarrow 0$ the gauge symmetry behaves as a global symmetry. This limit should, therefore, be blocked by quantum gravity, which is guaranteed if the WGC holds because then the cut-off of the effective theory also goes to zero as $g\rightarrow 0$. The formalism introduced in this paper allows us to show that there is a similar interpretation for the Swampland Distance Conjecture in our context. From the asymptotic expressions for the K\"ahler potential (\ref{Kpot3}) we see that the field $\R\;t$ develops a perfect shift symmetry at any infinite distance locus\footnote{The Nilpotent Orbit Theorem \cite{schmid} states that the
subleading contributions to the periods are exponentially suppressed with respect to the leading
one. This implies that the axion, identified with $\text{Re}\, t$, does not appear to leading order in the K\"ahler
potential, meaning that it enjoys an continuous global shift symmetry which is only
broken by exponentially suppressed terms, as explained in section \ref{sec:infinite_distance}. Notice that such a continuous global symmetry is not present at finite distance singularities, like the conifold.}. This is a global continuous symmetry. A more precise analogue to the WGC is reached if we dualise the axion $\R\;t$ to a 2-form field $B_2$, the kinetic term then takes the form
\be
\mathcal{L} \supset \frac{1}{f^2} \left|d B_2\right|^2 \;.
\ee
Here $f$ is the axionic decay constant obtained from the axion field metric $f^2=g_{t\bar t}$. At the infinite distance points, the metric vanishes so $f\rightarrow 0$ and the propagator of $B_2$ vanishes due to the infinite kinetic term. Therefore, the dynamics of the $B_2$ field decouples and we recover a global 2-form continuous symmetry. Similarly to the WGC, the Swampland Distance Conjecture ensures that the UV cut-off of the effective theory decreases as we approach the infinite distance locus, so that infinite distances cannot be described within a quantum field theory with a finite cut-off. %The particular relation between the cut-off and the field distance can also have interesting phenomenological implications for large field inflation, since large distances are in conflict with very high cut-offs.

Therefore, the appearance of the infinite tower of massless states at infinite distance can be understood as a quantum gravity obstruction to a global symmetry. However, we have also given another interpretation for the relation between the tower and the infinite distance which is that the infinite distance is itself emergent from integrating out the tower of states. In this sense we can also think of the global symmetry as emergent upon integrating out the tower of states, with a perfect global symmetry corresponding to the tower being infinite. In the next section, and as also proposed in \cite{Heidenreich:2017sim}, we will see that the magnetic WGC can also be understood in terms of integrating out the BPS states. Then again, we can think of the global symmetry at $g \rightarrow 0$ as being emergent in this sense. 

It is therefore natural to expect that a general statement is that the limit towards any global symmetry must be emergent, in the sense of integrating out states as studied here. In some sense this is expected since it is believed that there are no fundamental global symmetries in quantum gravity. However, the way that a global symmetry can emerge is made quite precise through the integrating out a tower of states procedure. And while the global symmetry limit is, of course, very exotic and can not be described with a quantum field theory, the important point is that the emergence is continuous upon approaching the limit. So it is a statement not just about the limit but about the approach to it. In this sense, for the field distance case it is a statement about the emergent nature of field space itself. For the gauge coupling it is a statement about the emergent nature of gauge fields. 

Let us emphasise though that, while the results of this work present some evidence towards such a picture, there remains much work to establish its validity more generally and firmly. 

%%%%%%%%%%%%%%%%%%%%%%%%%%%%%%%%%%%%%%%%%%%%%%%%%%%
\section{The gauge kinetic function and the Weak Gravity Conjecture\label{sec:WGC}}
%%%%%%%%%%%%%%%%%%%%%%%%%%%%%%%%%%%%%%%%%%%%%%%%%%%

Our focus so far has been primarily on the Swampland Distance Conjecture and distances in field space. In this section we extract the implications of our results for the Weak Gravity Conjecture \cite{ArkaniHamed:2006dz} and the gauge kinetic function. The key point is that the BPS states which have formed the focus of our work are charged under U(1) gauge symmetries. Therefore, they not only affect the moduli field space metric but also the gauge kinetic function. Our analysis will build on the results of section \ref{sec:grownorm} where we utilised the Nilpotent Orbit and $Sl_2$-Orbit theorems to determine the growth of the Hodge norm upon approaching infinite distance. Using this we will determine the asymptotic form of the gauge kinetic function upon approaching infinite distance. We will relate this behaviour to the BPS spectrum and will discuss how it emerges from integrating out the charged BPS states. Finally, we will relate it to the Weak Gravity Conjecture.

\subsection{Behaviour of the gauge kinetic function\label{sec:gauge}}

%As it is well known, the above BPS states are charged under the closed string U(1) gauge symmetries, and the gauge kinetic function is parametrised by the scalar manifold in such a way that ${\cal N}=2$ supersymmetry is preserved. 
The BPS states which are of interest to us are charged under U(1) gauge symmetries. Microscopically, in Type IIB string theory, the BPS states are D3-branes wrapping special Lagrangian cycles, the gauge fields arise from the closed-string RR field $C_4$ and the gauge kinetic function depends on the complex structure moduli of the Calabi-Yau threefold. 
%So far, the analysis has been focused on the field metric, but we can also study the behaviour of the gauge kinetic function when approaching different singular points of the complex structure moduli space of Calabi-Yau manifolds. 
%We will see that the electric gauge coupling always goes to zero at the singular points due to the appearance of charged massless states, but the asymptotic behaviour can be logarithmic or a power-law depending on whether the point is at finite or infinite distance, so that it involves a finite or infinite number of charged  fields becoming massless. 
The low energy ${\cal N}=2$ effective action takes the form
\beq
\mathcal{L}=\frac{R}{2} - g_{ij}\partial_\mu t^i\partial^\mu \bar t^j+\text{Im}\;\mathcal{N}_{IJ}F_{\mu\nu}^IF^{J,\mu\nu}+\text{Re}\;\mathcal{N}_{IJ}F_{\mu\nu}^I\left(\star F\right)^{J,\mu\nu} \;, \label{N2action}
\eeq
where $F_{\mu\nu}^I$ with $I=0,\dots, h^{2,1}(Y)$ are the field strengths of the electric U(1) gauge fields which together with the complex structure moduli $t^i$ complete $\mathcal{N}=2$ vector multiplets. The magnetic field strengths are defined as
\beq
\mathcal{G}_I=-\frac{\delta\mathcal{L}}{\delta F^I}=\text{Re}\;\mathcal{N}_{IJ} F^J-\text{Im}\;\mathcal{N}_{IJ}\star F^J\ .
\eeq
In the string theory setting at hand $F^I$ and $\mathcal{G}_I$ arise in the expansion of $F_5 = dC_4$ into 
the symplectic basis $(\alpha_K,\beta^L)$ introduced in \eqref{symplectic-basis} as 
\beq
   F_5 = F^I \wedge \alpha_I - \mathcal{G}_I \wedge \beta^I\ .
\eeq 
This implies that the distinction of electric and magnetic fields depends on the choice of $(\alpha_K,\beta^K)$ at the considered 
point in moduli space.  We will later use the  basis adapted to the type of singular locus that we approach. 

Let us next analyse the behaviour of the gauge kinetic matrix $\mathcal{N}_{KL}$ in (\ref{N2action}) near the singular points 
in moduli space. In order to do that we recall from \cite{Ceresole:1995ca} that $\mathcal{N}_{KL}$
is related to the Hodge-norm of the basis $(\alpha_K,\beta^L)$ as 
\beq
\left(\begin{array}{cc} \int \alpha_I \wedge * \alpha_J & \int \alpha_I \wedge * \beta^L\\  
\int  \beta^K \wedge * \alpha_J  & \int \beta^K \wedge * \beta^L\end{array}\right)
= - \left(\begin{array}{cc}  \text{Im}\,\mathcal{N} + \text{Re}\,\mathcal{N}(\text{Im}\,\mathcal{N})^{-1}\text{Re}\,\mathcal{N}&  \text{Re}\,\mathcal{N}(\text{Im}\,\mathcal{N})^{-1}\\  (\text{Im}\,\mathcal{N})^{-1}\text{Re}\,\mathcal{N}& (\text{Im}\,\mathcal{N})^{-1}\end{array}\right) \; .
\label{M}
\eeq
Using this identity it is easy to evaluate the growth of $\mathcal{N}_{KL}$ using the results 
of subsection \ref{sec:orbits_results}. 
Hence, we now use the basis $(\alpha_K,\beta^L)$ introduced for the respective singularities in \eqref{alpha-beta-d=3}, \eqref{alpha-beta-d=2}, and \eqref{alpha-beta-d=1}.  
At leading order in $\phi=\I\, t$ we can neglect the terms arising from $\text{Re}\;\mathcal{N}_{IJ}$, so we will set $\text{Re}\;\mathcal{N}_{IJ}=0$
in the following. The matrix \eqref{M} then decomposes into two components given in terms of $\text{Im}\,\mathcal{N}_{IJ}$ and its inverse. In the adapted basis we can write the leading behaviour as  
\beq
 ||\alpha_0||^2 \sim  \I\,{\cal N}_{00}  \ , \qquad  ||\alpha_A||^2  \sim  \I {\cal N}_{AA}  \ ,
\eeq
where $A=\{1,\alpha\}$. In turn, the behaviour of $ \I\,{\cal N}_{00}$ and $\I {\cal N}_{AA}$ as a function of $\phi$ for the different cases are given in table \ref{tab:growth}. Therefore, we can deduce the leading order behaviour of the gauge kinetic function. 
This is shown in table \ref{tab:gk}. We note that the growth of $\left(\I {\cal N}\right)_{aa}$  cannot be determined from this analysis without more information about the subleading terms in \eqref{growth-theorem}, so we will omit this component of the gauge kinetic function from our analysis from now on. Notice also that the component $\left(\I {\cal N}\right)_{11}$ only makes sense for $d=3$, while otherwise $A=\{ \alpha\}$. We can also give the behaviour in terms of the BPS mass of the states which become massless at the singularity locus, by using \eqref{expformodp}. Here $m_0$ refers to the lightest field in the tower, so $m_0\equiv \Lambda_0$ in figure \ref{masses}.

\begin{table}[h!]
\begin{center}
\begin{tabular}{| c | c | c | c | c |} 
\hline
 \rule[-.3cm]{0cm}{0.8cm} $d$ & $\left(\I {\cal N}\right)_{AA}\left(\phi \right)$ & $\left(\I {\cal N}\right)_{00}\left(\phi \right)$ & $\left(\I {\cal N}\right)_{AA}\left( m_0\right)$ & $\left(\I {\cal N}\right)_{00}\left( m_0\right)$ \\
\hline
 \rule[-.3cm]{0cm}{0.8cm} $0$ & $\phi$  & - & $\log m_0$  & - \\
\hline
 \rule[-.3cm]{0cm}{0.8cm} $1$ & $\phi$  & $\phi$ & $m_0^{-2}$ & $m_0^{-2}$ \\
\hline
 \rule[-.3cm]{0cm}{0.8cm} $2$ & $\phi$  & $\phi^{2}$ & $m_0^{-1}$  & $m_0^{-2}$\\
\hline
 \rule[-.3cm]{0cm}{0.8cm} $3$ & $\phi$  & $\phi^{3}$ & $m_0^{-2}$ & $m_0^{-6}$ \\
\hline
\end{tabular}
\end{center}
\caption{Table showing the leading behaviour of two key components of the gauge kinetic function. This is shown as a function of the one-parameter approach to the monodormy locus $\phi \rightarrow \infty$, and as a function of the mass of the BPS states which become massless on the monodromy locus. Infinite distance loci are classified by $d > 0$, while finite distance loci have $d=0$. The subscript in $\left(\I\, {\cal N}\right)_{AA}$ runs over $A=\{1,\alpha\}$ for $d=3$ but only over $A=\{\alpha\}$ for $d=0,1,2$.
} 
\label{tab:gk}
\end{table}

%The charge components ${\bf q}_0$ and $\delta {\bf q}$ have the properties that $S_1\left(  \delta {\bf q} ,  {\bf a}_0 \right) = 0$ while $S_1\left(  {\bf q}_0 ,  {\bf a}_0 \right) \neq 0$. 

%We can extract from the behaviour of ${\cal Q}^2$ as a function of $\phi$ the behaviour of gauge coupling function of the gauge fields under which the BPS states are charged. In particular, at leading order in $\phi$ we can neglect the topological part, so set $\text{Re}\;\mathcal{N}_{IJ}=0$, and the matrix $\cI$ decomposes into two components given in terms of $\text{Im}\,\mathcal{N}_{IJ}$ and its inverse. For a given charge ${\bf q}$, the leading order behaviour of ${\cal Q}_{\eta{\bf q}}^2$ gives the behaviour of $q^I \left( \text{Im}\,\mathcal{N} \right)_{IJ} q^J$. Therefore, ${\cal Q}^2$ is schematically giving the behaviour of the gauge coupling (squared) for the U(1)s ${\cal Q}^2 \sim g^2$.

From table \ref{tab:gk} we can see that, while the gauge coupling always goes to zero on an infinite monodromy locus, it does so as power law or logarithmically in the mass of the light states depending on whether we approach a point at infinite or finite distance respectively. In the next section we will give a physical interpretation of this behaviour in terms of integrating out an infinite or finite number of charged fields.

\subsection{Gauge kinetic function from integrating out states}

In section \ref{sec:gauge} we showed that the gauge kinetic function exhibits a logarithmic or power-law divergence, in the mass of the states which become massless, near singular points of the moduli space depending on whether the point is at finite or infinite distance. We propose that the origin of this difference in the asymptotic behaviour of the gauge coupling can be understood in terms of the properties of the charged fields becoming massless at the singularity. Specifically, we will analyse the quantum one-loop corrections to the gauge kinetic function coming from integrating out the charged fields, in a similar way that we did for the field space metric in section \ref{sec:integrating_out}. We will find that the behaviour in table \ref{tab:gk} can be reproduced precisely through this.
%We will also compare the magnetic cut-off suggested by the Weak Gravity Conjecture with the species bound for the tower of states that belongs to the infinite monodromy orbit.

The logarithmic divergence of the gauge kinetic function, the finite distance $d=0$ case in table \ref{tab:gk}, can be understood in terms of a single charged particle becoming massless at the singular point. This is well known for the conifold point \cite{Strominger:1995cz,Vafa:1995ta}. We consider, for simplicity, a single U(1) gauge field with gauge coupling $g$. Quantum one-loop corrections to the gauge coupling from integrating out a single charged fermion of charge $q$ and mass $m$ give
\beq
\label{gIR}
\frac{1}{g^2_{IR}}=\frac{1}{g^2(\mu=m)}=\frac{1}{g^2_{UV}}-\frac{q^2}{12\pi^{2}} \log{\frac{\Lambda_{UV}^2} {m^2}}\;.
\eeq
Here the infra-red value of the gauge coupling $g_{\rm IR}$, which depends on the energy scale at which it is evaluated $\mu$, is taken at the scale of the integrated out particle $m$. Below this scale, the effective theory only involves the U(1) gauge field and the running stops. The ultraviolet value of the gauge coupling $g_{\rm UV}$ is given at a cut-off scale $\Lambda_{UV}$. As we move in the moduli space the mass of the state goes to zero $m \rightarrow 0$. However, we have seen that the relevant UV scale $\Lambda_{UV}$ for our considerations in section \ref{sec:integrating_out} was the species scale (\ref{spesca}). Since there is only one state in the theory $S=1$, this is given by the Planck mass $M_p$, which we set to unity in table \ref{tab:gk}. More generally, we note that the K\"ahler potential stays finite at finite distance, which implies a finite UV scale. We therefore see that the inverse gauge coupling squared diverges logarithmically in the mass of the state which is becomes massless, reproducing the behaviour of the $d=0$ case in table \ref{tab:gk}.

When approaching an infinite distance singularity, two things change. First, we have to sum the contribution from all light particles running in the loop. As explained in section \ref{sec:integrating_out}, the number of stable particles $S$, below the scale at which gravity becomes strongly coupled, depends on the point in moduli space. Secondly, the UV cut-off also depends on the point in moduli space, so that it goes to zero at the singular locus. As in section \ref{sec:integrating_out}, we will this UV cut-off to be the species bound. The field dependence of both of these is given in (\ref{expformodp}).

The relevant one-loop quantum correction to the gauge kinetic function is given by
\beq
\label{1loopgco}
\text{Im\;}\mathcal{N}_{IJ}^{IR}\simeq \text{Im\;}\mathcal{N}_{IJ}^{UV}- \sum_k^{S}\left(\frac{8\,q_{k,I}q_{k,J}}{3\pi^{2}} \log{\frac{\Lambda_{UV}} {m_k}}\right) \;,
\eeq
where $q_{i,I}$ is the charge of the $k$-th particle with mass $m_k$ under the gauge field $F^I_{\mu\nu}$. 

We would now like to see if (\ref{1loopgco}) can reproduce table \ref{tab:gk}. Let us summarise the relevant results of section \ref{sec:orbits}. For $d=3$, the charge of the $k^{\rm th}$ state in the BPS tower can be written as
\be
\label{chdec3}
\ {\bf q}_k \equiv q_1\beta^1 + \left( q_0+k\, q_1\right)\beta^0+q_\alpha\beta^\alpha +q_a\beta^a \quad \text{with $q_1\neq 0$}  \;.
\ee
This tower of states corresponds to the infinite quotient monodromy orbit found in \eqref{final-d=3}. For $d<3$ such an infinite quotient monodromy orbit does not exist, but we can still have a tower of states generated  by a monodromy transformation different from the one about a generic point on the infinite distance locus, see section \ref{sec:intersection}. The charge of the $k^{\rm th}$ state in the monodromy orbit can be parametrised as
\be
\label{chdec1}
\ {\bf q}_k \equiv \left( q_0+k_0\right)\beta^0+ (q_\alpha + k_\alpha)\beta^\alpha +q_a\beta^a   \;,
\ee
where $q_a=0$ for the case $d=2$. The remaining properties of the tower regarding their mass behaviour are given in (\ref{expformodp}).

We can distinguish two cases when computing the one-loop corrections to the gauge kinetic function, whether the tower has the same charge or an increasing charge with respect to the gauge field. This implies two types of sums in \eqref{1loopgco} yielding
\bea
 \sum_k^{S} \left( q_0+k q_1\right)^2\log{\frac{\Lambda_{UV}} {m_k}} &\sim & \phi^d\;, \\
 \sum_k^{S} q_1^2 \log{\frac{\Lambda_{UV}} {m_k}}& \sim & \phi^{d/3}\;,
\eea
where we have used (\ref{expformodp}) for the UV cut-off scale and the number of particles $S$.
Note that the above sums are dominated by the states with high charges which, at infinite distance, always satisfy $\Lambda_{UV}/m_k\rightarrow 1$ since $m_k \sim S \Delta m \sim \Lambda_{UV}$. Therefore, the logarithm asymptotes to a constant leading to a final power law result\footnote{We encounter sums of the kind
\beq
\sum_k^S\log{\frac{S}{k}}\sim S+\dots\ ,\quad \sum_k^Sk^2\log{\frac{S}{k}}\sim \frac19 S^3+\dots
\eeq
which give the same parametric result regardless the presence of the logarithm in the sum.}.
We can then read off the behaviour of the elements of the gauge kinetic matrix relevant for table \ref{tab:gk}. These read
\bea
\text{Im\;}\mathcal{N}_{00} \sim  \phi^d\ , \quad 
\text{Im\;}\mathcal{N}_{AA} \sim  \phi^{d/3}\quad &\text{for $d=3$}\ ,\\
\text{Im\;}\mathcal{N}_{00} \sim  \text{Im\;}\mathcal{N}_{AA} \sim  \phi^{d}\quad &\text{for $d<3$}\ ,
\eea
where $A=\{1,\alpha\}$. We therefore have reproduced the table \ref{tab:gk}.
 In terms of the BPS mass of the states becoming massless, the gauge coupling goes to zero as a power law of the mass of the light states of the tower, precisely in the way obtained in the previous section. %Also note that for $\text{Im\;}\mathcal{N}_{00}$ we have used the fact that the sum is dominated by the states with high charges so that $m_k \sim S \Delta m \sim \Lambda_{UV}$ and so the logarithm asymptotes to a constant. 
This perfect matching for each component of the gauge kinetic function supports the identification of the monodromy orbit of BPS states as the infinite tower of states becoming massless at the infinite distance singularities. Notice that the only mismatch is in $ \text{Im\;}\mathcal{N}_{AA}$ for the case $d=2$. However, the charge space spanned by $\beta_\alpha$ could very well be empty, so there is not an associated gauge field $F_{\mu\nu}^A$. This is indeed the case for the locus $n=d=1$ in $\mathbb{P}^{\left(1,1,2,2,2\right)}$ detailed in section \ref{sec:intersection}. We leave a detailed investigation of this case for future work.

For completeness, let us comment on the contribution to the gauge coupling coming from integrating out a scalar field. The one-loop correction is simply given by
\beq
\frac{1}{g^2_{IR}}=\frac{1}{g^2_{UV}}-\frac{q^2}{48\pi^{2}} \log{\frac{\Lambda_{UV}^2} {m^2}}
\eeq
and the computation proceeds analogously to the fermionic case. The contribution to the gauge kinetic function will, therefore, have the same parametric dependence in terms of the mass of the particle up to possible numerical factors.

Let us finally comment on the relation of our results to the Weak Gravity Conjecture (WGC) \cite{ArkaniHamed:2006dz}.
At singular loci in moduli space, we find that the gauge coupling vanishes polynomially (logarithmically) fast in the mass of an infinite (finite) number of BPS states becoming massless \footnote{Our results for the rate at which the gauge coupling vanishes when approaching the singularities are also consistent with the bounds obtained in \cite{Montero:2017mdq} in the context of AdS/CFT.}. The presence of the light states can be understood in the context of the electric WGC. More concretely, the BPS states becoming massless are the states satisfying the WGC but, in our case, they do not form a lattice, as in other generalizations of the WGC to multiple U(1) gauge fields \cite{Heidenreich:2016aqi,Heidenreich:2015nta}. %(see also \cite{Andriolo:2018lvp} for similar ideas of having a tower of states satisfying the WGC). 
This is consistent with the fact that in supersymmetric settings the WGC is satisfied by BPS states which, therefore, saturate the WG bound \cite{ArkaniHamed:2006dz,Ooguri:2016pdq}. As explain in \cite{Palti:2017elp,Lust:2017wrl}, in $\mathcal{N}=2$ it is then essential to consider the contribution from scalar fields to the WGC bound as they contribute to the BPS bound.

The magnetic WGC states that the cut-off scale of the theory can be no higher than $g M_p$.  At finite distances in moduli space we find a finite number of light states and that the gauge coupling goes to zero on the monodormy locus logarithmically fast in the mass of the states. Therefore, the cut-off scale must also vanish logarithmically fast to satisfy the WGC. However, there is a much lower cut-off scale set by the mass scale of the states becoming light. The magnetic WGC does not imply that the scale of quantum gravity related physics is at $g M_p$ within an effective theory with a cut-off scale which is below $g M_p$. Only if one considers an effective theory with a cut-off scale above $g M_p$ does one reach an inconsistency. In this case, since the gauge coupling behaves logarithmically in the mass of the state, we see that we reach new physics exponentially fast before reaching the scale $g M_p$. Therefore, there is no sense in which quantum gravity related physics, such as an infinite number of states, must become light at finite distance. This is all consistent with the discussion in \cite{Lust:2017wrl} and in \cite{Heidenreich:2017sim}. Note that also, $g \rightarrow 0$ logarithmically fast, so in the case of finite number of light states, is still consistent with an emergent nature for that limit. 

At infinite distance we showed that a tower of states starting at $\Lambda_0$ reproduces the correct behaviour of the gauge couplings. The scale $\Lambda_0$ coincides with the gauge coupling related to $\I\, {\cal N}_{00}$ for $d<3$, and for $d=3$ for the gauge coupling associated to $\I\, {\cal N}_{11}$, as can be seen from the last column of table \ref{tab:gk}. For these cases we therefore observe that the magnetic WGC is implied by the idea that the gauge coupling behaviour emerges from integrating out the tower of charged BPS states. This matches the proposal in \cite{Heidenreich:2017sim}. 

In the case $d=3$ we find a mismatch between $\Lambda_0$ and the gauge coupling associated to $\I\, {\cal N}_{00}$. This is interesting in the sense that it shows how the emergence of weak gauge coupling from a tower of states need not imply the magnetic WGC. However, it is not a counter-example to the magnetic WGC because the monodromy orbit of BPS states is only a sub-set of the possible BPS states. In particular, one could have a tower of BPS states all with $S_1( {\bf q}_k,{\bf a}_0 )=0$ which would start at a mass scale $\phi^{-\frac32}$ and so match onto the gauge coupling. Such a tower would also lead to the same behaviour for the gauge coupling upon integrating out, but is not related to the monodromy orbit. It may be possible to relate it to a different monodromy orbit, as in section \ref{sec:intersection}, or more generally motivate it since it is a sub-space of the quotient space (\ref{quotient}). It is interesting to note that, by contrast, if instead we considered a tower of states with increasing charges under the gauge field associated to $\I\, {\cal N}_{11}$, so with $S_1 ( {\bf q}_k,{\bf a}_0 ) \sim k$, then integrating out such a tower would not match onto the behaviour of the gauge coupling.

\section{Summary\label{sec:conclusions}}

In this paper we studied infinite distance loci in the complex-structure moduli space of Calabi-Yau manifolds. The study was performed in the context of the Swampland Distance Conjecture which states that upon approaching infinite distance loci in moduli space there should exist an infinite tower of states whose mass decreases exponentially fast in the proper distance \cite{Ooguri:2006in}. Our proposal is to identify these as charged BPS states, which for the complex-structure moduli space in compactifications of type IIB string theory are D3-branes wrapping special Lagrangian three-cycles. They are charged under the closed-string Ramond-Ramond U(1) gauge symmetries.

We first introduced some existing mathematical results on infinite distance loci which allowed us to classify them algebraically. In general, a point in the moduli space was classified by the monodromy that the period vector undergoes when circling the point. A point at infinite distance, which means it is infinite distance along any path from some other point in the moduli space, has an infinite order monodromy around it. Further, it is classified by two integers, $n$ and $d$, which respectively determine the highest non-vanishing power of the logarithm of the monodromy matrix and the highest power which does not annihilate the period vector on the monodromy locus. 

The first key tool utilised is the Nilpotent Orbit Theorem of Schmid \cite{schmid}. This essentially determines the local form of the moduli space around any point, but is most powerful around loci of infinite monodromy. We applied it to extract the local form of the field space metric and the mass of BPS states. This lead directly to the result
\begin{itemize}
\item Approaching any locus of infinite distance, any BPS states which become massless on the locus become light at least exponentially fast in the proper distance.
\end{itemize}
This matches onto the Swampland Distance Conjecture proposal. 

It is a crucial point that an infinite tower of states should become massless at infinite distance. As we will see below this tower is central to much of the physics. We therefore determined the properties of this tower in as much detail as possible. The central difficulty is that the spectrum of BPS states in the theory is difficult to determine. Particularly so because it varies upon variations in complex-structure moduli space. In terms of the physics of the tower of states, this amounts to identifying the stable states in the tower. So while we know the mass of any would-be BPS state of a given charge we do not know if there is a BPS of that charge in the spectrum. Our proposal is to use the monodromy about the infinite distance point to identify a specific set of candidates for BPS states. Specifically, the monodromy action on the period vector has a natural action of the charges of would-be BPS states in the theory. Starting from a given charge, the monodromy action will transform it and acting repeatedly with the monodromy determines a {\it monodromy orbit} through the possible charges. At infinite distance the monodromy is of infinite order and so the monodromy orbit contains either one element, if the monodromy acts trivially on a charge, or an infinite number of charges. We then propose the following:
\begin{itemize}
\item Proposal: A candidate for an infinite tower of states which becomes massless at infinite distance is generated by an infinite monodromy orbit starting from a single BPS state.
\end{itemize}
More precisely, only a sub-set of the infinite monodromy orbit will be stable BPS states, and the number in this subset grows exponentially fast in the proper distance. We believe that this is a general phenomenon: the number of stable states in the tower increases exponentially fast as we approach the infinite distance locus. Note also that there could be other towers of particles becoming light and satisfying the Swampland Distance Conjecture.

We then identified the monodromy orbit and the set of stable BPS states more precisely. First, we decomposed any would-be BPS states which become massless on the monodromy locus into type I and type II states. Type I states become light exponentially fast in the proper distance while type II states becomes light as the exponential of an exponential in the proper distance. We then defined the quotient set of all massless would-be BPS states by type II states. The quotient means that we identify two states if their charge differs by a type II charge. We argue that this quotient is a good candidate for a subset of stable states at the singularity. The monodromy orbit generating BPS states is the embedding of the orbit into this quotient space, which we denote the {\it quotient monodromy orbit}. We then went on to study when this quotient monodromy orbit is infinite. This required introducing substantial mathematical technology based on Mixed Hodge Structures. Specifically, we utilised the $Sl_2$-Orbit theorem of Schmid \cite{schmid} and Deligne splittings of vector space of charges to show that 
\begin{itemize}
\item Infinite distance loci with $d=3$ always support an infinite quotient monodromy orbit induced by the monodromy about the infinite distance locus. 
\item Infinite distance loci with $d < 3$ do not support an infinite quotient monodromy orbit induced by the monodromy about the infinite distance locus. 
\end{itemize}
The first result is very encouraging and provides a specific identification of the tower of states for such loci. The second result is more puzzling since such loci are infinite distance but our proposed sub-set of BPS states are more difficult to identify. There are two important points to note about this. First, since we propose that the monodromy orbit identifies a sub-set of the BPS states it is only a sufficient, but not necessary, condition for having an infinite number of massless BPS states. The second point is that we show that there can still be an infinite quotient monodromy orbit but which is generated by a monodromy transformation about a different infinite distance locus in the moduli space which intersects the specific infinite distance locus. In this work we did not introduce the mathematical machinery to show that this happens generally, but do show it for interesting examples. There are, however, specific counter-examples to this proposal where infinite distance $d < 3$ loci exist which do not intersect any other monodromy loci. Specifically, in examples where the complex-structure moduli space is one (complex) dimensional. We suggest possible ways to understand such points in the context of our proposal, but leave a better understanding of them for future work.

At this point we make a distinction within our results. On the one hand, in our $\cN=2$ setting, we believe that the results so far have introduced a new perspective and several powerful techniques to perform a general analysis of infinite field distances and the existence of infinite BPS states in complex structure moduli space. We expect that, also in the discussed generality, our claims will survive further scrutiny only yielding further refinements.  On the other hand, the fact that we find patterns within this general framework naturally leads to more speculative and general proposals which we outline below. One of the reasons why the results are less established is that they require us to perform
perturbative computations in a theory with particle-number dependent cutoff to capture the impact of  
infinitely many modes potentially relevant in the loop. Another reason is that our analysis of the integrating out procedure is only a toy model since we utilise the IR expressions for the masses of the states rather than the UV form. It is therefore important to gather more evidence for these ideas. 

With this caution in mind, having identified significant details regarding a tower of BPS states which become exponentially light at infinite distance, we proposed an underlying microscopic physics explanation for why such a tower exists:
\begin{itemize}
\item Proposal: Infinite distances in moduli space arise {\it from} integrating out an infinite tower of states. 
\end{itemize}
It is well known that Calabi-Yau moduli spaces are quantum in the sense that they already have integrated out wrapped D3-brane states. The most famous example being the conifold singularity as associated to one of the wrapped D3-branes becoming massless. Our proposal is that the infinite distance loci, which are also singular, are the same in nature. So they can be thought of as arising from integrating out states, but this time an infinite number of them. Infinite distances are thereby seen as emergent. This idea was first tentatively proposed as a possibility in \cite{Ooguri:2006in}. We show that explicitly integrating out such a tower of states leads to a correction to the field space metric which precisely reproduces the logarithmic divergence of the proper field distance. This divergence implies in turn the exponential decrease in the mass of the states in the proper distance.\footnote{While writing this paper we were informed that in \cite{HRR-toappear} a similar conclusion was reached independently.} Further, the number of states required to be integrated out matches the growth in the number of stable states in the tower upon approaching infinite distance. While this is good evidence for this connection, it is important to note that this quantum induced infinite distance could be on top of a `classical' infinite distance already present. 

We also apply the techniques used to study the moduli space metric to the gauge kinetic function. We show that upon approaching an infinite distance locus the gauge kinetic function diverges exponentially fast in the proper distance, since it diverges as power law in the mass of the BPS states becoming light. This matches the behaviour proposed in \cite{Klaewer:2016kiy}.\footnote{Note the difference with respect to finite distance singularities, where there is only a finite number of particles becoming massless and the gauge kinetic function diverges logarithmically in the mass of the particles.} %For electric charged states which become massless at infinite distance,  the associated electric gauge coupling vanishes at infinite distance. 
We showed that this can be understood again from integrating out the BPS states which become light and are electrically charged under the gauge fields. Motivated by this match we therefore propose that a vanishing gauge coupling is also emergent in the same way:
\begin{itemize}
\item Proposal: Any vanishing gauge coupling limit $g \rightarrow 0$ arises {\it from} integrating out charged states starting from some ultraviolet scale, below $M_p$, where $g$ is finite. 
\end{itemize}
Note that this matches the fact that the region in moduli space where the gauge coupling vanishes, that is at infinite distance, is also argued to be emergent at the quantum level. 
It also matches the ideas proposed in  \cite{Harlow:2015lma,Heidenreich:2017sim} where the small gauge coupling is emergent upon integrating out states. We also presented evidence that the appropriate ultraviolet scale is actually the species scale. Note that $g \rightarrow 0$ either logarithmically, if a finite number of states are integrated out, or as a power law, if an infinite number of states are integrated out.
 
The relation between infinite distances in moduli space and vanishing gauge couplings also has a natural interpretation in terms of global symmetries. We showed that at any infinite distance locus there is a global symmetry in the form of a continuous shift symmetry for an axion field. Similarly, it is known that at vanishing gauge coupling the gauge $U(1)$ symmetry turns into a global $U(1)$ symmetry. Therefore, we expect that in a quantum gravity setting any limit which approaches a global symmetry should be emergent from integrating out a tower of states. Of course, the effective field theory entirely breaks down at the infinite distance locus, since the cut-off of quantum gravity goes to zero when we recover infinitely many massless states. But the important point is about the way we approach the limit  so that the emergence of the global symmetry is continuous. 
%Note that this phenomenon does not occur at finite distance singularities, where there is not continuous global symmetry emerging.

This relation between the cut-off of the effective theory, the field distance and the smallness of the gauge coupling can have interesting phenomenological implications. If we want to engineer an effective field theory valid up to a certain finite cut-off scale, these ideas imply a limit on how small gauge couplings and how large scalar field variations can be accommodated within the same effective theory  in order to be consistent with quantum gravity.

Our analysis is performed in a very general way, it applies to any generic infinite distance locus in any Calabi-Yau moduli space. However, there are many ways to extend it. First, still within the setting of the ${\cal N}=2$ complex-structure Calabi-Yau moduli space there are certain non-generic loci which are at the intersection of multiple infinite distance divisors. In such cases there are multiple local co-ordinates which diverge as opposed to the one parameter approach to the loci we studied. The mathematical technology for studying these is already available, since both the theorems of Schmid have a multi-parameter generalisation. It would therefore be interesting to study such loci in a similar way to the work in this paper. Similarly, while we have focused on the complex-structure moduli space of Calabi-Yau threefolds, much of the analysis can be generalised straightforwardly to Calabi-Yau fourfolds and implemented in the context of F-theory. The primary challenge in such a setting is the reduced supersymmetry which means it is more difficult to control the mass of the states without the BPS structure. 

Generalising further, it would be interesting to test the behaviour we found in other scalar field spaces in string theory. In particular in the K\"ahler moduli space of type IIB string theory and the mirror complex-structure moduli space in type IIA string theory. There, BPS states from wrapped branes are not particle-like in four dimensions but are extended objects. This suggests a generalisation of the Swampland Distance Conjecture where the infinite tower of states are not particles. There is also the open-string moduli space of D-branes in string theory which can be studied in a similar way. Indeed, recent results have showen how the open-string sector can be implemented in a structure which bears some similarities to our closed-string studies \cite{Herraez:2018vae}.

The ${\cal N}=2$ string theory moduli spaces we have studied are natural testing grounds for quantum gravity field spaces. Many of the results we found can be understood more generally away from an explicit string theory setting. Indeed, we formulated the proposals in such a general way. This suggests that it may be possible to find more evidence for the generality of these results by considering general quantum gravity physics, such as black hole physics. Indeed, some of the structure we obtained can be understood as in \cite{Klaewer:2016kiy} from black holes with scalar hair. It would be very interesting to develop such an approach and thereby build evidence for the generality of the physics.

\vspace{1cm}
\noindent
{\bf Acknowledgements:} 

\noindent
We would like to thank Frederik Denef, Michael Douglas, I\~naki Garc\'ia-Etxebarria, Matt Kerr, Dieter L\"ust, Fernando Marchesano and Miguel Montero for very helpful discussions and correspondence. 

\vspace{1cm}
\appendix

\noindent
{\Large \bf Appendices}

\section{Non-mutually stable BPS states\label{ap:qdq}}

In section \ref{sec:BPS} we argued that states of type I related by a monodromy transformation which differ by the charge of a massless state of type II are not mutually stable. A motivation for the microscopic physics comes from the distinguished triangles defined in \eqref{triangle}. In this picture, states ${\bf q}_B$ and $T {\bf q}_B$ will be mutually unstable if $S({\bf q}_A,{\bf q}_B)\neq 0$. Here we will prove that $S({\bf q}_A,{\bf q}_B)= 0$ if both states ${\bf q}_A$ and ${\bf q}_B$ are of type I. This motivates that the quotient monodromy orbit obtained upon modding out by type II states is a good candidate for a set of stable states becoming massless at the singularity. To show this, we will use the mixed Hodge structures introduced in section \ref{sec:orbits}. 

For convenience, let us recall \eqref{triangle} here,
\be
\delta {\bf q}_B=T {\bf q}_B - {\bf q}_B = S_0\left({\bf q}_A,{\bf q}_B \right) {\bf q}_A \;. 
\ee
This implies that the states  ${\bf q}_B$ and  $T{\bf q}_B$ differ by a charge $\delta {\bf q}_B$ proportional to $ {\bf q}_A $. Assuming ${\bf q}_B\in \cM_I$, we are interested in finding out for what states $ {\bf q}_A $ we can have $S_0\left({\bf q}_A,{\bf q}_B \right)\neq 0$, i.e. $S_0\left(\delta{\bf q}_B,{\bf q}_B \right)\neq 0$. Recall that $\delta {\bf q}=N {\bf q}+\frac12 N^2 {\bf q} +\frac16 N^3 {\bf q} $ where  ${\bf q}\in H^3(Y_3,Z)$.

Let us consider for simplicity $\delta {\bf q}=N {\bf q}$. The argument can be repeated for the terms $N^2 {\bf q}$ and $N^3 {\bf q}$ obtaining the same result. If ${\bf q}_B\in Re(P^{p,q})\subset\cM_I$ we have
\beq
S_0\left(N{\bf q}_B,{\bf q}_B \right)\neq 0\ \Rightarrow \ p+q=4
\label{A1}
\eeq
where we have used the orthogonality relations\eqref{Sj-orthogonality1-I} and  \eqref{Sj-orthogonality2-I} .
Since ${\bf q}_A$ is massless at the singularity, $\delta {\bf q}\in \cM$ which means $S_j\left(N{\bf q}_B,{\bf a}_0 \right)=0$ for $j\geq d/2$. However, if we want ${\bf q}_A$ to be a type I state we also need
$S_i\left(N{\bf q}_B,{\bf a}_0 \right)\neq 0$ for some $ i<d/2$.
Recalling that ${\bf q}_B\in \cM_I$ is also massless at the singularity and $S_i\left(N{\bf q}_B,{\bf a}_0 \right)=-S_{i+1}\left({\bf q}_B,{\bf a}_0 \right)$, the condition for having ${\bf q}_A\in \cM_I$ is modified to 
 \beq
S_i\left(N{\bf q}_B,{\bf a}_0 \right)\neq 0\ \ \text{for some} \ i<d/2-1
\label{A2}
\eeq
Using the orthogonality relations and the fact that ${\bf a}_0\in P^{3,d}$, this is satisfied for
 \beq
 S_i\left(NP^{p,q},P^{3,d}\right)\neq 0\ \Rightarrow \  p+q-2i+d=5
\label{A3}
\eeq
But it is easy to check that the conditions \eqref{A1}, \eqref{A2} and \eqref{A3} can never be satisfied simultaneously. Therefore, a non-vanishing product $S({\bf q}_A,{\bf q}_B)\neq 0$ with ${\bf q}_B\in \cM_I$ implies that $S_i\left({\bf q}_A,{\bf a}_0 \right)= 0$ for all $i\leq d$, which means that ${\bf q}_A$ can only be a type II state.

%%%%%%%%%%%%%%%%%%%%%%%%%%%%%%%%%%%%%%
\section{Some simple examples and classification results} \label{app:explicit_examples}

In order to illustrate the general concepts introduced in sections \ref{sec:infinite_distance} and \ref{sec:orbits}, in this appendix we provide some simple examples and review a classification of possible cases for the one-modulus case. 

Our examples will deal with three types of cases, denoted by I, II$_1$ and II$_2$, where the nilpotency indices of $N$ introduced in
(\ref{nidef}) and (\ref{def-d}) are 
\beq \label{three-cases}
  \text{I}: \ \ n=d=3\ ,\qquad  \text{II}_1:\ \ n=1\,,\ d=0\ , \qquad \text{II}_2:\ \ n=d=1\ . 
\eeq
Since we are dealing with one-modulus cases, one can use \eqref{One-modulus-relation} to conclude 
that the  two cases  I, II$_2$
are examples of monodromies around points at infinite distance, while the  case II$_1$ corresponds to 
a monodromy around a finite distance point. One can show, as we recall below, that in case II$_1$ the matrix $N$
has rank 1, while in case II$_2$ it has rank 2. The cases I and II$_1$ are not hard to realize geometrically. 
Evidently, case I corresponds to a maximally unipotent monodromy and hence arises at the large complex-structure 
point of Calabi-Yau threefolds.  The case II$_1$ arises at the conifold point. Finally, the case II$_2$ is 
the most unfamiliar one and we discuss it in more detail in section \ref{sec:orbits}. It arises 
at a so-called Tyurin degeneration \cite{TYU}. 

The examples also serve to illustrate a particular method for constructing the $W_i$. The starting point is the nilpotent matrix $N$ which satisfies $N^{n+1}=0$. Because it is nilpotent it can be completed into an $Sl_2$ triplet $\{Z,X_+,X_-\}$ where it plays the role of the lowering operator $N = X_-$. The generators satisfy the algebra
\bea
\left[Z,X_+\right] = 2 X_+ \;,\;\; \left[Z,X_-\right] = -2 X_- \;,\;\; \left[X_+,X_-\right] = Z \;.
\eea
Through this embedding into $Sl_2$ the spaces $W_i$ are defined as the eigenspaces generated by eigenvectors of $Z$ with eigenvalues less than or equal to $i-D$. In this example section we will consider matrix representations of the $Sl_2$ algebra. In general, the dimension of the matrices in the representation should be at least $\left(n+1\right)\times\left(n+1\right)$, but the full structure emerges in the cases when they are larger than this. It is known that the structure of the $W_i$ filtration is independent of the particular representation chosen for $Sl_2$. We will consider $4\times 4$ representations as illustrative examples. 

The $4\times 4$ representations are relevant for CY threefolds with exactly one complex structure modulus, 
i.e.~$h^{2,1}=1$. The real basis of three-forms introduced in \eqref{symplectic-basis} then only consists of four elements $(\alpha_0,\alpha_1,\beta^0,\beta^1)$.  
We will write all results in matrix form and introduce the standard unit vectors $e_1 = (1,0,0,0)^T$, $e_2= (0,1,0,0)^T$, $e_3= (0,0,1,0)^T$, and $e_4= (0,0,0,1)^T$. The $e_i$ are identified with the forms $(\alpha_0,\alpha_1,\beta^0,\beta^1)$. It is convenient 
to chose a different identification for the three cases \eqref{three-cases} and we will give the corresponding $\eta$ for each case.

\subsubsection*{An example filtration for case I:}

We consider first the case where $n=3$. In this case it is convenient to identify  
\beq \label{id_alpha}
   e_1 \cong \alpha_0\ ,\quad e_2 \cong \alpha_1\ ,\quad e_3 \cong \beta^1\ ,\quad e_4 \cong \beta^0\ ,
\eeq
such that the intersection matrix $\eta$ takes the form
\be
\eta = 
\left (
\begin{array}{cccc}
0 & 0 & 0 & 1 \\
0 & 0 & 1 & 0 \\
0 & -1 & 0 & 0 \\
-1 & 0 & 0 & 0  
\end{array}
\right) \ . \label{etacan}
\ee
An example monodromy matrix which leads to this case is the one associated to the large complex structure limit \footnote{Note that 
one can chose a basis such that $T$ is an integral matrix. Our choice is adapted to the integral basis chosen for the three-cycles.}
\be \label{exampleI}
T = 
\left (
\begin{array}{cccc}
1 & 0 & 0 & 0 \\
1 & 1 & 0 & 0 \\
-\frac12 & -1 & 1 & 0 \\
\frac16 & \frac12 & -1 & 1  
\end{array}
\right) \; ,\qquad N = 
\left (
\begin{array}{cccc}
0 & 0 & 0 & 0 \\
1 & 0 & 0 & 0 \\
0 & -1 & 0 & 0 \\
0 & 0 & -1 & 0  
\end{array}
\right) \;.
\ee 
where $N$  is constructed through $N=\log T$. Note that $N^{i+1}$ is of rank $3-i$.
We now construct the $Sl_2$ representation. We have for the generators 
\be
Z = 
\left (
\begin{array}{cccc}
3 & 0 & 0 & 0 \\
0 & 1 & 0 & 0 \\
0 & 0 & -1 & 0 \\
0 & 0 & 0 & -3  
\end{array}
\right) \;,\;\;
X_+ = 
\left (
\begin{array}{cccc}
0 & 3 & 0 & 0 \\
0 & 0 & -4 & 0 \\
0 & 0 & 0 & -3 \\
0 & 0 & 0 & 0  
\end{array}
\right) \;,\;\;
X_- = 
\left (
\begin{array}{cccc}
0 & 0 & 0 & 0 \\
1 & 0 & 0 & 0 \\
0 & -1 & 0 & 0 \\
0 & 0 & -1 & 0  
\end{array}
\right) \;.
\ee

The $W_j$ can then be constructed as the appropriate eigenvector subspaces. We summarize in the following by which of 
the basis vectors $e_i$ the $W_j$ are spanned. By definition \eqref{filtration} $W_6$ is the full vector space and hence spanned by 
$(e_1,e_2,e_3,e_4)$. The non-trivial part of the filtration reads
\be
W_5=W_4=  \text{span}(e_2,e_3,e_4) \;,\quad
W_3=W_2=  \text{span}(e_3,e_4) \;, \quad
W_1=W_0=  \text{span}(e_4) \; .
\ee
We can now also construct the $Gr_j \equiv W_j/W_{j-1}$. Clearly, one has $\text{dim} \, Gr_6 =
\text{dim}\, Gr_4= \text{dim} \, Gr_2=\text{dim} \, Gr_0= 1$, while all other $Gr_5, Gr_3, Gr_1$ are trivial. 
An equivalence class in $Gr_6$ can be represented by a vector $a^i e_i$, with $a^1 \neq 0$. The coefficients 
$a^2,a^3,a^4$ are not further restricted since these directions are identified with the trivial element in $Gr_6$. 
Furthermore, the $\cP_{3+i}$  are defined at the end of section \ref{sec:filt} as the kernels of $N^{i+1}$ acting on $Gr_{3+i}$. 
One easily checks that the only non-trivial $\cP_{3+i}$ is $\cP_6$, which is the one-dimensional quotient $Gr_{6}$ itself.
Considering the case $d=3$ we thus conclude that $\mathbf{a}_0$ is vector with 
a non-vanishing first entry. Further constraints arise 
from the fact that $\mathbf{a}_0$ is part of the $F^i_\infty$ filtration. We will give the allowed forms of $\mathbf{a}_0$ below.

%In this case $d=3$ and so we know from the $Sl_2$-orbit theorem that ${\bf a}_0$ has a non-trivial restriction to $P_6$. Subject to only this constraint, the most general ${\bf a}_0$ takes the form
%\be
%{\bf a}_0 =  \left( \begin{array}{c} \cdot \\ \cdot  \\ \cdot  \\ \cdot  \end{array} \right) \;.
%\ee
%We can now determine the action of $N$ on ${\bf a}_0$ as
%\be
%N {\bf a}_0 =  \left( \begin{array}{c} 0 \\ \cdot \\ \cdot  \\ \cdot  \end{array} \right) \;,\;\;
%N^2 {\bf a}_0 =  \left( \begin{array}{c} 0 \\ 0 \\ \cdot \\ \cdot  \end{array} \right) \;,\;\;
%N^3 {\bf a}_0 =  \left( \begin{array}{c} 0 \\ 0 \\ 0 \\ \cdot \end{array} \right) \;.
%\ee
%We therefore see that ${\bf a}_0 \in W_{6}$ and $N^{j+1}{\bf a}_0 \in W_{4-2j}$. 

\subsubsection*{An example filtration for case II$_1$:}

The second example filtration that we consider also uses identification \eqref{id_alpha} and $\eta$ of the form \eqref{etacan}.
The considered from of the monodromy matrix $T$ and associated $N$ are
\be \label{exampleII1}
T = 
\left (
\begin{array}{cccc}
1 & 0 & 0 & 0 \\
0 & 1 & 0 & 0 \\
0 & 0 & 1 & 0 \\
1 & 0 & 0 & 1  
\end{array}
\right) \; , \qquad
N = 
\left (
\begin{array}{cccc}
0 & 0 & 0 & 0 \\
0 & 0 & 0 & 0 \\
0 & 0 & 0 & 0 \\
1 & 0 & 0 & 0  
\end{array}
\right) \;.
\ee
Hence, $N$ is of rank $1$ and obeys $N^2=0$ such that $n=1$. This $N$ can be embedded in $Sl_2$ as
\be
Z = 
\left (
\begin{array}{cccc}
1 & 0 & 0 & 0 \\
0 & 0 & 0 & 0 \\
0 & 0 & 0 & 0 \\
0 & 0 & 0 & -1  
\end{array}
\right) \;,\;\;
X_+ = 
\left (
\begin{array}{cccc}
0 & 0 & 0 & 1 \\
0 & 0 & 0 & 0 \\
0 & 0 & 0 & 0 \\
0 & 0 & 0 & 0  
\end{array}
\right) \;,\;\;
X_- = 
\left (
\begin{array}{cccc}
0 & 0 & 0 & 0 \\
0 & 0 & 0 & 0 \\
0 & 0 & 0 & 0 \\
1 & 0 & 0 & 0  
\end{array}
\right) \;.
\ee
The $W_j$ are readily constructed. Since $n<3$ the filtration contains trivial 
parts, as discussed in \eqref{restricted_filtration}. More precisely, one has $W_6=W_5=W_4$ is 
the total space spanned by all $e_i$, while $W_1=W_0=W_{-1}$ are trivial. 
The non-trivial part of the filtration is
\be
W_3 = \text{span}(e_2,e_3,e_4)\ , \qquad 
W_2 = \text{span}(e_4)\ .
\ee
The $Gr_j$ are easily derived from these $W_j$. $Gr_4$ and $Gr_2$ are 
one-dimensional with non-trivial component along $e_4$ and $e_1$, respectively. 
$Gr_3$ is two-dimensional with non-trivial $e_2,e_3$-components, 
and all other $Gr_i$ are trivial. 
We can also construct the $\cP_{3+i}$  as the kernels of $N^{i+1}$ acting on $Gr_{3+i}$.
We find that $\cP_4$ is one-dimensional, while $\cP_3$ is two-dimensional, with all other 
$\cP_i$ being trivial. 
For $d=0$ we conclude that $a_0$ 
%is in $\cP_3$, i.e.~it 
has non-trivial entries along $e_2$ and $e_3$. Again one can further constrain the allowed $a_0$ as we discuss below. 

\subsubsection*{An example filtration for case II$_2$:}

In this case it is convenient to identify  
\beq \label{id_alpha}
   e_1 \cong \alpha_0\ ,\quad e_2 \cong \alpha_1\ ,\quad e_3 \cong \beta^0\ ,\quad e_4 \cong \beta^1\ ,
\eeq
such that the intersection matrix $\eta$ takes the form
\be
\eta = 
\left (
\begin{array}{cccc}
0 & 0 & 1 & 0 \\
0 & 0 & 0 & 1 \\
-1 & 0 & 0 & 0 \\
0 & -1 & 0 & 0  
\end{array}
\right) \ . \label{etacan2}
\ee
In this example we consider the monodromy $T$ and resulting matrix $N$ of the form
\be \label{exampleII2}
T = 
\left (
\begin{array}{cccc}
1 & 0 & 0 & 0 \\
0 & 1 & 0 & 0 \\
1 & 0 & 1 & 0 \\
0 & 1 & 0 & 1  
\end{array}
\right) \; , \qquad
N = 
\left (
\begin{array}{cccc}
0 & 0 & 0 & 0 \\
0 & 0 & 0 & 0 \\
1 & 0 & 0 & 0 \\
0 & 1 & 0 & 0  
\end{array}
\right) \;.
\ee
Hence, $N$ is of rank $1$ and obeys $N^2=0$ such that $n=1$. This $N$ can be embedded in $Sl_2$ as
\be
Z = 
\left (
\begin{array}{cccc}
1 & 0 & 0 & 0 \\
0 & 0 & 0 & 0 \\
0 & 0 & 0 & 0 \\
0 & 0 & 0 & -1  
\end{array}
\right) \;,\;\;
X_+ = 
\left (
\begin{array}{cccc}
0 & 0 & 1 & 0 \\
0 & 0 & 0 & 1 \\
0 & 0 & 0 & 0 \\
0 & 0 & 0 & 0  
\end{array}
\right) \;,\;\;
X_- = 
\left (
\begin{array}{cccc}
0 & 0 & 0 & 0 \\
0 & 0 & 0 & 0 \\
1 & 0 & 0 & 0 \\
0 & 1 & 0 & 0  
\end{array}
\right) \;.
\ee

Again we easily construct the $W_j$, by first noting that $n<3$ such that filtration contains trivial 
parts \eqref{restricted_filtration}. One has $W_6=W_5=W_4$ is 
the total space spanned by all $e_i$, while $W_1=W_0=W_{-1}$ are trivial. 
The non-trivial part of the filtration is
\be
W_3 = 
W_2 = \text{span}(e_3,e_4)\ .
\ee
The $Gr_j$ are easily derived from these $W_j$. $Gr_4$ and $Gr_2$ are 
two-dimensional with non-trivial component along $e_1,e_2$ and $e_3,e_4$, respectively. All other $Gr_i$ are trivial. 
We can also construct the $\cP_{3+i}$  as the kernels of $N^{i+1}$ acting on $Gr_{3+i}$.
We find that $\cP_4$ is one-dimensional, while $\cP_3$ is two-dimensional, with all other 
$\cP_i$ being trivial. For $d=0$ we conclude that $a_0$ 
%is in $\cP_3$, i.e.~it 
has non-trivial entries along $e_2$ and $e_3$. Again one can further constrain the allowed 
$a_0$ as we discuss below. 

\subsubsection*{Local classifications in the one-modulus case}

While the presented examples seem only to represent specific choices of monodromy matrices, they actually 
provide the key examples appearing in a classifications of allowed filtrations and vectors $\mathbf{a}_0$.  
Considering $4\times 4$ representations it is possible to get quite far without assuming a particular form for $T$ 
but only using general constraints. In particular, since the $Sl_2$ algebra is invariant under unitary transformations 
one can use this freedom to eliminate components in the most general matrices. 
Using this, and other constraints, an analysis of the one-modulus case was performed in full generality in \cite{GGK}. 

It was found that there are precisely three classes of $N$ for one modulus, of which \eqref{exampleI}, \eqref{exampleII1} and \eqref{exampleII2} are special examples. 
These are summarised in table \ref{classification-onemodulus}, where the first column gives $\eta$ and hence allows the match with the basis $(\alpha_K,\beta^K)$
define in \eqref{symplectic-basis}.

\begin{table}[h!]
\begin{tabular}{|c |c|c|c|c| c|} 
\hline
case & prop. $N$ & form of $\eta$ & form or $N$ & form of $\mathbf{a}_0$ & constants\\
\hline \hline
\rule[-1.1cm]{0cm}{2.4cm} I 
& $\begin{array}{c} N^4 = 0,\\ 
           N^3 \neq 0\phantom{.}
           \end{array}$ 
& $\left (\!\!
\begin{array}{cccc}
0 & 0 & 0 & 1 \\
0 & 0 & 1 & 0 \\
0 & -1 & 0 & 0 \\
-1 & 0 & 0 & 0  
\end{array}
\!\!
\right) $ &
$\left (\!\!
\begin{array}{cccc}
0 & 0 & 0 & 0 \\
a & 0 & 0 & 0 \\
e & b & 0 & 0 \\
f & e & -a & 0  
\end{array}\!\!
\right)$ 
&
$\left(\!\! \begin{array}{c} 1 \\ 0 \\ \frac{f}{2a}\\ \pi \end{array} \!\!\right)$
&
$\begin{array}{c}
a,b,f \in \mathbb{Z}, \\
a\neq 0 , b> 0\ , \\
e \in \mathbb{Z}[\frac{1}{2}],\\
%e+\frac{ab}{2}, \frac{a^2 b}{6} \in \mathbb{Z}, \\
\pi \in \mathbb{C}
\end{array}$
\\
\hline
\rule[-1.1cm]{0cm}{2.4cm} II$_1$ 
& 
$\begin{array}{c}
N^2=0, \\
\text{rk}(N)=1 
\end{array}$
&  $\left (\!\!
\begin{array}{cccc}
0 & 0 & 0 & 1 \\
0 & 0 & 1 & 0 \\
0 & -1 & 0 & 0 \\
-1 & 0 & 0 & 0  
\end{array}\!\!
\right) $
&
$ \left (\!\!
\begin{array}{cccc}
0 & 0 & 0 & 0 \\
0 & 0 & 0 & 0 \\
0 & 0 & 0 & 0 \\
a & 0 & 0 & 0  
\end{array} \!\!\right) 
$
&
$\left( \!\!\begin{array}{c} 0 \\ 1 \\ \tau \\ \delta - \tau \gamma \end{array} \!\!\right)$
&
$\begin{array}{c}
a \in \mathbb{Z}, 
a\neq 0 , \\
\tau \in \mathbb{C},\\
 \text{Im}\tau \neq 0,\\
\gamma, \delta \in \mathbb{R} 
\end{array}$
\\
\hline
\rule[-1.1cm]{0cm}{2.4cm} II$_2$ 
&  
$\begin{array}{c}  N^2=0, \\
\text{rk}(N)=2 
\end{array}$
&
$\left (\!\!
\begin{array}{cccc}
0 & 0 & 1 & 0 \\
0 & 0 & 0 & 1 \\
-1 & 0 & 0 & 0 \\
0 & -1 & 0 & 0  
\end{array}\!\!
\right) $
&
$\left (\!\!
\begin{array}{cccc}
0 & 0 & 0 & 0 \\
0 & 0 & 0 & 0 \\
a & 0 & 0 & 0 \\
0 & c & 0 & 0  
\end{array}\!\!
\right)$ 
&
$\left(\!\! \begin{array}{c} 1 \\ i \sqrt{\frac{a}{c}} \\ 0 \\ \gamma \end{array}\!\! \right)$
&
$\begin{array}{c}a\geq c > 0,\\
\gamma \in \mathbb{C} \end{array}$\\
\hline

\end{tabular}
\caption{Classification of occurring infinite order monodromies and vectors $\mathbf{a}_0$ for the one-modulus case  \cite{GGK}.} 
\label{classification-onemodulus}
\end{table}

\subsubsection*{Global classifications in the one-modulus case}

Having summarised the classification of monodromy about a local point in the one-parameter moduli space in table \ref{classification-onemodulus},
we now discuss some global aspects. 
In fact, it is possible to classify also the global structure of the moduli space for such one-parameter Calabi-Yau threefolds
that have a moduli space $\mathbb{P}^1/\left\{0,1,\infty \right\}$.
This was performed in \cite{DM} and here we summarize their results. The moduli space $\mathbb{P}^1/\left\{0,1,\infty \right\}$ is parameterised by $z$ and has three monodromy points. The monodromy about $z=0$ is maximally unipotent, so $n=3$, and is therefore a large-complex structure limit. The monodromy about $z=1$ is of rank 1 and is the conifold locus. 

The monodromy matrices about these loci satisfy the relation $T_{0}T_{1}T_{\infty}=\mathbb{1}$.  They can therefore be specified by any two elements, which take the form
\be
T_0 = 
\left (
\begin{array}{cccc}
1 & 0 & 0 & 0 \\
1 & 1 & 0 & 0 \\
0 & m & 1 & 0 \\
0 & 0 & 1 & 1  
\end{array}
\right) \;,\;\;
T_1 = 
\left (
\begin{array}{cccc}
1 & -a & -1 & -1 \\
0 & 1 & 0 & 0 \\
0 & 0 & 1 & 0 \\
0 & 0 & 0 & 1  
\end{array}
\right) \;.
\ee
The symplectic form for contractions in the basis where the monodromy matrices takes not the above form (\ref{etacan}) or \eqref{etacan2}. 
Rather, by an appropriate choice of basis, it is given by
\be
\eta = 
\left (
\begin{array}{cccc}
0 & -a & -1 & -1 \\
a & 0 & 1 & 0 \\
1 & -1 & 0 & 0 \\
1 & 0 & 0 & 0  
\end{array}
\right) \label{etanoncan} \;.
\ee

There are 14 possible cases which are labelled by the integer choices for $a$ and $m$ and are given in \cite{DM}, Table 1. Of these, there are 3 special cases where the monodromy about $z=\infty$ is such that $N_{\infty}$ is of rank 2 and so $n=d=1$. These are given by\footnote{Note that the $N_{\infty}$ for these cases are given by $N_{\infty}=\left(T_{\infty}\right)^p-\mathbb{1}$ where $p=4,6,3$ respectively.}
\be
\left\{m,a\right\} = \left\{4,4\right\},\;\left\{1,2\right\},\;\left\{9,6\right\}\;. \label{TDcases}
\ee
The Calabi-Yau realisations of these monodromy loci are, for example, 
the mirrors of $\mathbb{P}^5_{1,1,1,1,2,2}\left[4,4\right]$, $\mathbb{P}^5_{1,1,2,2,3,3}\left[6,6\right]$, $\mathbb{P}^5\left[3,3\right]$ respectively. These geometries 
have been analysed in various works, see e.g.~\cite{Walcher:2009uj}. %One shows that at these special $n=d=1$ points, known as Tyurin degenerations,  the Calabi-Yau splits into a union of two Fano threefolds. 
 All the cases in (\ref{TDcases}) are known as Tyurin degenerations. Such degenerations were studied in \cite{DHT}. 
 
 We conclude this section by stressing that indeed all cases I, II$_1$ and II$_2$ are realised geometrically. Hence, in 
 order to analyse the conjecture about infinite distance and the existence of light states we have to investigate the 
 two infinite distance cases I, II$_2$. This is the task of section \ref{sec:orbits}.

\bibliographystyle{jhep}
\bibliography{refs-swamp}

\end{document}